\documentclass[a4paper,11pt]{article}
\usepackage{graphicx} 
\pdfoutput=1 
\usepackage{subcaption}
\usepackage{xcolor}
\usepackage{jcappub} 
                     
\usepackage{hanging}
\usepackage{orcidlink}
\usepackage[T1]{fontenc} 
\usepackage{multirow} 

\newcommand{\bq}{\boldsymbol q}
\newcommand{\bx}{\boldsymbol x}
\newcommand{\bk}{\boldsymbol{k}}
\newcommand{\bPsi}{\boldsymbol{\Psi}}

\newcommand{\ihmpc}{\,h{\rm Mpc}^{-1}}

\usepackage{bm}

\newcommand{\velocileptors}{\texttt{velocileptors}}
\newcommand{\folps}{\texttt{Folps$\nu$}}
\newcommand{\pybird}{\texttt{PyBird}}

\title{A comparison of effective field theory models of redshift space galaxy power spectra for DESI 2024 and future surveys}

\author[1,2]{{M.~Maus}\orcidlink{0000-0002-9020-911X},}
\emailAdd{mark.maus@berkeley.edu}
\author[3]{{Y.~Lai},}
\emailAdd{y.lai1@uq.net.au}
\author[4,5, 10]{{H.~E.~Noriega}\orcidlink{0000-0002-3397-3998},}
\emailAdd{henoriega@icf.unam.mx}
\author[5]{{S.~Ramirez-Solano},}
\emailAdd{sadiramirez@estudiantes.fisica.unam.mx}
\author[6,4]{{A.~Aviles}\orcidlink{0000-0001-5998-3986},}
\author[7]{{S.~Chen}\orcidlink{0000-0002-5762-6405},}
\author[4]{{S.~Fromenteau},}
\author[8,9,10]{{H.~Gil-Mar\'in}\orcidlink{0000-0003-0265-6217},}
\author[3]{{C.~Howlett}\orcidlink{0000-0002-1081-9410},}
\author[5]{{M.~Vargas-Maga\~na}\orcidlink{0000-0003-3841-1836},}
\author[1,2]{{M.~White}\orcidlink{0000-0001-9912-5070},}
\author[11]{{P.~Zarrouk}\orcidlink{0000-0002-7305-9578},}
\author[2]{{J.~Aguilar},}
\author[13]{{S.~Ahlen}\orcidlink{0000-0001-6098-7247},}
\author[14]{{O.~Alves},}
\author[15]{{S.~Brieden}\orcidlink{0000-0003-3896-9215},}
\author[16]{{D.~Brooks},}
\author[17]{{E.~Burtin},}
\author[2]{{T.~Claybaugh},}
\author[18]{{S.~Cole}\orcidlink{0000-0002-5954-7903},}
\author[19]{{K.~Dawson},}
\author[20]{{M.~Icaza-Lizaola}\orcidlink{0000-0002-2547-3184},}
\author[5]{{A.~de la Macorra}\orcidlink{0000-0002-1769-1640},}
\author[17]{{A.~de~Mattia},}
\author[16]{{P.~Doel},}
\author[2,21]{{S.~Ferraro}\orcidlink{0000-0003-4992-7854},}
\author[22]{{N.~Findlay}\orcidlink{0009-0007-0716-3477},}
\author[23,24]{{J.~E.~Forero-Romero}\orcidlink{0000-0002-2890-3725},}
\author[9,22,25]{{E.~Gaztañaga},}
\author[2]{{S.~Gontcho A Gontcho}\orcidlink{0000-0003-3142-233X},}
\author[26]{{C.~Hahn}\orcidlink{0000-0003-1197-0902},}
\author[27,28,29]{{K.~Honscheid},}
\author[30]{{M.~Ishak}\orcidlink{0000-0002-6024-466X},}
\author[2]{{A.~Kremin}\orcidlink{0000-0001-6356-7424},}
\author[2]{{M.~Landriau}\orcidlink{0000-0003-1838-8528},}
\author[11]{{L.~Le~Guillou}\orcidlink{0000-0001-7178-8868},}
\author[31,32]{{M.~Manera}\orcidlink{0000-0003-4962-8934},}
\author[33,32]{{R.~Miquel},}
\author[34]{{E.~Mueller},}
\author[22]{{S.~Nadathur}\orcidlink{0000-0001-9070-3102},}
\author[35,6]{{G.~Niz}\orcidlink{0000-0002-1544-8946},}
\author[17,2]{{N.~Palanque-Delabrouille}\orcidlink{0000-0003-3188-784X},}
\author[36,37,38]{{W.~J.~Percival}\orcidlink{0000-0002-0644-5727},}
\author[2,39,21]{{C.~Poppett},}
\author[40]{{F.~Prada}\orcidlink{0000-0001-7145-8674},}
\author[41]{{M.~Rezaie}\orcidlink{0000-0001-5589-7116},}
\author[42,17]{{A.~Rocher}\orcidlink{0000-0003-4349-6424},}
\author[43]{{G.~Rossi},}
\author[44]{{E.~Sanchez}\orcidlink{0000-0002-9646-8198},}
\author[2]{{D.~Schlegel},}
\author[45,14]{{M.~Schubnell},}
\author[46]{{D.~Sprayberry},}
\author[14]{{G.~Tarl\'{e}}\orcidlink{0000-0003-1704-0781},}
\author[47]{{S.~Yuan}\orcidlink{0000-0002-5992-7586},}
\author[22,48]{{R.~Zhao}\orcidlink{0000-0002-7284-7265},}
\author[2]{{R.~Zhou}\orcidlink{0000-0001-5381-4372},}
\author[48]{{H.~Zou}\orcidlink{0000-0002-6684-3997},}
\affiliation{Affiliations are in Appendix \ref{sec:affiliations}}

\abstract{In preparation for the next generation of galaxy redshift surveys, and in particular the year-one data release from the Dark Energy Spectroscopic Instrument (DESI), we investigate the consistency of a variety of effective field theory models that describe the galaxy-galaxy power spectra in redshift space into the quasi-linear regime using 1-loop perturbation theory. These models are employed in the pipelines \texttt{velocileptors}, \texttt{PyBird}, and \texttt{Folps$\nu$}.  While these models have been validated independently, a detailed comparison with consistent choices has not been attempted. After briefly discussing the theoretical differences between the models we describe how to provide a more apples-to-apples comparison between them. We present the results of fitting mock spectra from the \texttt{AbacusSummit} suite of N-body simulations provided in three redshift bins to mimic the types of dark time tracers targeted by the DESI survey. We show that the theories behave similarly and give consistent constraints in both the forward-modeling and ShapeFit compressed fitting approaches. We additionally generate (noiseless) synthetic data from each pipeline to be fit by the others, varying the scale cuts in order to show that the models agree within the range of scales for which we expect 1-loop perturbation theory to be applicable. This work lays the foundation of Full-Shape analysis with DESI Y1 galaxy samples where in the tests we performed, we found no systematic error associated with the modeling of the galaxy redshift space power spectrum for this volume.}

\begin{document}
\maketitle
\flushbottom
\newpage

\section{Introduction}
\label{sec:intro}


The evolution of structures in the Universe is a powerful observational tool that holds a wealth of information about its initial conditions and expansion history. Large Scale Structure (LSS) can be probed with spectroscopic surveys that use galaxies, quasars, the Ly$\alpha$ forest, etc.\ as tracers of the underlying matter distributions. The completed Sloan Digital Sky Survey (SDSS; \cite{SDSS2000}) consisting of the Baryon Oscillation Spectroscopic Survey (BOSS; \cite{Dawson13,Reid16,Alam17,SDSSIII2015}) and the SDSS-IV extended Baryon Oscillation Spectroscopic Survey (eBOSS; \cite{eBOSS:2020yzd,Bautista18}) is the largest, public galaxy redshift survey to date.  However, it has only scanned a small fraction of the observable Universe. Therefore, much of the information on fundamental physics and the Universe's history we can potentially gain from LSS observations lies ahead of us with upcoming surveys such as the Dark Energy Spectroscopic Instrument (DESI; \cite{DESI,DESI2016_2,DESI2022,DESI2023,DESI2023_2}), Euclid \cite{Euclid}, and beyond \cite{Schlegel22}. 

In parallel with the design of new instruments is also the development of more sophisticated analysis pipelines to interpret the increasingly more constraining data. In galaxy clustering analyses, the most common approach is to employ methods based on cosmological perturbation theory.  Such theories have a history dating back decades (see e.g.\ ref.~\cite{Bernardeau02} for an early review and references to the original literature) but have seen continual development to the present day (see e.g.\ ref.~\cite{Ivanov_EFT22} for a recent review).  Modern perturbation theories include contributions from non-linear dynamics \cite{Bernardeau02}, broadening of features by long-wavelength fields\footnote{This frequently goes under the name ``IR resummation'', because it involves summing the contributions of long-wavelength perturbations to higher order than other terms.  A comparison of the Eulerian and Lagrangian approaches can be found in Appendix A of ref.~\cite{Vlah15}, and the close numerical correspondence of the approaches can be found in Fig.~3 of ref.~\cite{ChenVlahWhite20b}.} \cite{Bharadwaj96,Meiksin99,ESW07,Smith08,Mat08a}, redshift-space distortions \cite{Kaiser87,Hamilton98}, galaxy bias \cite{Kaiser84,Desjacques16} and `counterterms' to mitigate sensitivity to small-scale physics \cite{Ivanov_EFT22}.  Such models can be formulated in either a Lagrangian or Eulerian framework and are able to self-consistently describe a wide range of cosmological observables. 

There are four main versions of these perturbative models that have been implemented in efficient codes and are typically applied to data.  These will be introduced in some detail below.  In principle, these codes all agree on a `unique' prediction for the low-order statistics of biased tracers in redshift space, for both configuration-space and Fourier-space statistics.  However, in practice, these models are normally applied to data or simulations with different datasets or assumptions about window functions, nuisance parameters, fitting range, priors and other details.  This makes understanding the degree of agreement between the models difficult, and can erroneously give the impression that the models are inconsistent in some way.  In order to dispel this misconception, and in anticipation of the upcoming Y1 data release from DESI \cite{DESI2023b.KP1.EDR,DESI2024.I.DR1,DESI2024.II.KP3,DESI2024.III.KP4,DESI2024.V.KP5,DESI2024.IV.KP6,DESI2024.VI.KP7A,DESI2024.VII.KP7B,DESI2024.VIII.KP7C} for which these models will be employed, it is important to understand how differences in constraints between modeling pipelines are driven by internal priors and to show that the different codes perform similarly under the same assumptions. The purpose of this paper is to compare the constraints on $\Lambda$CDM and compressed parameters generated by three Fourier-space models that will be used on upcoming DESI data analyses, and show that under consistent assumptions these models do in fact agree in their constraints. This issue has been partially investigated for the \pybird, \texttt{class-pt}, and \texttt{CLASS-OneLoop} models in Refs.~\cite{Simon23,Linde24}. However, those models are based on the same theory and differ only in their prior choices. The models we compare here (\velocileptors, \folps, and \pybird) additionally differ in their theoretical frameworks and resummation schemes. We also compare the behavior of our models within two different fitting techniques, where one (``Full-modeling'') directly varies the parameters within a given cosmological model (e.g. $\Lambda$CDM) while the other (``ShapeFit'') uses a single reference linear power spectrum and then varies compressed parameters that control the shape information and late-universe dynamics. The comparison in this paper focuses on Fourier space (power spectrum) models, but a similar comparison is underway that compares the predictions of these models, transformed to configuration space, to those of the purely configuration space model \texttt{EFT-GSM} \cite{KP5s5-Ramirez}. 

The outline of the paper is as follows.
In \S\ref{sec:data} we describe the mock galaxy catalogs that are used in the bulk of this work to test the agreement of the models on complex datasets for which the `true' cosmology is known.
In \S\ref{sec:theory} we first give an overview of the Eulerian and Lagrangian Perturbation Theory (EPT and LPT) formalisms before introducing each model under consideration. We note that detailed descriptions of these models can be found in their respective individual papers and only a brief summary of their relevant features are discussed here. We then describe two parametrization choices (minimal and maximal freedom) for the bias terms that we will employ in our comparisons and then conclude the section with a short summary of the key differences between the models. Then, in \S\ref{sec:methods} we briefly describe the Full-modeling and ShapeFit fitting methods. In \S\ref{sec:results}, we present results both on Abacus simulations data as well as on noiseless theoretical mock power spectra created by the models themselves.
Finally, we present our conclusions in \S\ref{sec:conclusions}.

\section{Data}
\label{sec:data}

In this paper, we present a comparison between theories using mock data generated from N-body simulations and from the theories themselves.  The N-body data represent plausible and complex galaxy populations that the models should all be able to fit with the advantage of knowing the `true' cosmology from which they were drawn.  Data generated from the theories themselves is essentially noiseless and thus allows us to compare the theories in great detail.

The galaxy clustering simulations we use are from the \texttt{AbacusSummit} \cite{Maksimova21} suite, which were created on the \texttt{Summit} supercomputer, located at the Oak Ridge Leadership Computing Facility, using the \texttt{Abacus} \cite{Garrison21} N-body code.  We focus on the 25 realizations with a fixed cosmology, each in a cubic box geometry of 8 [$h^{-3}$Gpc$^3$] in volume.  In order to reduce sample variance `noise' in the power spectra we average over the 25 realizations and work with the mean $P_\ell(k)$, corresponding to a volume of 200 [$h^{-3}$Gpc$^3$]. In this paper we focus on HOD samples calibrated to reproduce the power spectra of the luminous red galaxies (LRG) at a redshift of $z=0.8$, emission line galaxies (ELGs) at $z=1.1$, and QSOs at $z=1.4$ \cite{ELGdesi,LRGQSOdesi}. The mock data power spectra are shown in the left panels of Fig.~\ref{fig:mock_data} for each of the three tracers. We only use the monopole ($P_0(k)$) and quadrupole ($P_2(k)$) data, as we only focus our comparison within the standard $\Lambda$CDM model for which the hexadecapole does not significantly influence constraints\cite{KP5s2-Maus,KP5s3-Noriega,KP5s4-Lai}. In the right panels we show comparisons of predictions from each of the three EFT models with $\Lambda$CDM parameters fixed to the true input values of the mocks. These will be discussed further in \S\ref{sec: results_theory}

In our fits, we assume a Gaussian likelihood, and so we need an estimate of the covariance matrix.  Our covariance matrix is obtained by Monte-Carlo using 1000 ``effective Zel'dovich approximation'' mocks (\texttt{EZmocks}; \cite{Chuang2015}). Since it is numerically computed from mocks, we do not assume the covariance to be Gaussian and thus there are nonzero off-diagonal terms. In principle we can also compute analytic covariances for these samples following the methods of Ref.~\cite{Wadekar2020}, where assumptions about Gaussianity are relevant, but we do not do so here.  
As we are fitting the combined volume of 25 Abacus realizations, the covariance matrix for a single-box volume can be rescaled by a factor of up to 1/25 before we expect to see statistically significant shifts in the parameters just from fluctuations in the (mean) data vector.  While the implied statistical precision in this case exceeds the convergence that has been demonstrated for the N-body simulations themselves \cite{Angulo22,Grove22}, we currently see no evidence of a bias that projects into the subspace of the cosmological parameters. We discuss results from fitting data with both rescaled and unrescaled covariance, and specify which is used with the labels $V_{25}$(200 [$h^{-3}$Gpc$^3$]) and $V_1$(8 [$h^{-3}$Gpc$^3$]) respectively.

We also employ the theoretical models themselves as noiseless ``data'', to be fit by the other modeling codes. This can aid in analyzing the behavior of bias, stochastic, and counterterms of each model when the true values are known. In addition, we have more freedom in scaling up/down the covariance rather than being limited by the number of mocks and possible inaccuracies in the simulations.

\section{Theory and Models}
\label{sec:theory}

\subsection{Overview of EPT and LPT}

As stated in the introduction, cosmological perturbation theory comes in two flavors, Eulerian (EPT) and Lagrangian (LPT). In the Eulerian formulation, the cold dark matter and baryons are treated as a perfect, pressure-less fluid which obeys the continuity, Euler and Poisson equations. Defining the overdensity, $\delta = \rho/\bar{\rho} -1$, and velocity, $\mathbf{v}$, fields these equations become \cite{Bernardeau02,Ivanov_EFT22,Dodelson03}:
\begin{align}
    \partial_{\tau}\delta + \nabla\cdot[(1+\delta)\boldsymbol{v}] &= 0 \\
    \partial_{\tau}\boldsymbol{v}+\mathcal{H}\boldsymbol{v}+ \boldsymbol{v}\cdot\nabla\boldsymbol{v} &= -\nabla\Phi \\
    \nabla^2\Phi &= \frac{3}{2}\mathcal{H}^2\delta,
\end{align}
where $\mathcal{H}$ is the conformal Hubble parameter. These equations can be solved perturbatively by expanding $\delta$ and the velocity divergence, $\theta = \nabla\cdot\boldsymbol{v}$, in powers of their lowest-order (linear) solutions:
\begin{align}
    \delta(\bk,z) = \sum_{n=1}^{\infty}a^n\delta^{(n)}(\bk), 
    \quad , \quad
    \theta(\bk,z) = \mathcal{H}\sum_{n=1}^{\infty}a^n\theta^{(n)}(\bk),
\end{align}
in which
\begin{align}
    \delta^{(n)}(\bk) = \int\frac{d^3\bk_1\cdots d^3\bk_n}{(2\pi)^{3n}}(2\pi)^3\delta^D(\sum\bk_i -\bk)\delta^{(1)}(\bk_1)\cdots\delta^{(1)}(\bk_n)F_n(\bk_1,...,\bk_n,\bk),
\end{align}
and similarly for $\theta^{(n)}$.  In these equations, we have followed the standard practice of approximating the time-dependence of each term as $a^n$ and factoring this out of the $\delta^{(n)}$ and $\theta^{(n)}$.  This will be an excellent approximation for our purposes \cite{Donath+:2020,Fasiello+:2016}.

By contrast, Lagrangian Perturbation Theory treats cold dark matter as collisionless particles whose observed (Eulerian) coordinates, $\bx$, are related to their initial (Lagrangian) positions, $\bq$, via the displacement field such that $\bx(\bq)= \bq + \bPsi(\bq)$. Nonlinear dynamics is then governed by a second order differential equation in the displacement field: 
\begin{align}
    \partial_\tau^2\bPsi + \mathcal{H}\partial_\tau\bPsi = -\nabla_{\bx}\Phi(\bq+\bPsi(\bq))
    \quad . 
\end{align}
where $\tau$ is the conformal time.  The matter overdensity is then derived from number conservation as
\begin{align}
    1+\delta(\bx) = \int d^3\bq\ \delta^{\rm D}(\bx-\bq-\bPsi)
    \quad\Rightarrow\quad
    1+\delta(\bk) = \int d^3\bq\ e^{i\bk \cdot (\bq + \Psi(\bq))}.
    \label{eq: m_cons}
\end{align}
Expanding the above exponential as a Taylor series, as well as expanding the displacement field $\bPsi = \bPsi^{(1)}+\bPsi^{(2)}+\bPsi^{(3)}+\cdots$ gives expressions for $\delta^{(n)}(\bk)$ that agree with the Eulerian case at each order. That the Eulerian and Lagrangian approaches agree to all orders in perturbation theory, while the behavior of the two systems that they model disagree\footnote{In a collisional fluid when two fluid elements cross a shock converts kinetic energy into thermal energy.  In a collisionless system of particles, the crossing of trajectories implies multi-streaming, and the velocity field becomes multivalued at a given $\mathbf{x}$.} when trajectories of fluid elements or particles cross, is a hint that we will need to include beyond-PT contributions to our model \cite{McQWhi16}.

Since the galaxy redshift surveys target galaxies as tracers of the matter density field we need a `bias model' to connect the two \cite{Desjacques16}.  We follow standard practice and express the galaxy overdensity as a functional of the matter field that can include all combinations allowed by the symmetries.  Specifically, we use a Taylor expansion, with bias coefficients that are free parameters when modeling data. In the Eulerian prescription, the galaxy biasing is a function of the non-linear matter field at $\mathbf{x}$:
\begin{align}
    \delta_{\rm g}(\bx) = c_1\delta_{\rm m}(\bx) + c_2(\delta_{\rm m}^2(\bx) - \left\langle\delta_{\rm m}^2(\bx)\right\rangle) + c_{\rm s}(s^2(\bx) - \left\langle s^2(\bx)\right\rangle) + ...
\end{align}
The Lagrangian case is expanded in the initial density field at Lagrangian coordinate $\bq$:
\begin{align}
    1+\delta_{\rm g}(\bk) = \int d^3\bq\ e^{i\bk \cdot (\bq + \Psi)} F[\delta^{(1)}(\bq)],
    \label{eq:Ldelta}
\end{align}
with the bias functional $F[\delta^{(1)}(\bq)]$, written as $F(\bq)$ for convenience, given by
\begin{align}
    F(\bq) = 1 +  b_1\delta_0 + \frac{1}{2}b_2(\delta_0(\bq)^2 - \left\langle\delta_0^2\right\rangle)+b_s(s_0^2(\bq) - \left\langle s^2\right\rangle).
    \label{eq:Lbias}
\end{align}

Finally, the ``effective'' field theories employed in the models include additive corrections that correspond to the sensitivity to small-scale physics that can not be described perturbatively. Couplings of short wavelength modes are included in the form of counterterms ($\sim k^2P_{\rm lin}$) and stochastic terms within each model. 

With these preliminaries understood, we now turn to the different models that we shall compare in this paper.


\subsection{Velocileptors}

\begin{table}[t!]
\centering                          
\begin{tabular}{c|c|cc|c}        
Full-Modeling & ShapeFit & \multicolumn{2}{c|}{LPT Bias}  & Stoch/Counter \\ 
              &          & Min. F. & Max. F.         &               \\ \hline \hline
H$_0$        & $f\sigma_8$ & \multicolumn{2}{c|}{$(1+b_1)\sigma_8$} & $\alpha_0$ \\
$\mathcal{U}[55,79]$ & $\mathcal{U}[0,2]$ & \multicolumn{2}{c|}{$\mathcal{U}[0.5,3.0]$} & --- \\
\hline
$\omega_{\mathrm{b}}$ &  $\alpha_{\parallel}$ & \multicolumn{2}{c|}{$b_2$} & $\alpha_2$ \\
$\mathcal{N}[0.02237,0.00037]$ & $\mathcal{U}[0.5,1.5]$ & $\mathcal{N}[0,10]$ & $\mathcal{U}[-15,15]$ & --- \\
\hline
$\omega_{\mathrm{cdm}}$ &  $\alpha_{\perp}$ & \multicolumn{2}{c|}{$b_s$} & SN${}_0$ \\
$\mathcal{U}[0.08,0.16]$ & $\mathcal{U}[0.5,1.5]$ & 0 & $\mathcal{U}[-15,15]$ & --- \\
\hline
$\log(10^{10} A_\mathrm{s})$& $m$ & \multicolumn{2}{c|}{$b_3$} & SN${}_2$ \\
$\mathcal{U}[2.03,4.03]$ & $\mathcal{U}[-3.0,3.0]$& 0 & $\mathcal{U}[-15,15]$ & --- \\
\end{tabular}
\caption{Velocileptors LPT priors on parameters used in the Full-Modeling ($\Lambda$CDM) and ShapeFit fitting methods. The $\Lambda$CDM model involves H$_0$, $\Omega_{\mathrm{b}}$,$\omega_{\mathrm{cdm}}$, $\log(10^{10} A_\mathrm{s})$ and all of the bias, stochastic, and counterterms. The ShapeFit method fits $f\sigma_8$, $\alpha_{\parallel}$, $\alpha_{\perp}$, $m$ as well as the same bias, stochastic and counterterms.  The entries  $\mathcal{U}[{\rm min,max}]$ and $\mathcal{N}[\mu,\sigma]$ refer to uniform and Gaussian normal distributions, respectively. For the bias terms we show both minimal and maximal freedom cases, defined in \S~\ref{sec: bias}. }    
\label{tab: Vel_priors} 
\end{table}

The first model we consider is the fully resummed, 1-loop, LPT pipeline implemented in \velocileptors\footnote{\href{https://github.com/sfschen/velocileptors/}{github.com/sfschen/velocileptors}} \cite{Chen20,Chen21}. The redshift space power spectrum is computed using the biasing scheme of eq.~\ref{eq:Lbias} and the overdensity of eq.~\ref{eq:Ldelta} such that:
\begin{align}
    P_{\rm s}^{\rm LPT}(\bk) = \int d^3\bq \left\langle e^{i\bk \cdot (\bq + \Delta_{\rm s})}F(\bq_1)F(\bq_2) \right\rangle_{\bq = \bq_1-\bq_2},
    \label{eq:VPint}
\end{align}
where $\Delta_{\rm s} = \Psi_{\rm s}(\bq_1) - \Psi_s(\bq_2)$ and the ``s'' subscripts denote the redshift space displacement fields.  We use the ``Full LPT'' model in \velocileptors\ which 
resums both the long-wavelength displacement \textit{and} velocity contributions.

The integrals in eq.~\ref{eq:VPint} are done using FFTs (see Appendix D of \cite{Chen20}).  The model also includes stochastic contributions and counter terms. These terms enter the power spectrum with $\sigma_n$ and $\alpha_n$ coefficients, respectively, such that the final LPT power spectrum is:
\begin{align}
    P_{\rm s}(\bk) = P_{\rm s}^{\rm LPT}(\bk) + k^2(\alpha_0 + \alpha_2\mu^2 + \alpha_4\mu^4)P_{\rm s, Zel}(\bk) + R_{\rm h}^3(1 + \sigma_2k^2\mu^2+\sigma_4k^4\mu^4),
\end{align}
where $P_{\rm s,Zel}(\bk)$ refers to the linear Zeldovich approximation and $\mu$ is the cosine of the angle between $\bk$ and the line-of-sight (assumed fixed). For the stochastic contributions we use the parametrization $\text{SN}_0 = R_{\rm h}^3$, $\text{SN}_2 = R_{\rm h}^3\sigma_2$, and $\text{SN}_4 = R_h^3\sigma_4$ where $R_{\rm h}$ is a characteristic scale of halo formation. In Table~\ref{tab: Vel_priors} we report the priors used on parameters in our \velocileptors\ model that enter the monopole and quadrupole components of the power spectra. 
We note that when excluding the hexadecapole from our analysis, we set $\alpha_4 = \text{SN}_4 = 0$.

\velocileptors\ also includes a model based on EPT, with the terms collected in a way informed by LPT but with resummation handled in a different way to the LPT module.  This makes it an interesting counterpoint to the LPT model and the ``natively EPT'' models we discuss below. This EPT model is constructed from the LPT expressions but with the IR scale, $k_{\rm IR} \rightarrow 0$ and bias parameters remapped to the Eulerian prescription of ref.~\cite{McDonald_2009}. The IR resummation scheme then involves applying a damping factor to the BAO signal of the power spectrum prediction by separating it from the smooth part via a wiggle/no-wiggle split using the sine transform method of Refs.~\cite{Hamann2010,Wallisch2018}.


\subsection{PyBird} 

The next model is based on 1-loop Eulerian perturbation theory.
\pybird\footnote{\href{https://github.com/pierrexyz/pybird}{github.com/pierrexyz/pybird}}\ (The Python code for Biased tracers in redshift space) employs an effective field theory formalism that is very similar to \texttt{CLASS-PT} \cite{ClassPT_2020,Chudaykin:2020ghx}, with the two models' galaxy bias and counterterm parameterizations being related by simple linear transformations (see Ref.~\cite{Simon23} for details). The derivation for the EFT in \pybird\ can be found in ref.~\cite{Perko2016,d_Amico_2020}, so we will not repeat it here. The redshift-space galaxy power spectrum up to one-loop order is given by \cite{d_Amico_2020, Perko2016}:
\begin{align}
    P_{\rm g}(k, \mu) &= Z_1(\mu)^2 P_{\mathrm{lin}}(k) + 2 \int \frac{d^3 q}{(2\pi)^3} Z_2 (\boldsymbol{q}, \boldsymbol{k-q}, \mu)^2 P_{\mathrm{lin}}(|\boldsymbol{k-q}|)P_{\mathrm{lin}}(q) \nonumber \\
    & + 6 Z_1(\mu)P_{\mathrm{lin}}(k) \int \frac{d^3 q}{(2\pi)^3} Z_3(\boldsymbol{q}, \boldsymbol{-q}, \boldsymbol{k}, \mu) P_{\mathrm{lin}}(q) + 2 Z_1(\mu)P_{\mathrm{lin}}(k)\times\nonumber \\
    &\times\left(c_{\rm ct}\frac{k^2}{k_{\rm M}^2} + c_{\rm r,1} \mu^2 \frac{k^2}{k_{\rm M}^2} + c_{\rm r,2} \mu^4 \frac{k^2}{k_{\rm M}^2}\right) + \frac{1}{\overline{n}_{\rm g}}\left(c_{\epsilon, 1} + c_{\epsilon, 2}\frac{k^2}{k_{\rm M}^2} + c_{\epsilon, 3} f\mu^2 \frac{k^2}{k_{\rm M}^2}\right)
    \label{eq: GPS}
\end{align}
where \(P_{\mathrm{lin}}\) is the linear power spectrum, \(c_{\rm ct}, c_{\rm r, 1}, c_{\rm r, 2}\) are the counterterms, \(c_{\epsilon, 1}, c_{\epsilon, 2}, c_{\epsilon, 3}\) are the stochastic terms, \(f\) is the linear growth rate of structure, \(\overline{n}_{\rm g}\) is the number density of galaxies in the survey, and the redshift space kernels are given in eq.~(2.2) of ref.~\cite{d_Amico_2020}, following \cite{Bernardeau02,Scoccimarro99}. By default the scale that suppresses higher order derivatives in the bias expansion is set to $k_{\rm M} = 0.7\,h\, \mathrm{Mpc}^{-1}$. In addition to the counter terms and stochastic terms, \pybird\ has 4 bias parameters: \(b_1, b_2, b_3, b_4\). This is the bias parameterization described in Ref.~\cite{Perko2016}, often referred to as the ‘West Coast’ (WC) parameterization in the literature \cite{Nishimichi20,Holm23,Simon23}. For the stochastic terms, the main difference between \pybird\ and \velocileptors\ is that \pybird\ has an additional \(k^2/k_{\rm M}^2\) term but does not include \(k^4 \mu^4\). For the counter terms, \velocileptors\ uses the Zeldovich approximation while \pybird\ uses the redshift space linear power spectrum \(Z_1(\mu) P_{\mathrm{lin}}(k)\). When the difference between the two power spectra is small, \(c_{\rm ct} = \alpha_0, c_{\rm r, 1} = \alpha_2\ \mathrm{and}\ c_{\rm r, 2} = \alpha_4\).  For later use we define \(c_{\epsilon, \mathrm{mono}} = c_{\epsilon, 1} + (2f/3) c_{\epsilon, 2}\), and \(c_{\epsilon, \mathrm{quad}} = (4f/15) c_{\epsilon, 3}.\).

\begin{table}[t!]
\centering                          
\begin{tabular}{c|c|c|c}        
Full-Modeling & ShapeFit & Bias & Stoch/Counter \\ \hline \hline
H$_0$ &  $f\sigma_8$ & $b_1$ & $c_{\rm ct}$ \\
$\mathcal{U}[55.36,79.36]$ & $\mathcal{U}[0,1]$ & $\mathcal{U}[0.0,4.0]$ & --- \\
\hline
$\omega_{\mathrm{b}}$ &  $\alpha_{\parallel}$ & $b_2$ & $c_{\rm r, 1}$ \\
$\mathcal{N}[0.02237,0.00037]$ & $\mathcal{U}[0.9,1.1]$ & $1$ & --- \\
\hline
$\omega_{\mathrm{cdm}}$ &  $\alpha_{\perp}$ & $b_3$ & $c_{\rm r,2}$ \\
$\mathcal{U}[0.08,0.16]$ & $\mathcal{U}[0.9,1.1]$ & $1$ & $0$ \\
\hline
$\log(10^{10} A_\mathrm{s})$& $m$ &$b_4$ & $c_{\epsilon, 1}$ \\
$\mathcal{U}[2.0364,4.0364]$ & $\mathcal{U}[-1.0,1.0]$& $\mathcal{U}[-15.0,15.0]$ & --- \\
\hline
$w$& & & $c_{\epsilon, \mathrm{mono}}$ \\
$\mathcal{U}[-1.3, -0.7]$ & & & --- \\
\hline
& & & $c_{\epsilon, \mathrm{quad}}$ \\
& & & ---
\end{tabular}
\caption{The priors of Shapefit and Full-Model parameters for \pybird\. We use \(\mathcal{U}\) (\(\mathcal{N}\)) to denote the uniform (Gaussian) prior. The first number is the lower bound (mean), and the second number is the upper bound (standard deviation). This configuration is the minimum freedom (``MinF") configuration. We use this configuration to match with other pipelines in DESI. The value for \(b_2, b_3\), and \(b_4\) are obtained through the co-evolution relation. The detailed derivation is given in section \ref{sec: bias}. For the maximum freedom configuration, we put an infinite flat prior on \(b_3\) and a $\mathcal{U}[-10, 10]$ prior on \(b_3\) and \(b_4\). }    
\label{tab: pybird_priors} 
\end{table}

\subsection{FOLPS$\nu$}

Our final EPT model, \folps\footnote{\href{https://github.com/henoriega/FOLPS-nu}{https://github.com/henoriega/FOLPS-nu}. With JAX: \href{https://github.com/cosmodesi/folpsax}{https://github.com/cosmodesi/folpsax}.} \cite{Noriega:2022nhf}, is a Python code based on a perturbative model, with the added value of accounting for the presence of massive neutrinos \cite{Aviles:2020cax, Aviles:2021que,Noriega:2022nhf}.  As for the previous models, \folps\ is based on EPT up to one-loop. The redshift space power spectrum is given by:

%
\begin{align}\label{eq:P_EFT}
P^\text{EFT}_{\rm s}(k,\mu) &= P_{\delta\delta}(k) + 2 f_0 \mu^2 P_{\delta\theta}(k) + f_0^2 \mu^4 P_{\theta\theta}(k) + A^{\rm TNS}(k,\mu) + D(k,\mu) \nonumber\\
&\quad + (\alpha_0 + \alpha_2 \mu^2 + \alpha_4 \mu^4 ) k^2 P_{\rm lin}(k)  \nonumber\\
&\quad + P_{\rm shot} \big[\alpha^{\rm shot}_0 + \alpha^{\rm shot}_{2} (k\mu)^2 \big].
\end{align}
where each of these terms is defined in detail in ref.~\cite{Noriega:2022nhf}. A key feature of \folps\ is that it employs beyond-EdS kernels (called \texttt{fk}-kernels \cite{Rodriguez-Meza:2023rga,Aviles:2021que}) to properly account for the effects of the free-streaming scale introduced by massive neutrinos rather than making the $\omega_\nu\ll \omega_{\rm cdm}$ approximation made in the other codes.
%
%
%
%
%
The first line on the right hand side of eq.~(\ref{eq:P_EFT}) gives the perturbative, non-linear, redshift-space power spectrum as for the other models.  The second and third lines are the counterterms and stochastic terms, respectively.
We employ the biasing scheme first introduced in \cite{McDonald_2009, McDonald:2006mx}, and generalized to cosmologies with additional scales (such as the neutrino free streaming scale) in \cite{Aviles:2020cax}.  For simplicity, we adopt the approximate bias treatment outlined in Appendix A.1 in ref.~\cite{Aviles:2020wme}. 

\folps\ also accounts for large scale bulk flows via the IR-resummations technique of refs.~\cite{Senatore:2014via,Ivanov:2018gjr}, and computes the multipoles using the IR-resummed EFT power spectrum, as specified in equations~(2.20) and~(2.24) in Ref.~\cite{KP5s3-Noriega}.

In this work, we keep the total mass of neutrinos fixed at its fiducial value $M_\nu = 0.06\, \text{eV}$, for which differences in results between EdS- and \texttt{fk}-kernels are small. \folps\ demonstrates its full potential in the neutrino mass constraints, as explored in ref. \cite{KP5s3-Noriega} using synthetic data mimicking DESI Year-1 and Year-5 tracers and error bars, revealing a significant enhancement in the neutrino mass constraints of around 14\% when switching from the usual EdS kernels to the accurate \texttt{fk}-kernels.

\begin{table}[t!]
\centering                          
\begin{tabular}{c|c|cc|c}        
Full-Modeling & ShapeFit & \multicolumn{2}{c|}{Bias}  & Stoch/Counter \\ 
              &          & Min. F. & Max. F.         &               \\ \hline \hline
H$_0$        & $f\sigma_8$ & \multicolumn{2}{c|}{$b_1$} & $\alpha_0$ \\
$\mathcal{U}[50,90]$ & $\mathcal{U}[0,1]$ & \multicolumn{2}{c|}{$\mathcal{U}[0, 10]$} & --- \\
\hline
$\omega_{\mathrm{b}}$ &  $\alpha_{\parallel}$ & \multicolumn{2}{c|}{$b_2$} & $\alpha_2$ \\
$\mathcal{N}[0.02237,0.00037]$ & $\mathcal{U}[0.8,1.4]$ & \multicolumn{2}{c|}{$\mathcal{U}[-50, 50]$} & --- \\
\hline
$\omega_{\mathrm{cdm}}$ &  $\alpha_{\perp}$ & \multicolumn{2}{c|}{$b_{s^2}$} & $\alpha^{\rm shot}_0$ \\
$\mathcal{U}[0.05,0.20]$ & $\mathcal{U}[0.8,1.4]$ & $=-\tfrac{4}{7}(b_1-1)$ & \textit{Uninformative} & --- \\
\hline
$\log(10^{10} A_\mathrm{s})$& $m$ & \multicolumn{2}{c|}{$b_{\rm 3nl}$} & $\alpha^{\rm shot}_2$ \\
$\mathcal{U}[2.0,4.0]$ & $\mathcal{U}[-3.0,3.0]$& $=\tfrac{32}{315}(b_1-1)$ & \textit{Uninformative} & --- \\
\end{tabular}
\caption{\folps\, priors in both Full-Modeling and ShapeFit analyses for minimal (Min. F.) and maximal (Max. F.) freedom settings. We utilize uninformative priors for the EFT counterterms $\alpha_0$ and $\alpha_2$, as well as for the stochastic variables $\alpha^{\rm shot}_0$ and $\alpha^{\rm shot}_2$. However, we analytically marginalize over these parameters.}    
\label{tab: folps_priors} 
\end{table}




\subsection{Minimal and Maximal freedom bias parametrizations}
\label{sec: bias}


One of the difficulties in comparing the theories above is that they employ different conventions for the bias parameters that relate fluctuations in the matter density to galaxy overdensities. While the translation is known in principle, in practice it is not easy to find direct mappings from the biasing scheme of one model to that of all the others as one code can group terms in ways that cannot be reproduced with simple transformations of bias parameters in the other codes. This also makes it more difficult to be fully consistent in the priors applied to bias parameters in each model. We can mitigate this problem, however, by comparing our models within two opposing parameterizations, which we call minimal and maximal freedom. For the maximal freedom case, all four bias parameters of each model are allowed to vary freely, with very wide uninformative priors. While this does not mean that all models perform identically, it ensures that any differences are kept small.

We define the minimal freedom, or ``coevolution'' case by assuming that at initial times the bias is purely given by the linear ($b_1$) and second-order bias ($b_2$), and that non-local bias ($b_s$) as well as third-order bias ($b_3$) terms emerge at later times only through non-linear evolution due to gravity. Since the Lagrangian picture defines the biases at initial coordinates $\bq$, we assume specifically that the Lagrangian biases $b_s^{\rm L}$ and $b_3^{\rm L}$ are fixed to zero, while $b_1^{\rm L}$ and $b_2^{\rm L}$ are allowed to vary. In the Eulerian picture, the bias parameters are defined at the observed positions and time and thus depend on the combination of initial (Lagrangian) values and the evolution of the density fields. Therefore, when the Lagrangian $b_s^{\rm L}$ and $b_3^{\rm L}$ biases are zero, the corresponding Eulerian biases are directly proportional to $b_1^{\rm L} = b_1^{\rm E} - 1$. While we can define a ``minimal freedom'' parameterization within each model that involves only two bias parameters, these are not the exact same parameters in each model but rather linear combinations of each other. Nonetheless, the minimal freedom or ``coevolution'' parameter choice allows us to reduce the number of terms responsible for the different behavior between models.

Between the \velocileptors\ LPT and EPT models, a linear mapping exists at one-loop order to relate the two bias schemes \cite{ChenVlahWhite20}.  These bias relations imply that coevolution in the velocileptors EPT model involves setting 
\begin{equation}
    b_2^{\rm E} = b_2^{\rm L} + \frac{8}{21}b_1^{\rm L}
    \quad , \quad
    b_s^{\rm E} = -\frac{2}{7}(b_1^{\rm E}-1 ) 
    \quad , \quad
    b_3^{\rm E} = b_1^{\rm E}-1. 
\end{equation}
Meanwhile, in \folps\, one fixes the tidal $b_{s^2}$ and 3rd order $b_{\rm 3nl}$ biases to coevolution via \cite{Chan:2012jj,Baldauf:2012hs,Saito:2014qha}
\begin{equation}
b_{s^2} = -\frac{4}{7} (b^{\rm E}_1-1) \quad \text{and} \quad b_{3 \rm nl} = \frac{32}{315} (b^{\rm E}_1-1).
\end{equation}
The connection with the \textsc{PyBird} parameters is slightly more complex, and is most easily accomplished using the ``monkey bias'' formalism of \cite{Fujita_2020}.  Using $b_3^{\rm L}=b_s^{\rm L}=0$ we have 
\begin{align}
    \tilde b_1 &= b_1^{\rm v} + 1 \nonumber \\
    \tilde b_2 &= \frac{7}{2}\left(\frac{2}{7} + b_s^{\rm L}\right) = 1  \nonumber \\
    \tilde b_3 &= \frac{7(42 - 145b_1^v-21b_3^v+630b_s^{\rm L})}{441} = \frac{294-1015(b_1 - 1)}{441} \nonumber \\
    \tilde b_4 &= -\frac{7}{5}(b_1-1)-\frac{7}{10}b_2^{\rm v} 
    \label{eq:conversion}
\end{align}
After the conversion, the only two free parameters are \(b_1\) and \(b_2^{\rm v}\). To simplify the equations further, instead of putting priors on \(b_2^{\rm v}\) and then finding \(b_4\) we put a flat prior on \(b_4\) directly. The range of this flat prior is determined by the prior on \(b_1\) and \(b_2^{\rm v}\) in \velocileptors.

\subsection{Overview of key model differences}
\label{sec: diff}
While the models have been derived by different groups, starting from different assumptions and proceeding in different ways, they arise from a systematic procedure for capturing the effects of dynamics, redshift-space distortions and biasing.  The end result should thus be consistent predictions for the redshift-space power spectrum of biased tracers once we account for different conventions.  Any remaining differences should arise only at 2-loop order or be due to the addition of higher-order counterterms\footnote{Different parameterizations can also correspond to different priors, but we try to minimize this issure through the minimal and maximal freedom choices}.  With this in mind, in this section, we try to summarize the key differences between the models.

Perhaps the easiest differences to see are that the LPT flavor of \velocileptors\ has its counterterms of the form $k^2P_{\rm Zel}$\footnote{In Ref.~\cite{KP5s2-Maus} a reparameterization of the counterterms is used that changes changes slightly the definition of $\alpha_n$ parameters and additionally uses $k^2P_{\rm lin}$. This gives identical results as it doesn't change the theory but slightly modifies how priors are chosen on $\alpha_n$. This is not important here because we choose infinite priors on counterterm parameters for all models}, while the other models use $k^2P_{\rm lin}$.  These differ at high-$k$ by terms that are formally 2-loop order.  \pybird\ includes an extra, isotropic stochastic term of the form $k^2/k_{\rm M}^2$ that the other models do not include.

The resummation of long-wavelength modes is handled quite differently in the different codes.  In \velocileptors\ the long-wavelength components of the displacements are exponentiated while the short-wavelength components are treated perturbatively.  The split is governed by a parameter $k_{\rm IR}$.  For reasonable volumes, the choice of $k_{\rm IR}$ has little effect, however, if errors corresponding to the (physically unrealizable) $200\,h^3\mathrm{Gpc}^{-3}$ volume are used we see differences.  This is expected, as the implied errors are small enough that 2-loop effects cannot be ignored in any of the models.  A similar procedure is done for \pybird.  By contrast, the EPT variant of \velocileptors\ and the \folps\ model split the linear theory spectrum into a ``wiggle'' and ``no-wiggle'' piece, and then damp only the ``wiggle'' component. The results for the final sum are quite insensitive to the precise decomposition. The dependence on the precise wiggle/no-wiggle decomposition methods (e.g the methods described in Refs.~\cite{EH1998,Hinton2017,Wallisch2018}) has been tested in Ref.\cite{KP5s4-Lai}, while tests on the choice of IR cutoff scale are presented in Refs.~\cite{Chen21,KP5s4-Lai}.

\section{Fitting methods}
\label{sec:methods}

There are two primary methods for fitting power spectra that we consider here. The first, ``ShapeFit'' (SF; \cite{Briedan21}), extends the standard template fitting used in many previous galaxy clustering analyses.  For the standard template fitting it is assumed that the shape of the power spectrum is well-determined by early Universe measurements (e.g.~CMB anisotropies) and late-time effects do not alter the shape of this spectrum.  A fiducial cosmology, and thus a fiducial linear power spectrum, $P_{\rm lin}$, is picked and held fixed in the fits. The parameters being varied, corresponding to changes in ``late time cosmology'' are the amplitude $f\sigma_8$ and the distance scalings, 
\begin{align}
    \alpha_{\parallel} = \frac{H^{\rm ref}(z)}{H(z)}\left(\frac{r_{\rm d}^{\rm ref}}{r_{\rm d}}\right), \quad \quad \alpha_{\perp}=\frac{D_{\rm A}(z)}{D^{\rm ref}_{\rm A}(z)}\left(\frac{r_{\rm d}^{\rm ref}}{r_{\rm d}}\right),
\end{align}
where $H(z)$, $D_{\rm A}(z)$, and $r_{\rm d}$ are the comoving Hubble parameter, angular diameter distance, and sound horizon scale at drag epoch. These distance scalings are driven by both the Alcock-Paczinsky effect (AP;\cite{Alcock79}), which is the anisotropic distortion when the true cosmology differs from the fiducial in the conversion of angles and redshift to physical distances, and the changes in $r_{\rm d}$ with varying cosmology. 
Varying these parameters effectively changes the power spectrum without the need to recompute $P_{\rm lin}$ in its entirety.  One disadvantage of the template fit is that one sacrifices information contained in the shape of $P_{\rm lin}$, and so obtains sub-optimal constraints on parameters affecting its shape when not combining with e.g.\ CMB measurements.  The ShapeFit method aims to maintain some of the simplicity of the standard template fit, while also capturing some of the information from the shape of $P_{\rm lin}$ via two additional compressed parameters $m$ and $n$, defined through
\begin{equation}
    P^{\prime}_{\rm lin}(\bk) = P_{\rm lin}(\bk)\ \exp\left\{ \frac{m}{a}\tanh \left[a\ln\left(\frac{k}{k_{\rm p}}\right) \right] + n\ln\left(\frac{k}{k_{\rm p}}\right) \right\}. 
    \label{eq: plin_sf}
\end{equation}
 The scale-dependent $m$ parameter above modulates the shape of the power spectrum and aims to mimic the effect that different $\omega_m$ and $\omega_b$ values have on the shape of $P_{\rm lin}(\bk)$. We use for the amplitude $a=0.6$ and pivot scale $k_{\rm p} = 0.03 \ \ihmpc$, as suggested in Ref.~\cite{Briedan21}. These constants are chosen in order for changes in $m$ to best replicate how $\omega_m$ and $\omega_b$ influence $P_{\rm lin}(\bk)$. The scale-independent $n$ parameter changes the slope in a way that exactly reproduces the effect of different $n_{\rm s}$ when computing $P_{\rm lin}(\bk)$, i.e. $n = n_{\rm s} - n_{\rm s}^{\rm ref}$. 
The inclusion of the shape parameters therefore makes ShapeFit more constraining than the standard template fit, which is particularly evident when only LSS data are analysed, not in combination with CMB data or priors. It is one of the primary methods that will be used in analyzing power spectra/correlation function data in DESI, especially when studying the impact of systematic effects. All models in this paper implement the ShapeFit method by sampling in the parameters $(f\sigma_8,\alpha_{\parallel},\alpha_{\perp},m)$\footnote{In this paper we keep $n$ fixed to its fiducial value. This parameter is directly related to the spectral tilt $n_{\rm s}$. In full-shape analyses, $n_{\rm s}$ is a weakly constrained parameter and historically has been kept fixed to the e.g. CMB-measured value. While the increased constraining power of DESI and beyond allows us to vary $n_{\rm s}$ going forward, an informative prior is still needed, and for DESI its width is chosen such that constraints on other parameters are unaffected\cite{DESI2024.V.KP5}. Since this paper focuses on the comparison between different effective field theories, we do not find the inclusion of such a prior-dominated parameter to provide any additional insights. However, the effect of varying it is explored in companion DESI papers\cite{KP5s2-Maus,KP5s3-Noriega,KP5s5-Ramirez} and the relevance of projection effects when varying $n$ (for SF) and $n_{\rm s}$ (for Full-Modeling) will be discussed in Ref.~\cite{DESI2024.V.KP5}.} with the same $P_{\rm lin}(\bk)$ in order to obtain the observable power spectrum multipole predictions. In \velocileptors\ a Taylor series emulator is used in order to improve the evaluation of theory multipoles, while \folps\ and \pybird\ use as default settings an approximation for the loop integrals~\cite{Briedan21}:
\begin{align}
    \int d^3\bq P^{\prime}_{\rm lin}(q) F_n(...) \approx \left(\frac{P^{\prime}_{\rm lin}(k)}{P_{\rm lin}(k)} \right) \int d^3\bq P_{\rm lin}(q) F_n(...),
\end{align}
with $F_n$ kernels defined in the beginning of \S\ref{sec:theory}. Under this approximation the loop integrals only need to be computed once and then are simply rescaled as $m$ is varied. This allows for fast evaluation of theory multipoles in lieu of an emulator. The robustness of the emulator for \velocileptors\ is shown in Appendix E in Ref.~\cite{KP5s2-Maus} and the stability of the above approximation used for \pybird\ and \folps\ is tested in Appendix B of Ref.~\cite{KP5s3-Noriega}. Since both the emulator and the approximation show neglibible differences in constraints on Abacus mocks when compared to direct computations, we do not expect these implementation choices to contribute to any disagreements between EFT models.

The second method used in our comparison is based on a more direct approach and involves choosing a cosmological model, e.g.\ $\Lambda$CDM, and directly varying the underlying parameters of this model, predicting the data and comparing the prediction to the measurement. This direct-fitting approach, hereafter referred to as Full-Modeling (FM), captures information from both the early and late universe as the parameters of the model, ($H_0$, $\omega_\mathrm{b}$, $\omega_\mathrm{cdm}$ and $\log(10^{10}A_\mathrm{s})$ in our examples) control the shape of $P_{\rm lin}$ as well as the late-time dynamics/geometry. While this method does involve generating a new linear power spectrum at every step of a Markov Chain Monte Carlo (MCMC), the use of emulators can speed up the computation enough that differences in convergence time between FM and compressed parameters are insignificant. The Full-modeling method also takes into account the AP effect through the scaling parameters $q_{\parallel,\perp}$ are related to $\alpha_{\parallel,\perp}$ but absent the factors of $r_{\rm d}^{\rm ref}/r_{\rm d}$ because the changes in BAO scale are implicit when varying cosmological parameters. The purely geometrical $q_{\parallel,\perp}$ parameters are not sampled in this modeling method but are derived and applied to the power spectrum predictions as the cosmological parameters are varied. The actual implementation of the AP distortion is consistent in all models and involves re-scaling the power spectrum to observed coordinates using $q_{\parallel,\perp}$ (for FM) or $\alpha_{\parallel,\perp}$(for FM) via:
\begin{align}
    P^{\rm obs}_s(\bk_{\rm obs}) = q_\perp^{-2} q_\parallel^{-1} P_s(\bk)
    \quad , \quad
    k^{\rm obs}_{\parallel, \perp} =  q_{\parallel, \perp}\  k_{\parallel, \perp}.
    \quad
\end{align}

Finally, for all models and modeling methods, we compute the linear theory power spectrum, $P_{\rm lin}$ from \textsc{CLASS} \citep{Diego_Blas_2011} or \textsc{CAMB} \citep{Lewis_2011} before constructing the 1-loop EFT power spectrum multipoles.
\section{Results}
\label{sec:results}

\subsection{Fixed-cosmology fits}

\begin{figure}
\begin{center}
\resizebox{\columnwidth}{!}{\includegraphics{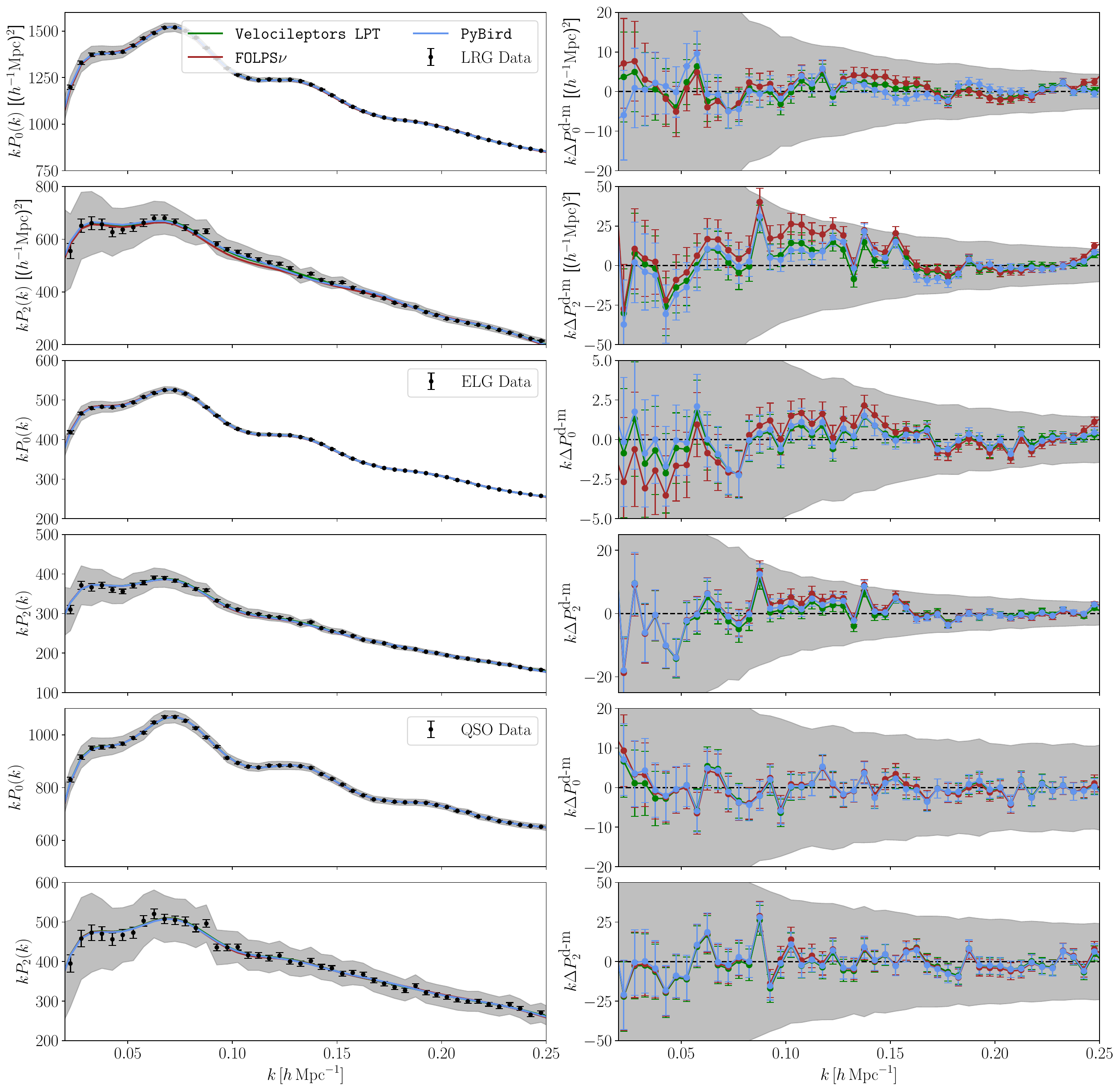}}
\caption{Abacus LRG, ELG, and QSO mock data along with each Fourier space model's prediction with $\Lambda$CDM parameters fixed to truth but allowing the nuisance parameters to vary to best fit the data. The error bars on the data reflect the rescaled covariance with 200 ($h^{-1}$Gpc)$^3$ volume, while the shaded region shows the covariance corresponding to a single cubic box with 8 ($h^{-1}$Gpc)$^3$ volume, which is more similar to the constraining power of a physical survey.  
\label{fig:mock_data}}
\end{center}
\end{figure}

\begin{figure}
\captionsetup[subfigure]{labelformat=empty}
\begin{subfigure}{.5\textwidth}
\centering
\includegraphics[height=8cm]{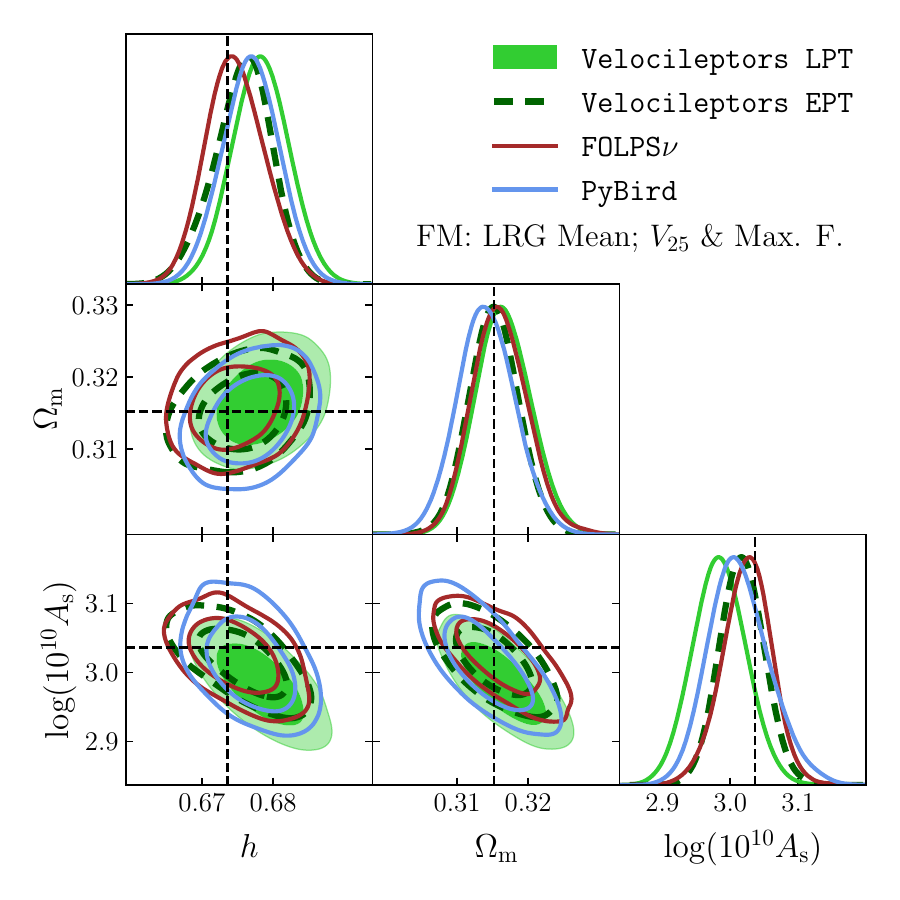}
\caption{Full-Modeling}
\end{subfigure}%
\begin{subfigure}{.5\textwidth}
\centering
\includegraphics[height=8cm]{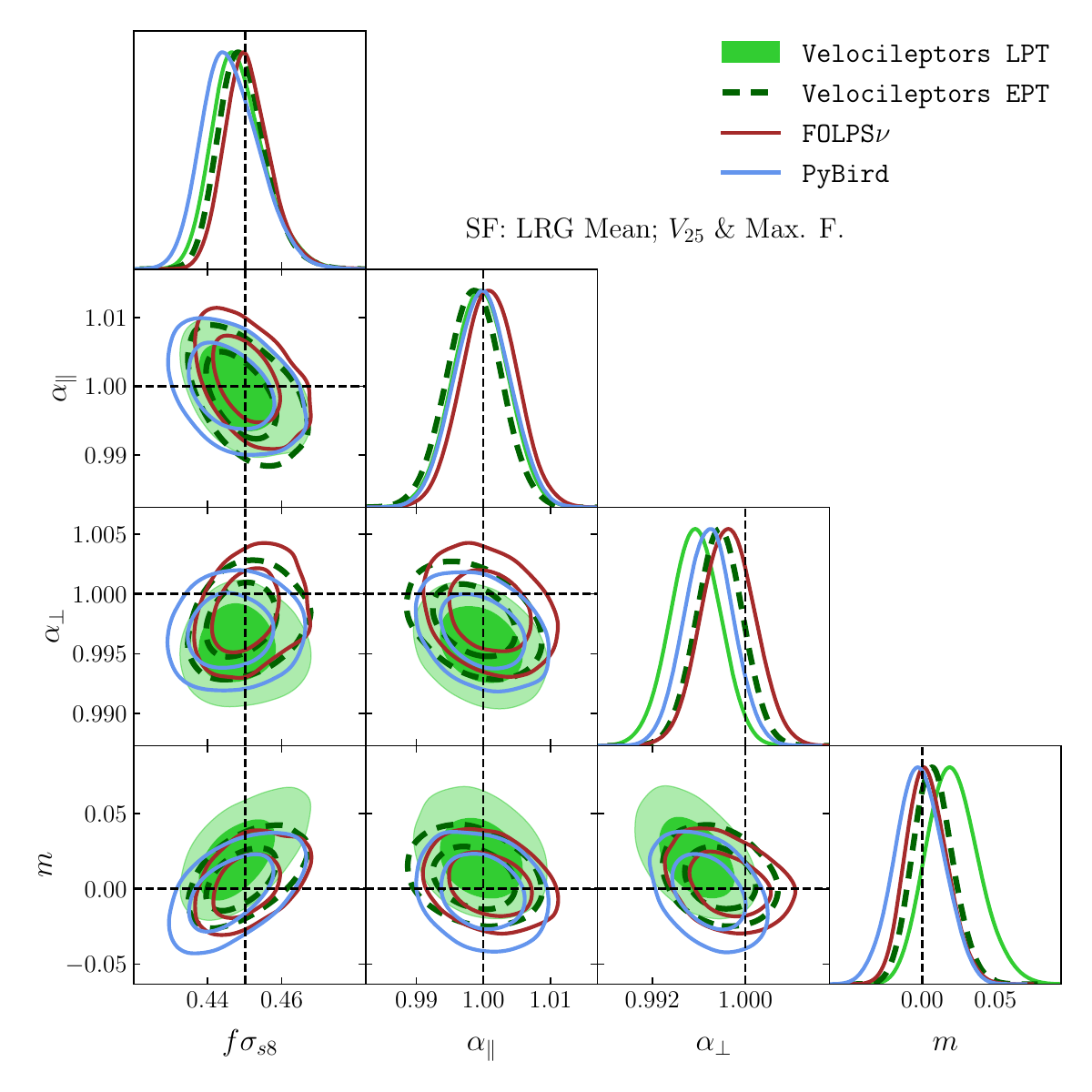}
\caption{ShapeFit}
\end{subfigure}%
\caption{Comparison between the three Fourier space models using the Full-Modeling method (left) and ShapeFit method (right), fitting the mean of the Abacus LRG ($z=0.8$) simulations. In both fits we show the constraints in the ``maximal'' freedom parametrization where the four bias parameters are allowed to vary independently in each model. The results shown use the full covariance with the 1/25 rescaling from the single-box volume. In both cases we use scale cuts of $0.02\leq k \leq 0.18 \ \ihmpc$ for all models.
}
\label{fig: comp_pre}
\end{figure}

We begin by showing in Fig.~\ref{fig:mock_data} the LRG ($z=0.8$), ELG ($z=1.1$), and QSO  ($z=1.4$) simulated data that we are fitting with our models. In the left panels we show the monopole and quadrupole data for each tracer type along with theoretical models generated with each pipeline with $\Lambda$CDM parameters fixed to truth and nuisance parameters varied to best match the data up to $k_{\rm max}=0.25 h^{-1}$Mpc. We allow all four bias parameters to vary, and use the covariance rescaled to the 200 ($h^{-1}$Gpc)$^3$ volume of the 25 mock realizations. The error bars of the data reflect this covariance, but we also show shaded regions corresponding to the error bars of the data from a single cubic box volume of 8 ($h^{-1}$Gpc)$^3$ that more closely resembles the constraining power of a typical survey. In the right-hand panels of Fig.~\ref{fig:mock_data} we show the difference between the data and the models at each $k$-bin. It is instructive to see how the nonlinear terms in the models behave when the linear power spectrum and cosmology dependence are fixed. We observe similar trends from each of the models, most noticeably in the LRG quadrupole, where all three of them show the same bump in $\Delta P_2^{(\mathrm{data-model})}$ at $k\sim0.1h^{-1}$Mpc. Additionally, all three models seem to diverge from the data the same way near $k\sim0.25h^{-1}$Mpc, which sets the range of validity of these models. For both of the LRG multipoles as well as the monopole ELG curves, we see that the \velocileptors\ and \pybird\ predictions overlap pretty closely, whereas the \folps\ model is slightly more offset from the data for some $k$ ranges.

\subsection{Tests on Mock data}

We show in Fig.~\ref{fig: comp_pre} a comparison between each model's posteriors fitting to the LRG mocks with free parameters and priors described in Tables~\ref{tab: Vel_priors}-\ref{tab: folps_priors}, with the ``Maximal freedom'' choice of bias parameters. While all models give constraints within around ~1$\sigma$ of the true values, there are noticeable disagreements between the models for both the ShapeFit and Full-modeling methods. These shifts between models can be driven by differences in the handling of counter or stochastic terms, bias parametrization, IR-resummation schemes, or even numerical effects in the code that may become important in extremely constraining data sets. Indeed, we note that the covariance used here reflects a survey volume of 200 ($h^{-1}$Gpc)$^3$, which far exceeds that of any physically achievable survey on our past lightcone to these redshifts. At such a volume, the data has a statistical error on the order of a third of a percent at each point in the $k$ data vector, a level of precision at which we could well be dominated by systematic uncertainties in the N-body simulations. On the other hand, our goal in performing tests of our models on the \textsc{Abacus} simulations is to detect systematic uncertainties in our models that will be relevant in the DESI-Y5 data; and given the size of the Y5 footprint this is not achievable without a very large simulation volume. We therefore present the ``full volume'' results with the understanding that some hard-to-quantify level of discrepancy between the models and true values is expected, while focusing on ensuring that our models agree at the level of precision attainable at volumes relevant to a physical survey. For this reason, we use Fig.~\ref{fig: comp_pre} simply to demonstrate the level of agreement we currently have at the full volume, but for the remainder of this paper will restrict our attention to fits using the covariance of a single box (V = 8 ($h^{-1}$Gpc)$^3$), which is more comparable to the volumes spanned by a realistic survey. However, the data vectors we fit to remain the ones computed from the average of 25 abacus boxes in order to reduce the noise. 


\begin{figure}
\captionsetup[subfigure]{labelformat=empty}
\begin{subfigure}{.5\textwidth}
\centering
\includegraphics[height=8cm]{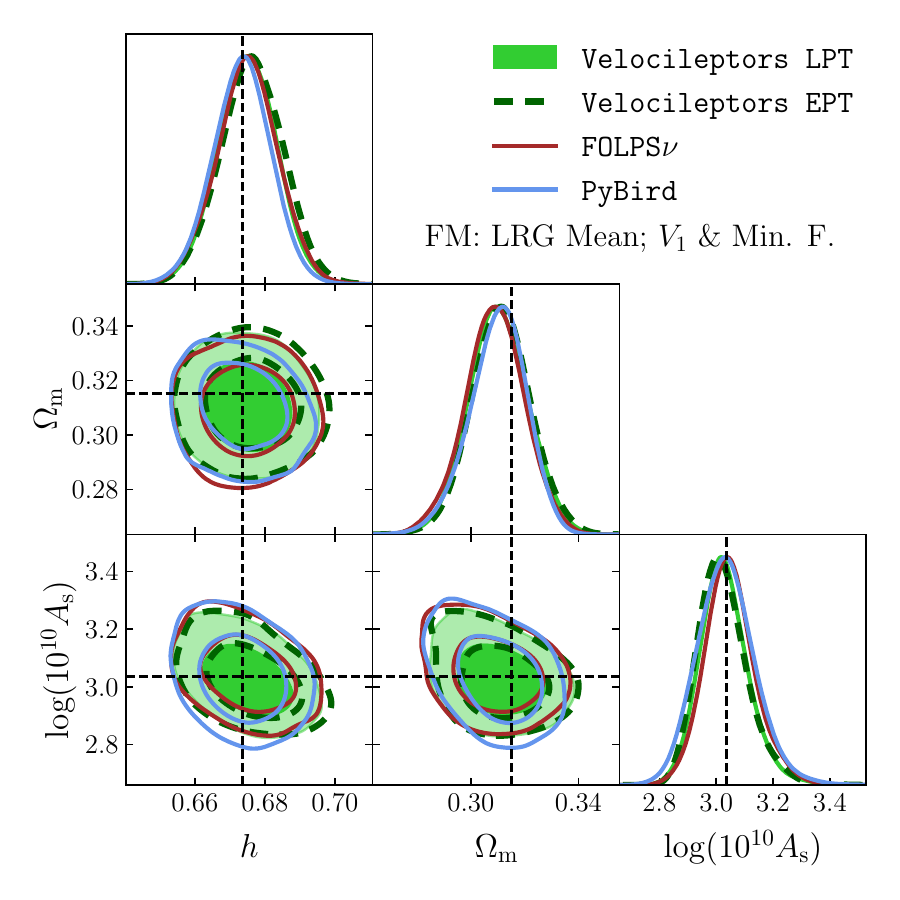}
\end{subfigure}%
\begin{subfigure}{.5\textwidth}
\centering
\includegraphics[height=8cm]{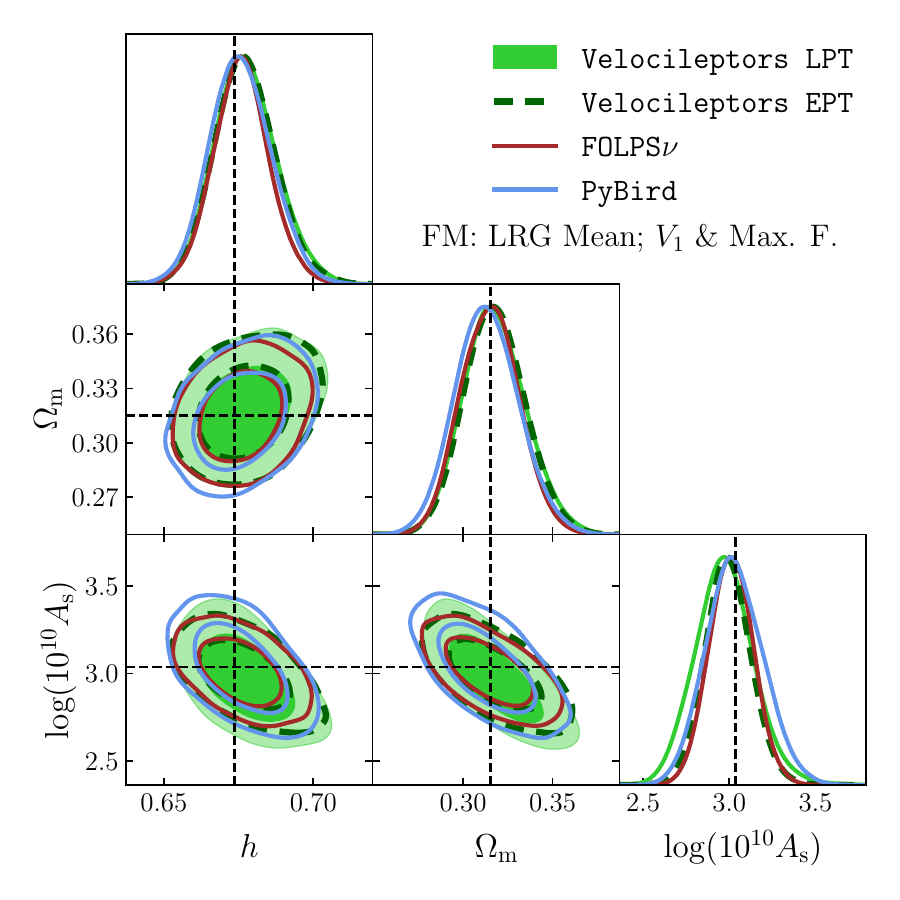}
\end{subfigure}%
\\
\begin{subfigure}{.5\textwidth}
\centering
\includegraphics[height=8cm]{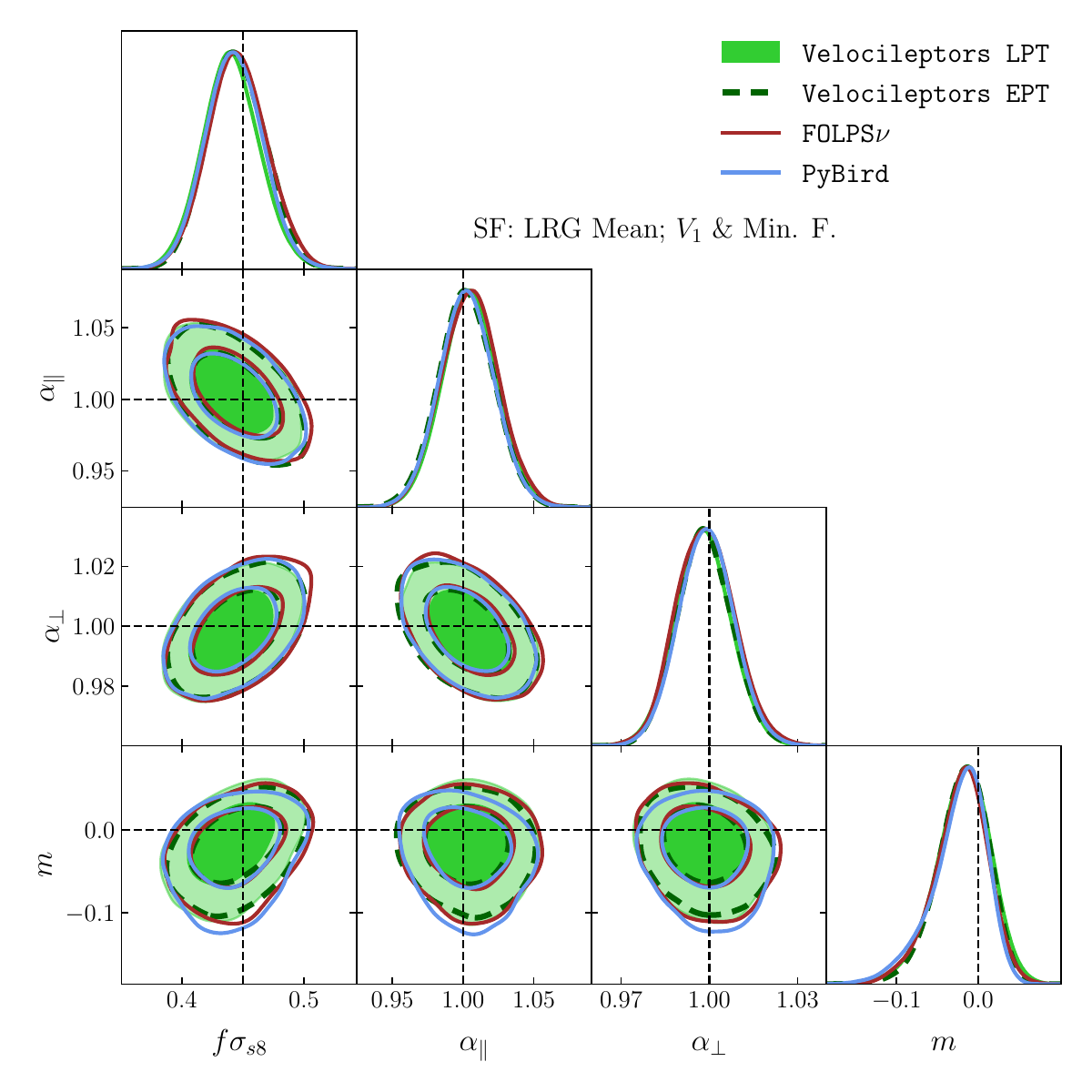}
\end{subfigure}%
\begin{subfigure}{.5\textwidth}
\centering
\includegraphics[height=8cm]{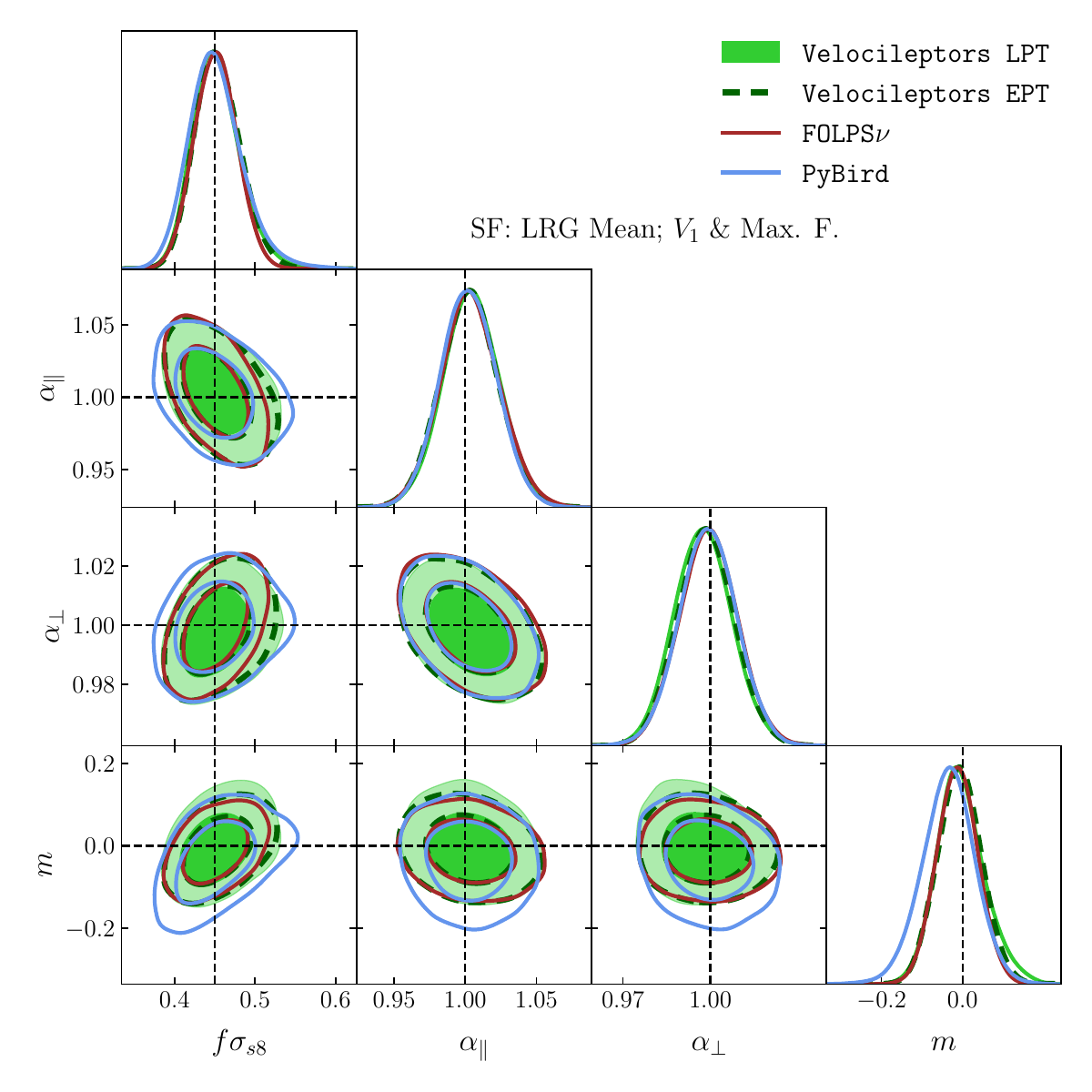}
\end{subfigure}%

\caption{Comparison between \folps, \velocileptors\ (LPT and EPT), and \pybird\ full-modeling (top row) and ShapeFit (bottom row) constraints on the mean of the mock LRG data, respectively. In the left panels we show the ``minimal'' freedom parametrization, in which Eulerian $b_{s2}$ and $b_{3\rm nl}$ parameters are fixed to coevolution and the corresponding Lagrangian parameters fixed to zero. In the right panels we employ the ``maximal'' freedom parameter choice for which all bias parameters vary independently with uniform priors. These fits used the single-box covariance volume without any rescaling. We use a scale cut of $0.02\leq k \leq 0.18 \ \ihmpc$ for all fits shown.
}
\label{fig: LRG_V1}
\end{figure}

As discussed in \S~\ref{sec: bias}, the bias terms and how they are defined can differ between theories, making it difficult to convert between biasing schemes and by extension use consistent priors. Choices made by the user about which parameters to vary and which to fix also affect the volume of the parameter space and can narrow/widen constraints purely from a degrees-of-freedom standpoint. This can greatly complicate the comparisons as the baseline settings preferred for each model as described in Refs.~\cite{KP5s2-Maus,KP5s3-Noriega,KP5s4-Lai} are quite different: the \folps\, model is often used with a coevolution parametrization in which the tidal and third order parameters, b$_{s^2}$ and b$_{3\rm nl}$, are determined by $b_1$, so only two bias parameters are varying freely. Meanwhile, the \velocileptors\ results often have Lagrangian b$_3$ fixed with the remaining three free. For PyBird it is common to fix $b_4$ to anticorrelate with $b_2$.  We thus compare the constraints between Fourier space models while taking more care to standardize the amount of freedom afforded to the bias terms. We show in Fig.~\ref{fig: LRG_V1} comparisons in the Full-Modeling and ShapeFit constraints on LRG data, respectively, with both ``maximal'' (Max.\ F) and ``minimal'' (Min.\ F) freedom settings, which are defined in \S~\ref{sec: bias}. We see that in both cases there is very good agreement between the two \velocileptors\ models, \folps, and \pybird\ in the ShapeFit and Full-Modeling methods. We find that the models generally agree better in the minimal freedom case, and this is a more well-defined parametrization that can be applied consistently to each theory. In the maximal freedom case, we observe slight differences in the widths of the $\log A_{\rm s}$ posteriors in Full-modelling. In ShapeFit the \pybird\ model has a wider posterior in the $m$ parameter, and explores more negative m values than the other theories.

\begin{figure}
\captionsetup[subfigure]{labelformat=empty}
\begin{subfigure}{.5\textwidth}
\centering
\includegraphics[height=8cm]{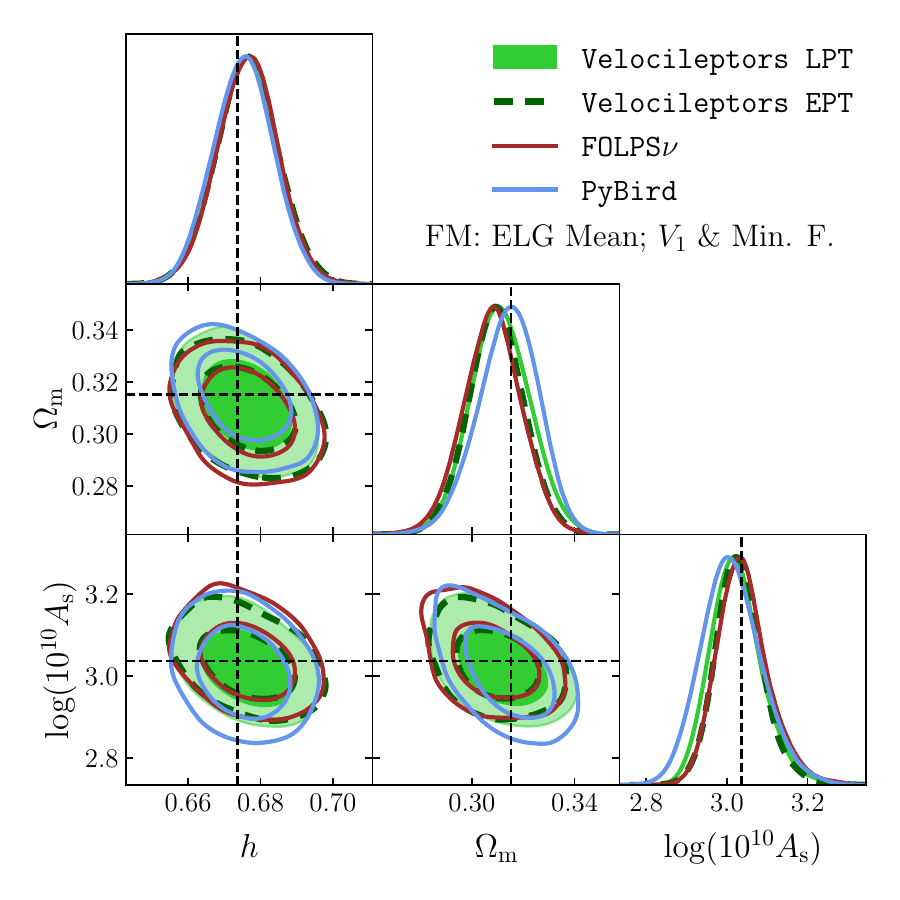}
\caption{ELG ($z=1.1$)}
\end{subfigure}%
\begin{subfigure}{.5\textwidth}
\centering
\includegraphics[height=8cm]{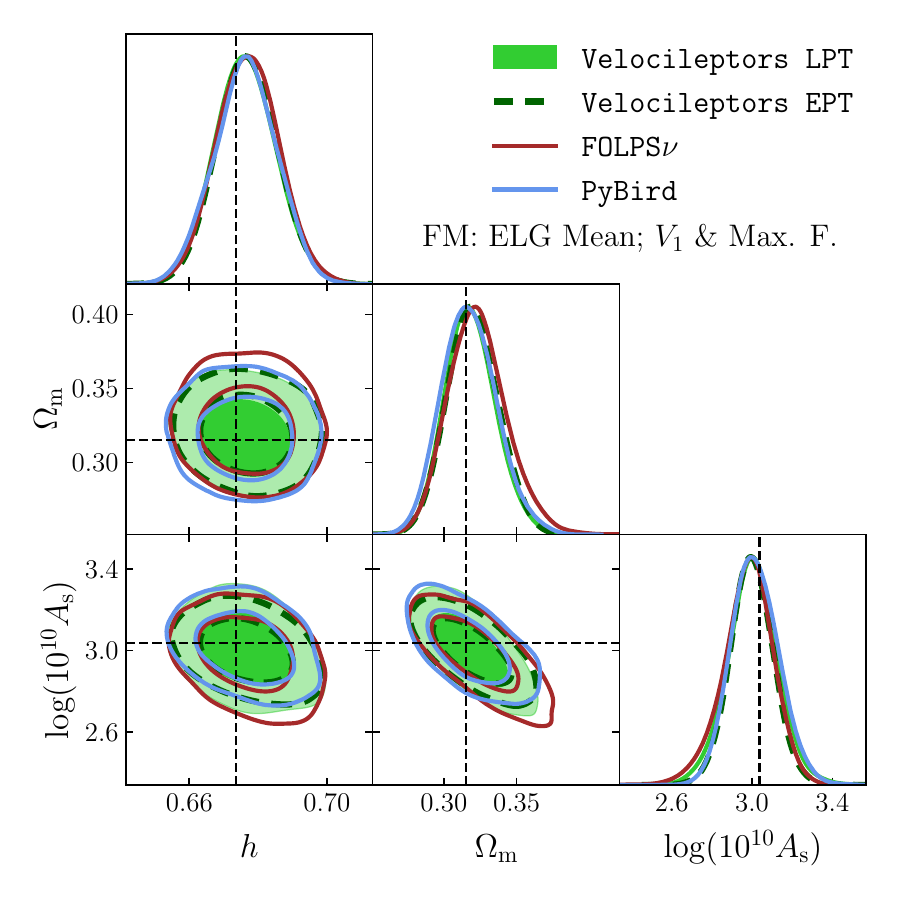}
\caption{ELG ($z=1.1$)}
\end{subfigure}%
\\
\begin{subfigure}{.5\textwidth}
\centering
\includegraphics[height=8cm]{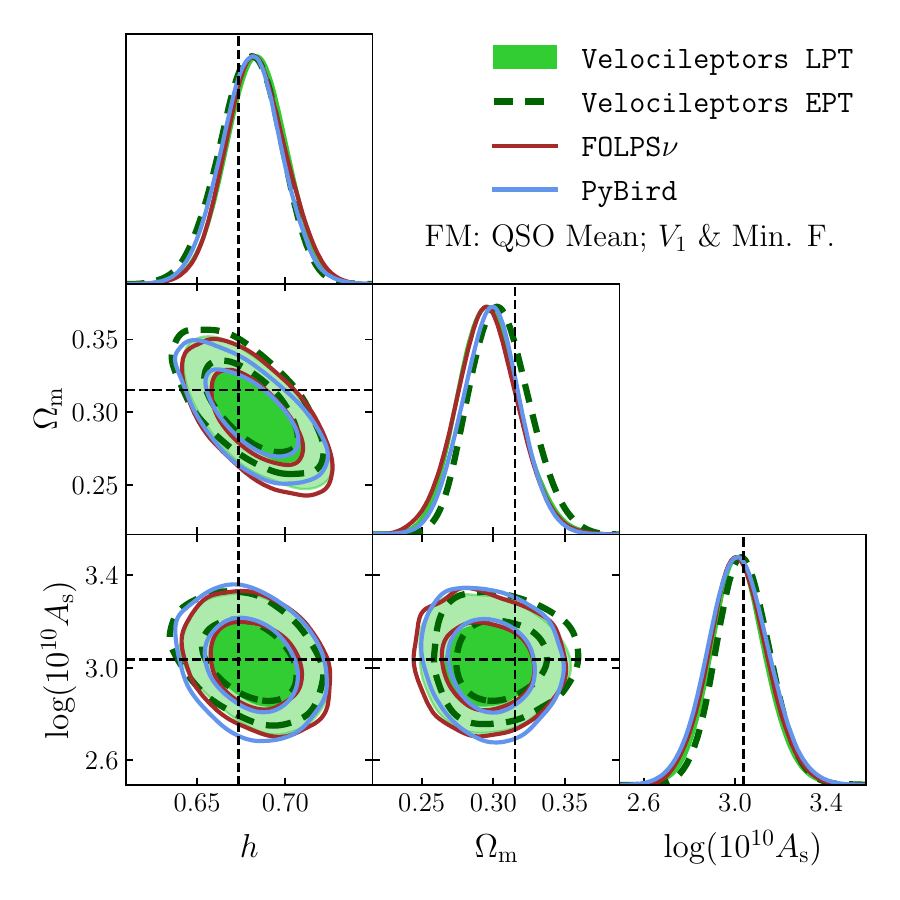}
\caption{QSO ($z=1.4$)}
\end{subfigure}%
\begin{subfigure}{.5\textwidth}
\centering
\includegraphics[height=8cm]{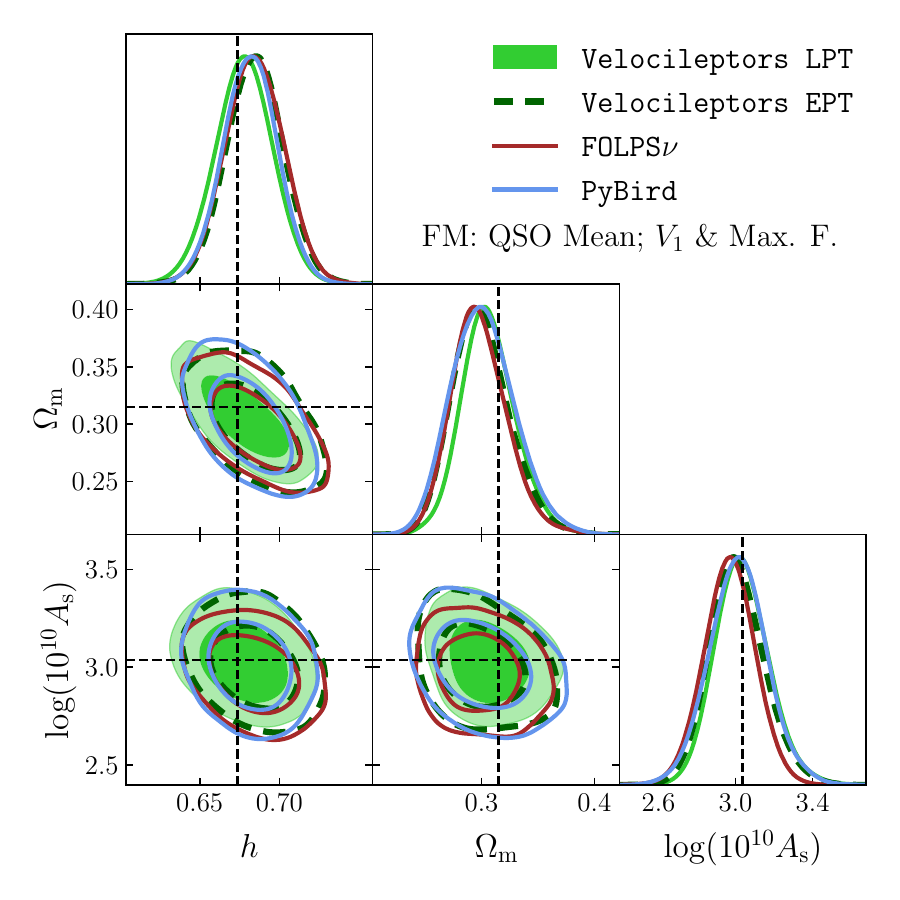}
\caption{QSO ($z=1.4$)}
\end{subfigure}%
\caption{Comparison between \folps\, \pybird\ and \velocileptors\ (LPT and EPT) full-modeling constraints on the mean of the mock ELG and QSO data, respectively. In the left column we show the ``minimal'' freedom parametrization and the right column shows the ``maximal'' freedom. These fits use the single-box covariance volume without any rescaling. We use a scale cut of $0.02\leq k \leq 0.18 \ \ihmpc$ for all fits shown.
}
\label{fig: ELG_QSO_FM}
\end{figure}

\begin{figure}
\captionsetup[subfigure]{labelformat=empty}
\begin{subfigure}{.5\textwidth}
\centering
\includegraphics[height=8cm]{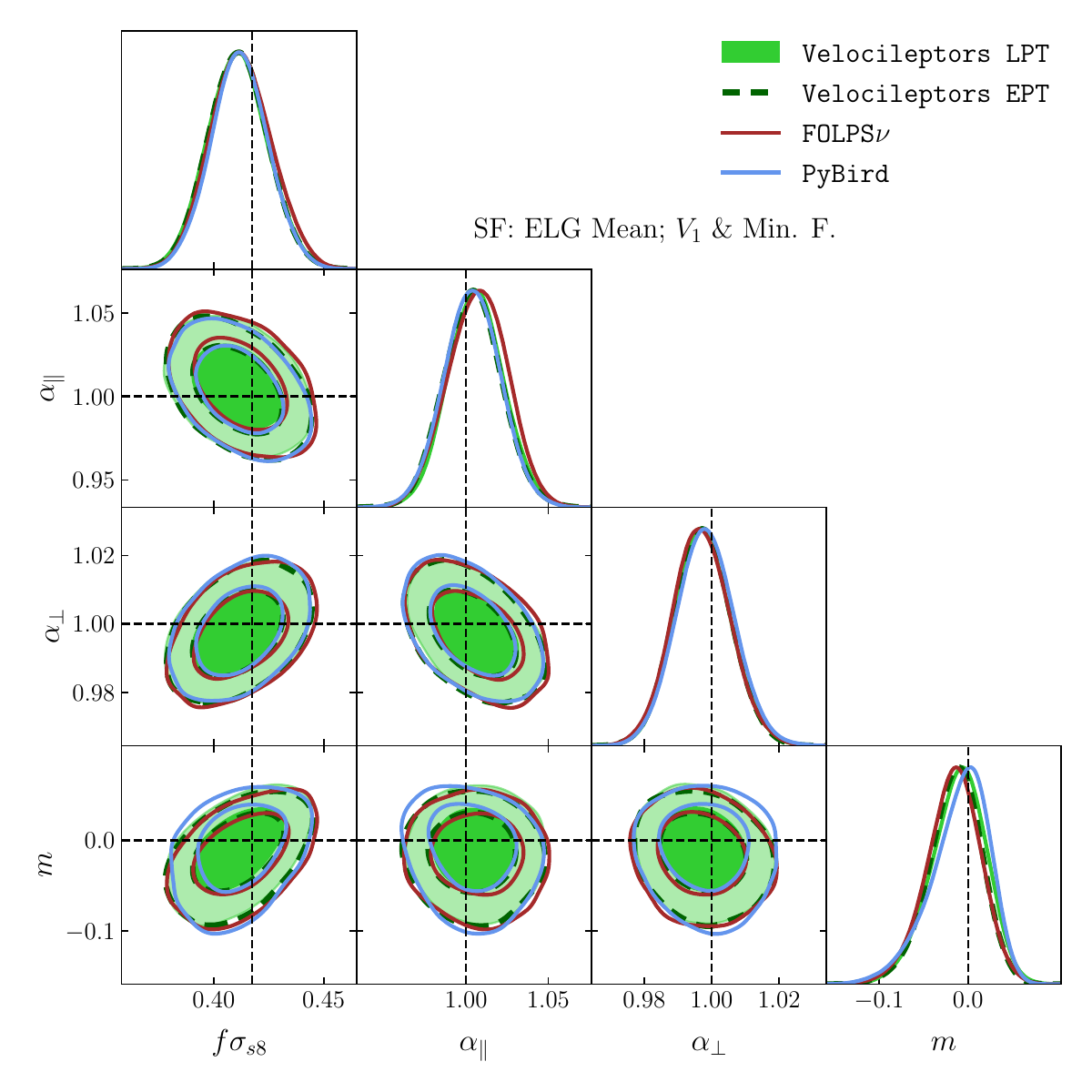}
\caption{ELG ($z=1.1$)}
\end{subfigure}%
\begin{subfigure}{.5\textwidth}
\centering
\includegraphics[height=8cm]{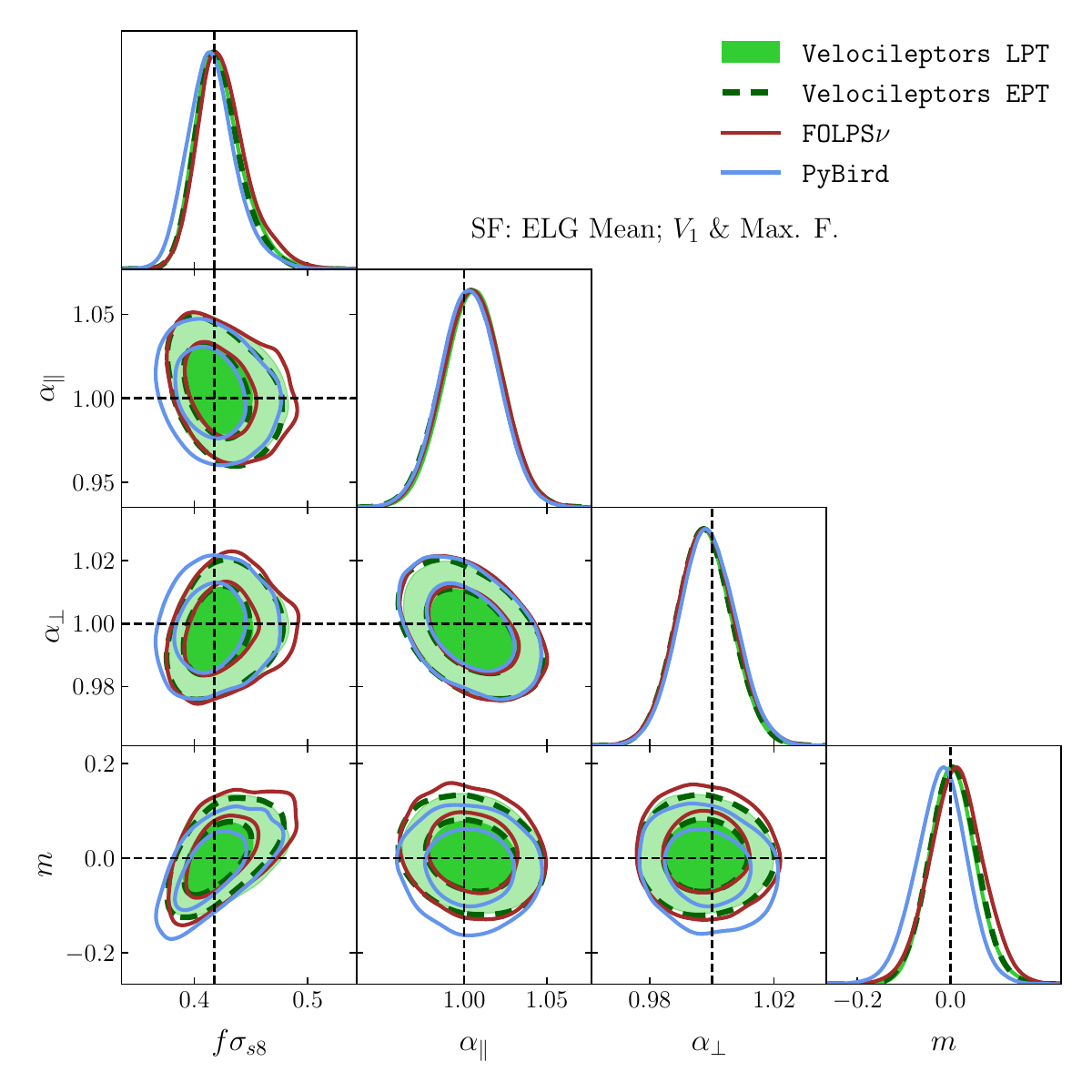}
\caption{ELG ($z=1.1$)}
\end{subfigure}%
\\
\begin{subfigure}{.5\textwidth}
\centering
\includegraphics[height=8cm]{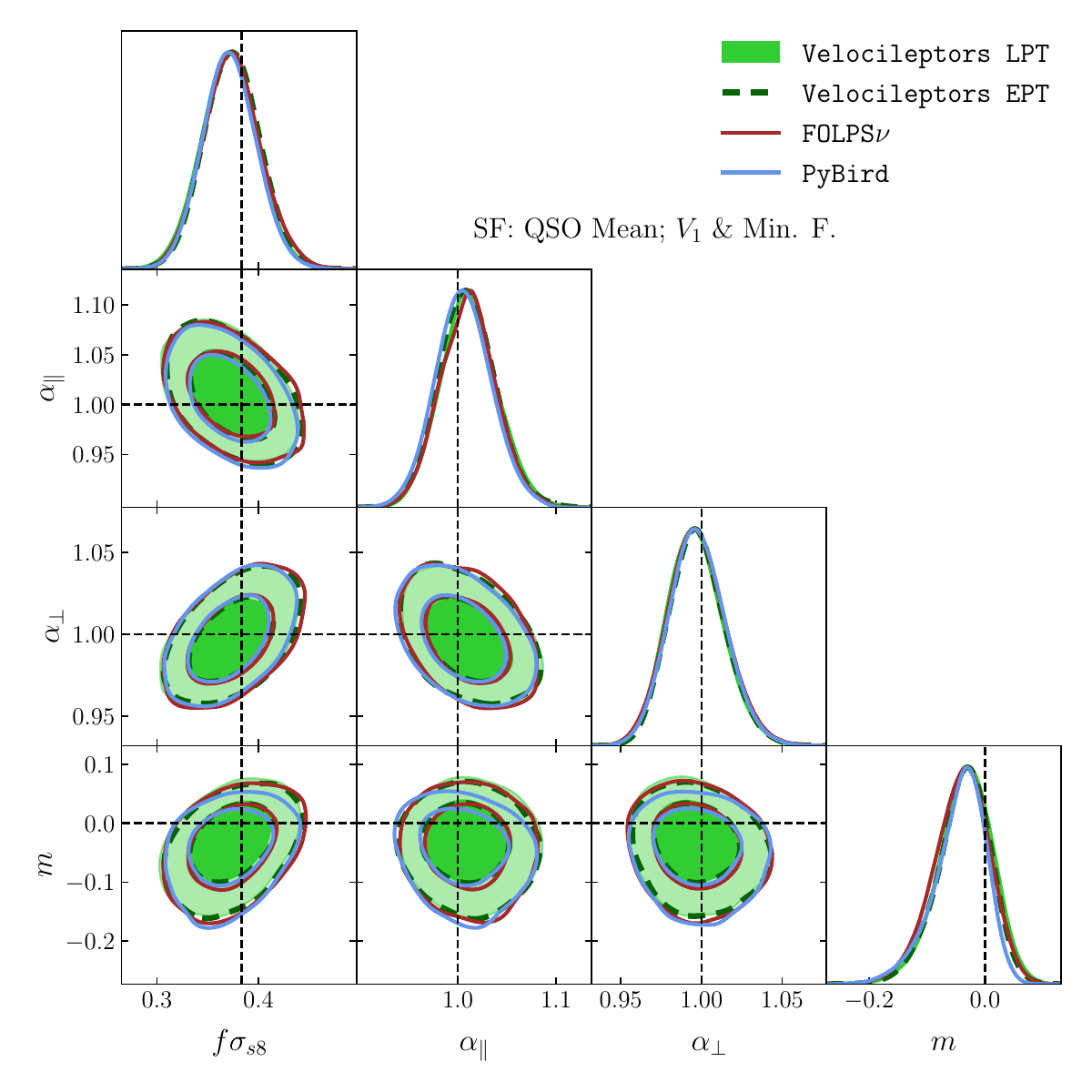}
\caption{QSO ($z=1.4$)}
\end{subfigure}%
\begin{subfigure}{.5\textwidth}
\centering
\includegraphics[height=8cm]{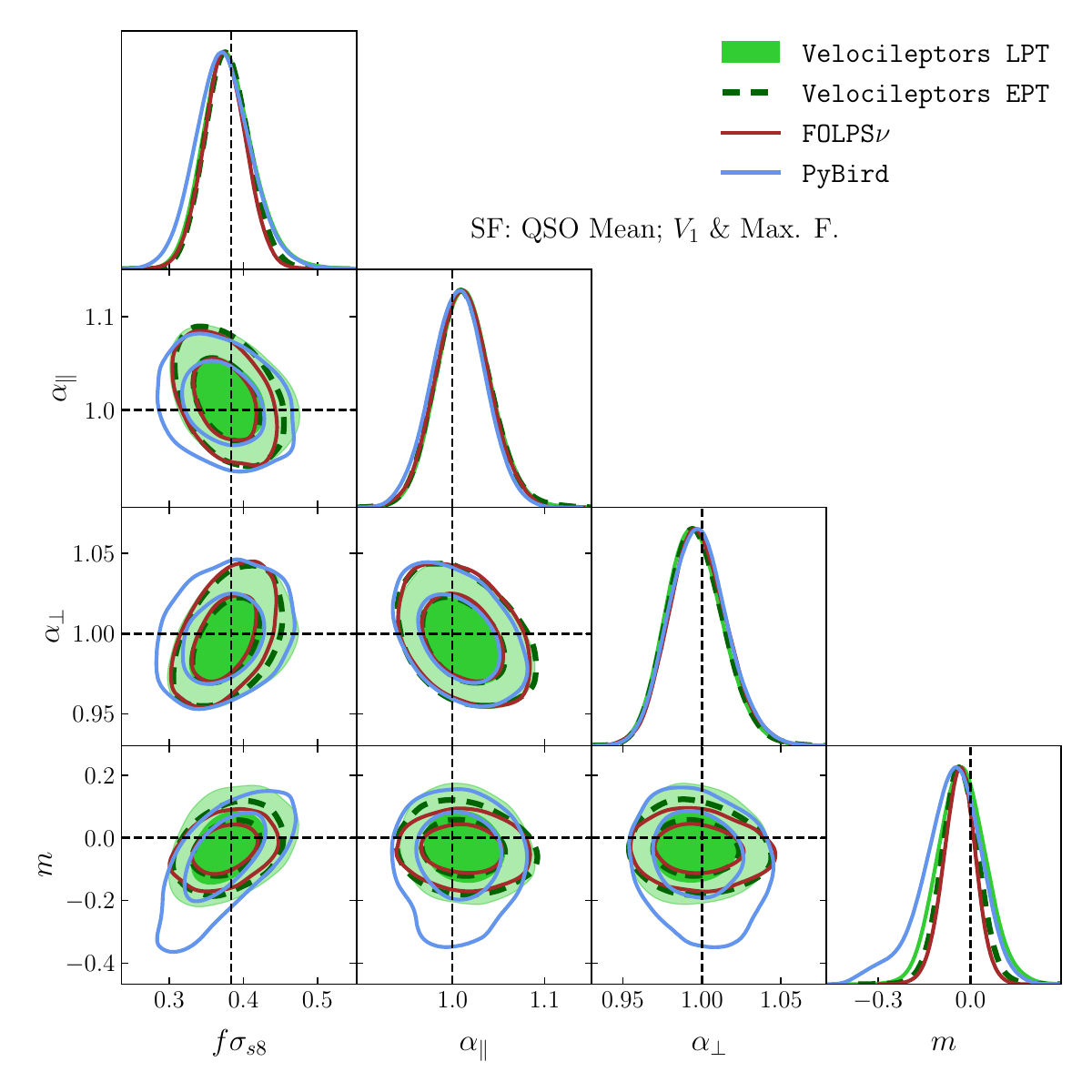}
\caption{QSO ($z=1.4$)}
\end{subfigure}%
\caption{Comparison between \folps\, \pybird\ and \velocileptors\ (LPT and EPT) ShapeFit constraints on the mean of the mock ELG and QSO data, respectively. In the left columns we show the ``minimal'' freedom parametrization and on the right we show the ``maximal'' freedom case. These fits use the single-box covariance volume without any rescaling. We use a scale cut of $0.02\leq k \leq 0.18 \ \ihmpc$ for all fits shown.
}
\label{fig: ELG_QSO_SF}
\end{figure}

\begin{figure}
\captionsetup[subfigure]{labelformat=empty}
\begin{subfigure}{0.5\textwidth}
\centering
\includegraphics[width=1\textwidth]{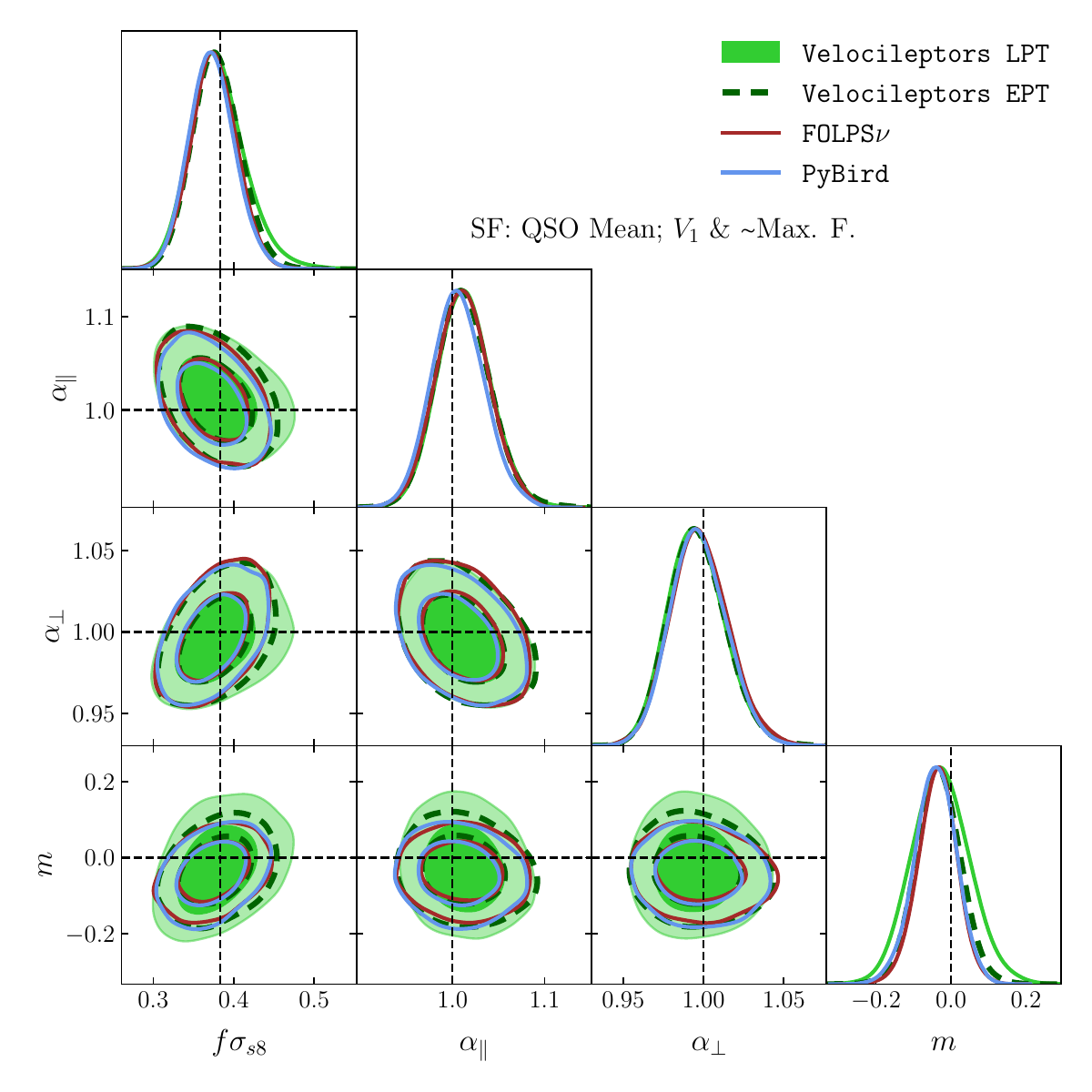}
\end{subfigure}%
\begin{subfigure}{0.5\textwidth}
\centering
\includegraphics[width=1\textwidth]{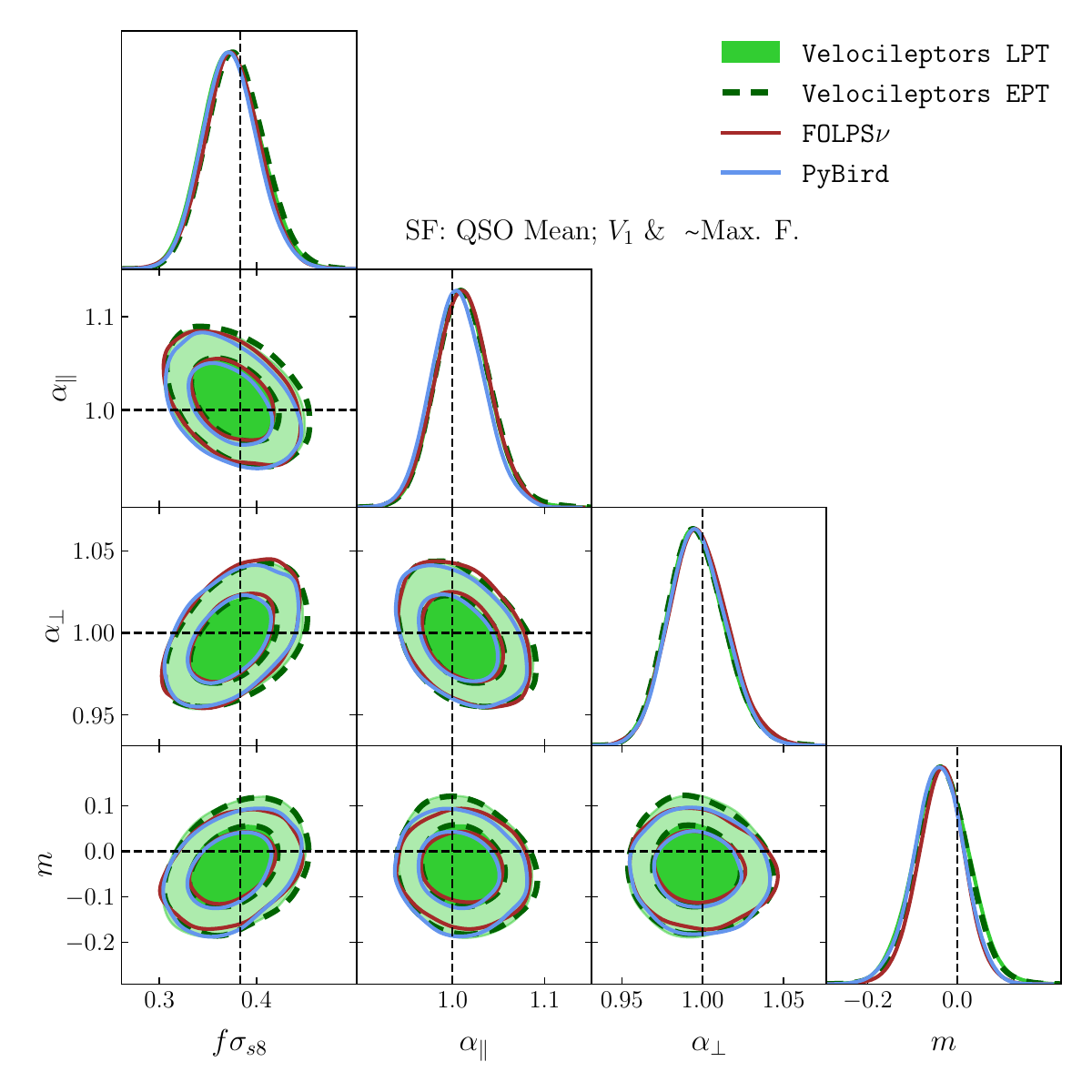}
\end{subfigure}%
\caption{ShapeFit constraints on the QSO data in the maximal freedom case, but introducing priors on some of the bias parameters for \pybird\ and \velocileptors\ LPT. In the left plot a Gaussian prior $\mathcal{N}[0.,10.]$ is applied to the $b_3$ parameter of the \pybird\ model only, while the other models still have uniform priors on all parameters. In the right plot we additionally introduce Gaussian priors $\mathcal{N}[0.,5.]$ on both Lagrangian $b_s$ and $b_3$ parameters in the \velocileptors\ LPT model. We use a scale cut of $0.02\leq k \leq 0.18 \ \ihmpc$ for all fits shown.
\label{fig: b3_priors}}
\end{figure}

We next show the results for the ELG and QSO tracers for Full-modeling and ShapeFit methods in Figs.~\ref{fig: ELG_QSO_FM} and \ref{fig: ELG_QSO_SF}, respectively. For minimal freedom, we observe consistent constraints on the QSO data between pipelines for both modeling methods. However, when fitting the ELG data, the \pybird\ model constrains the power spectrum shape slightly differently from the other models. We observed this through a small shift in $\Omega_{\rm m}$ and $m$ constraints for Full-modeling and ShapeFit respectively. In the maximal freedom case, we see better consistency between theories in Full-Modeling than in ShapeFit. Similarly to the LRG tracer, we find that the \pybird\ model has more freedom in exploring larger negative values of the $m$ parameter. These extreme values for $m$ correspond to nonphysical $\Lambda$CDM models and are therefore disfavored in the Full-modeling fit. We find that in \pybird\ these large negative values of $m$ are accompanied by large negative $b_3$ (around $-100$), resulting in canceling effects to the power spectrum such that a moderately high likelihood is still obtained. In order to prevent this effect we introduce a Gaussian prior on $b_3$ centered on zero with a width of 10, and show the result in the left hand plot of Fig.~\ref{fig: b3_priors}. We find that with a more informative prior on $b_3$, the tail of the $m$ posterior distribution goes away and the \pybird\ model agrees more closely with the others. We still find a slightly wider posterior in the $m$ parameter for \velocileptors\ LPT compared to the other models. This can also be attributed to the amount of freedom allowed in the bias parameters, in this case both Lagrangian $b_s$ and $b_3$. If we include Gaussian priors $\mathcal{N}[0,5]$ on both $b_s^{\rm L}$ and $b_3^{\rm L}$, the constraints on ShapeFit parameters tighten enough that \velocileptors\ LPT matches the constraining power of the other models, as shown in the right-hand side of Fig.~\ref{fig: b3_priors}. 
One can make the case that this is no longer a ``maximal'' freedom scenario as we have added several informative priors in order to get the best agreement between all models. We note however that models with bias parameters in the tens or hundreds are not theoretically self-consistent, and we would not be able to trust our results or theory if the posteriors peaked at such values for a given data set. In our case, we do not observe any preference (i.e.\ shifts in the posterior peaks) towards extreme values when allowing more freedom in the bias parameters, but these regions are still allowed by the likelihood. We are therefore justified in cutting out those regions via Gaussian priors. If such priors turn out to be informative, then this should be noted, i.e.\ the data alone are not able to fully constrain the model without additional information.  Our priors should be as broad as possible, and consistent with the model and the data. We refer readers to the appendix of Ref.~\cite{KP5s2-Maus} for a detailed discussion of projection effects and priors relevant to DESI fullshape modeling, along with additional references cited therein that encounter similar issues in other areas of cosmology.




\subsection{Tests on noiseless theoretical model}
\label{sec: results_theory}
 
We next turn to fits of data that are generated from our theoretical models. This ``data'' is shown in Fig.~\ref{fig:mock_data} and is generated by fixing the $\Lambda$CDM parameters to their true Abacus values and then varying the nuisance parameters to best fit the mock data up to $k_{\rm max} = 0.25 h^{-1}$Mpc. These data vectors are noiseless, which allows for cleaner comparisons between the theories without being obscured by cosmic variance or systematics of the simulations. While the previous section compared the performance of the different models on simulations, i.e.\ their ability to constrain parameters to their truth, the purpose of this section is to quantify the agreement of theories. To that end, we cycle through each of the theoretical data sets and perform fits using the two pipelines that were not used to produce said data. We perform these tests as a function of $k_{\rm max}$ in order to observe how the predictions of the models diverge as higher $k$-modes are included in the fits. We generally expect the models to agree better at lower $k \lesssim 0.1 \ihmpc$ (larger scales) as the physics at these scales is well-described by linear theory, which is the same in all of the models. The models differ in their handling of higher-order nonlinear terms, which are more important at higher $k \gtrsim 0.1 \ihmpc$. For these tests, we use the single-box covariance that was also used for the Abacus mocks in the previous section, and we restrict our attention to the minimal freedom parameterization. These tests have been repeated with maximal freedom but the results show the same behavior, so we restrict our discussion to the minimal freedom case for simplicity.

\begin{figure}
\captionsetup[subfigure]{labelformat=empty}
\begin{subfigure}{1\textwidth}
\centering
\includegraphics[width=1\textwidth]{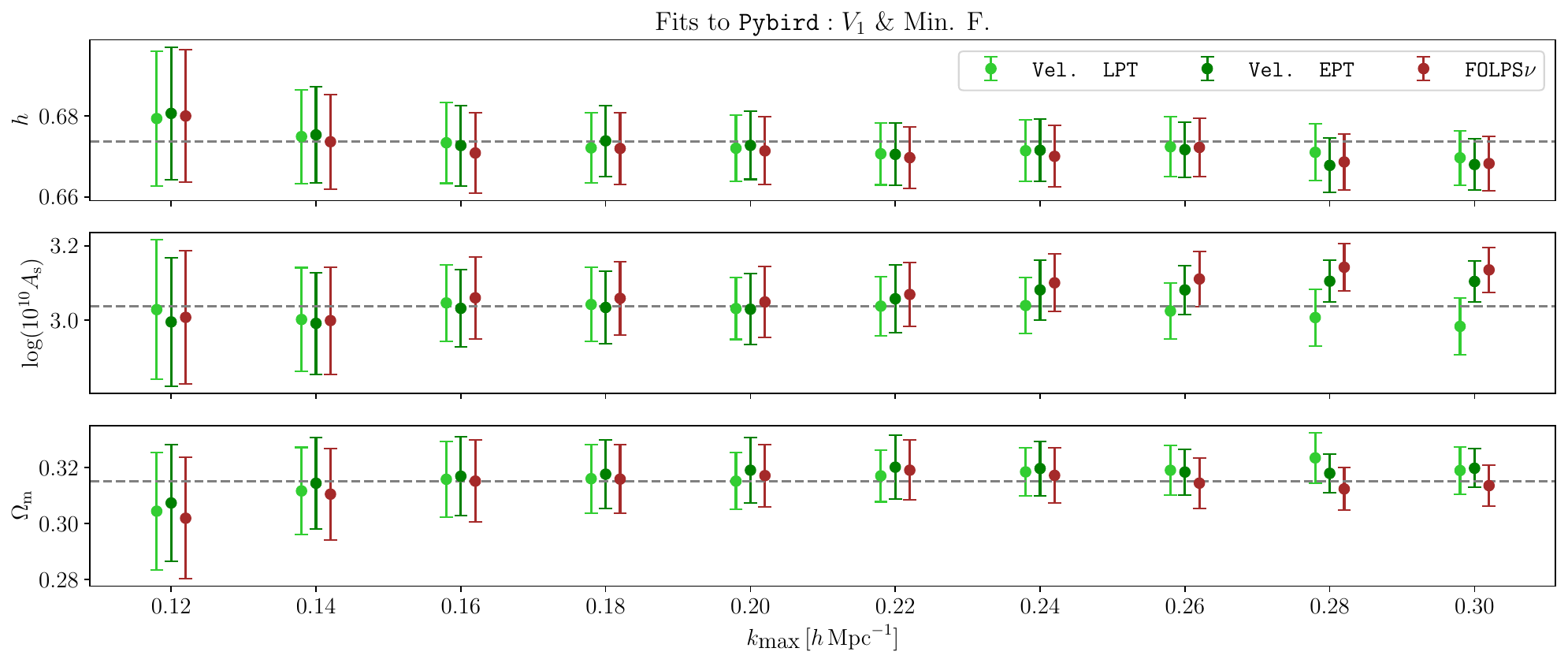}
\end{subfigure}%
\\
\begin{subfigure}{1\textwidth}
\centering
\includegraphics[width=1\textwidth]{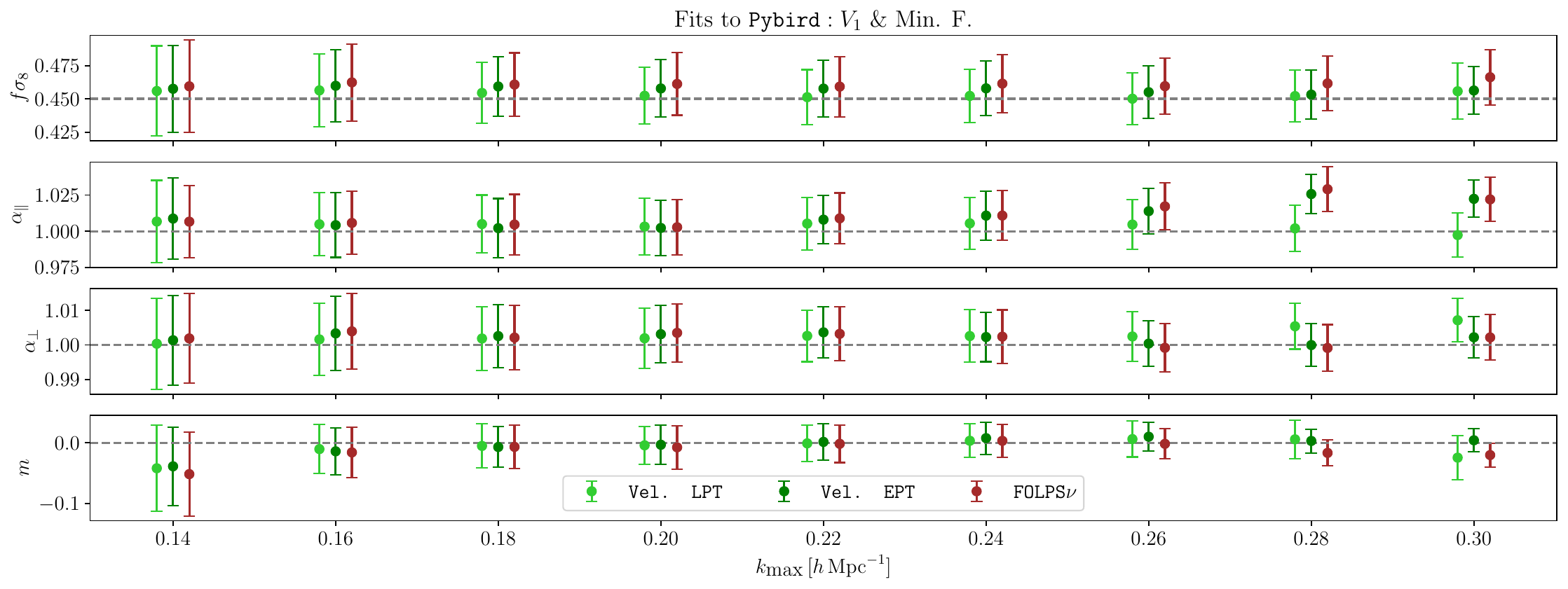}
\end{subfigure}%
\caption{Comparison of 1D constraints of Full-Modeling and ShapeFit fits from velocileptors (EPT and LPT) and \folps\, to the theoretical model generated by PyBird. The theoretical model has $\Lambda$CDM fixed to the true abacus cosmology and nuisance parameters shifted to best-fit the LRG mock data. 
\label{fig: kmax_Pymodel_LRG}}
\end{figure}

In Fig.~\ref{fig: kmax_Pymodel_LRG} we show comparisons of constraints between the LPT and EPT velocileptors and \folps\, models while fitting to the theoretical spectra generated by PyBird to imitate LRG mocks. We present both the Full-Modeling and ShapeFit results. For both methods, the models agree up to a $k_{\rm max}\sim0.24 h^{-1}$Mpc (both with each other and with the PyBird data). For Full-Modeling, see \folps\, shift upwards by $\gtrsim 1\sigma$ from truth in $\log A_{\rm s}$ at $k_{\rm max}\sim0.26 h^{-1}$Mpc, while in the ShapeFit method the velocileptors LPT and \folps\, models disagree with each other in $\alpha_{\parallel}$ by $\gtrsim 1\sigma$ only at $k_{\rm max}\sim0.28 h^{-1}$Mpc and above. However, both velocileptors EPT and \folps\, already deviate from the true PyBird value by $\sim 1\sigma$ at $k_{\rm max}\sim0.26 h^{-1}$Mpc. We also note that the velocileptors EPT model follows the trends of \folps\, in $\alpha_{\parallel,\perp}$ at higher $k_{\rm max}$ rather than agreeing more closely with velocileptors LPT, which does not deviate from the truth by more than $1\sigma$ for any Full-Modeling or ShapeFit parameter until $k_{\rm max}\sim0.3 h^{-1}$Mpc. 

\begin{figure}
 	\begin{center}
 	\includegraphics[width=\linewidth]{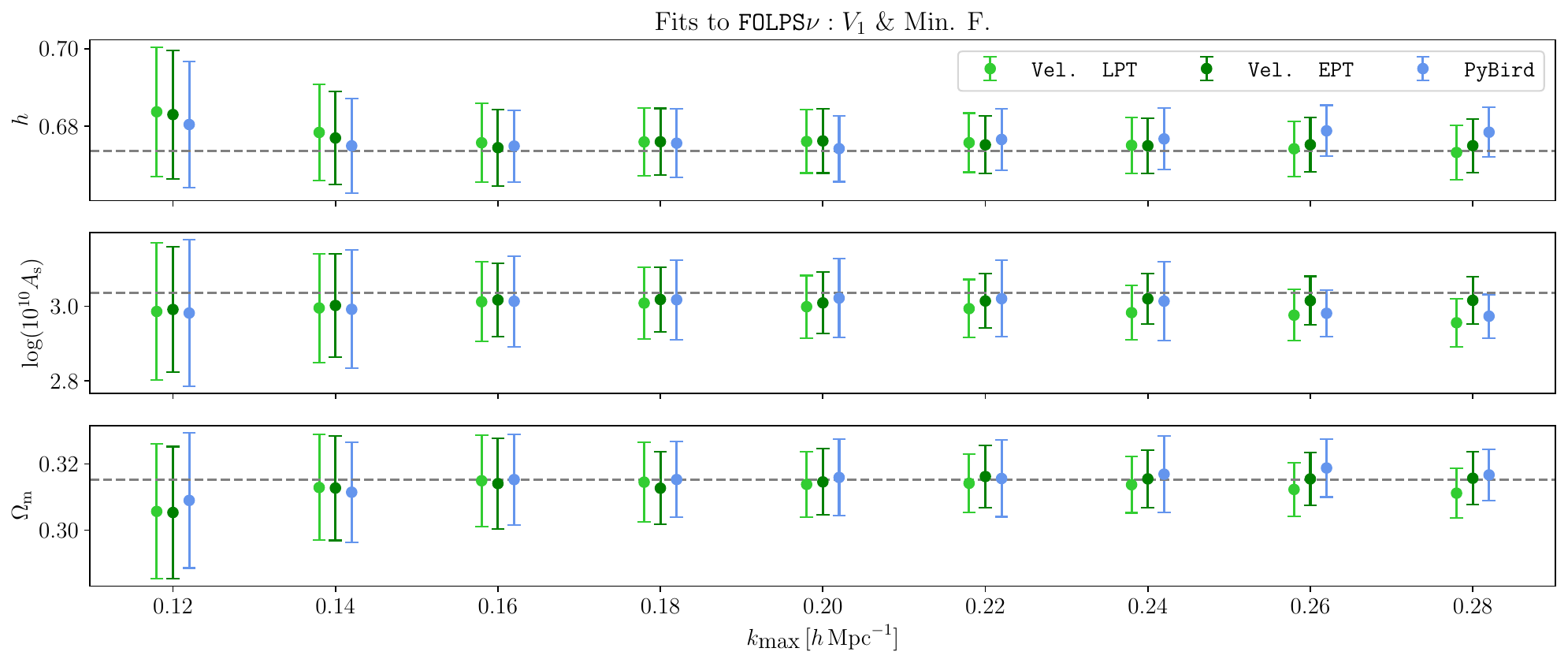}
 	\caption{Comparison of 1D constraints of Full-Modeling  fits from velocileptors (EPT and LPT) and \pybird\, to the theoretical model generated by \folps. The theoretical model has $\Lambda$CDM fixed to the true abacus cosmology and nuisance parameters shifted to best-fit the LRG mock data. 
 	\label{fig: kmax_Fmodel_LRG}}
 	\end{center}
 \end{figure}

\begin{figure}
    \begin{center}
    \includegraphics[width=\linewidth]{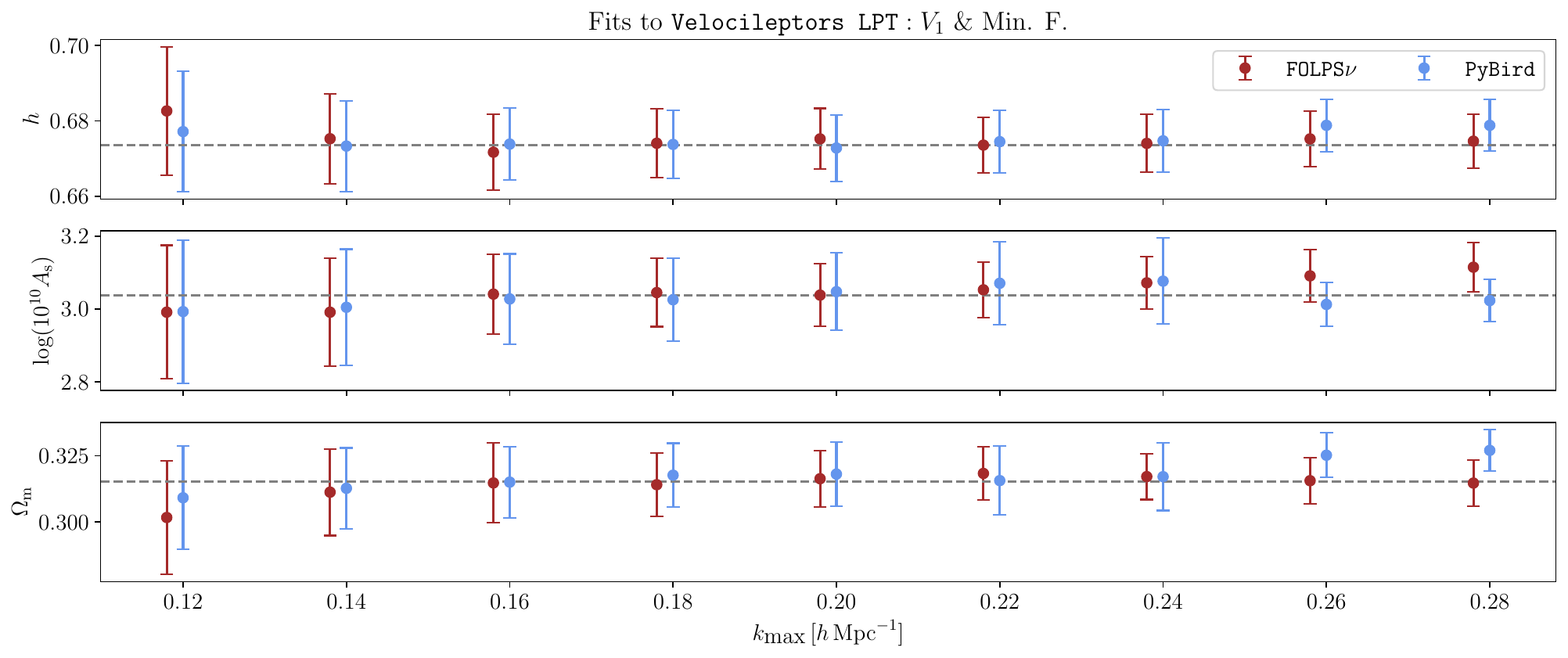}
    \caption{Comparison of 1D constraints of Full-Modeling from \pybird\ and \folps\, to the theoretical model generated by \velocileptors\ LPT. The theoretical model has $\Lambda$CDM fixed to the true abacus cosmology and nuisance parameters shifted to best-fit the LRG mock data. 
    \label{fig: kmax_Vmodel_LRG}}
    \end{center}
\end{figure}

We next repeat this test with noiseless data generated by the \folps\ model in Fig.~\ref{fig: kmax_Fmodel_LRG}, focusing only on Full-modeling for simplicity. In this case, the \velocileptors\ LPT constraint on $\log A_{\rm s}$ begins to deviate from the truth by 1$\sigma$ at $k_{\rm max}\gtrsim0.26 \ihmpc$, while the EPT model remains unbiased for the whole range of $k_{\rm max}$. For the \pybird\ model, we observe similar offsets in $\log A_{\rm s}$ in the same direction as the \velocileptors\ LPT results. Finally, we show in Fig.~\ref{fig: kmax_Vmodel_LRG} the comparison of Full-Modeling constraints from fitting to the data vector generated by the \velocileptors\ LPT model. For the $k_{\rm max} \leq 0.28 \ihmpc$ fits the \pybird\ model is consistent with the \velocileptors\ LPT data whereas the \folps\ model begins to deviate by 1$\sigma$ in $\log A_{\rm s}$ for $k_{\rm max} \gtrsim 0.26 \ihmpc$. From Figs.~\ref{fig: kmax_Pymodel_LRG}-\ref{fig: kmax_Vmodel_LRG} we notice a trend that at lower $k$ all models behave similarly while at higher $k$ the \velocileptors\ LPT and \pybird\ models are more consistent with each other, whereas \folps\ agrees better with \velocileptors\ EPT. We can also see this to some extent in the power spectrum predictions shown in the right-hand panels of Fig.~\ref{fig:mock_data}, where the \velocileptors\ LPT and \pybird\ curves match each other more closely than \folps. This may be due to the differences in IR-resummation schemes between the models. As noted in \S~\ref{sec: diff}, \pybird\ and \velocileptors\ LPT use a similar resummation procedure of exponentiating long-wavelength displacements while perturbatively expanding the short-wavelength modes, while \velocileptors\ EPT and \folps\ both use the wiggle-no wiggle split procedure. Our results suggest that the similarities between \velocileptors\ LPT and \pybird\ (and \folps\ with \velocileptors\ EPT) at high $k$ are related to the IR-resummation methods employed by the models (see ref.~\cite{Chen21} for further discussion). 

In Appendix~\ref{app: ELG_QSO_kmax} we perform similar tests as those described here for theoretical data vectors produced from fits to the ELG and QSO Abacus mocks. Our conclusions mimic those above, and show that our models agree with one another in the range of scales\footnote{The dependence on scale cuts of each model using Abacus mocks is presented and discussed in Refs.~\cite{KP5s2-Maus,KP5s3-Noriega,KP5s4-Lai,KP5s5-Ramirez}} appropriate for Full-Shape analyses using 1-loop perturbation theory.  

Finally, we find the maximum likelihood values from \velocileptors\ LPT and EPT when fit to the theoretical data sets generated by \folps\ and \pybird\ to $k_{\rm max}=0.18 \, \ihmpc$. For each FM and SF parameter, the average shift between best-fit and truth provides a rough estimate for the theoretical-systematic error of the models that can be compared to the statistical errors, $\sigma$. For FM we obtain systematic errors of $\lesssim 0.02\sigma$ in $\Omega_{\rm m}$, $H_0$ and $\log(10^{10}A_{\rm s})$. For SF we find systematic errors of $\lesssim 0.05\sigma$ for $f\sigma_{s8}$, $\alpha_{\parallel}$ and $\alpha_{\perp}$, while in $m$ we get a systematic error of $0.17\sigma$.

\section{Conclusion}
\label{sec:conclusions}

Analyses of two-point statistics from previous galaxy redshift surveys (e.g.\ BOSS and eBOSS) using effective field theories (EFT) of Large Scale Structure have often obtained constraints on cosmological parameters, such as $\sigma_8$, that are in mild tension with one another. These differences can usually be attributed to the settings of pipelines used and modeling choices (see e.g. \cite{Carrilho23,Holm23,Simon23,Linde24}) but could have given the erroneous impression that different effective field theories are not well understood and inconsistent. In anticipation of the Y1 data release of the Dark Energy Spectroscopic Instrument, this paper aims to show that various EFT models/pipelines used for fitting galaxy power spectra in redshift space are actually consistent at the level of precision of a physical survey-- as long as one handles with care the freedom afforded to bias/counter/stochastic terms in the models and is consistent with choice of scale cuts. We focus our attention on three such models: \velocileptors, \folps, and \pybird\, and show that, despite relying on different theoretical frameworks (Eulerian vs Lagrangian perturbation theories, EFT corrections, IR resummation schemes), the predictions from these models are in close agreement with one another up to scales of $k_{\rm max} \simeq 0.25\ihmpc$ and at the level of precision of 8 ($h^{-1}$Gpc)$^3$ survey volumes. Compared to the full DESI footprint the simulation volume used in this comparison is larger than those of the BGS ($0.1<z<0.4$) and two of the LRG ($0.4<z<0.6$,$0.6<z<0.8$) redshift bins, whose volumes are rougbly 2,3,and 5 ($h^{-1}$Gpc)$^3$ respectively. 
The third DESI LRG bin ($0.8<z<1.1$), both ELG bins ($0.8<z<1.1$, $1.1<z<1.6$), and the QSO ($0.8<z<2.1$) are all larger with volumes of ~10, 20, and 60 ($h^{-1}$Gpc)$^3$, and the total comoving volume up to $z= 2.1$ is around 70 ($h^{-1}$Gpc)$^3$\cite{DESI2023a.KP1.SV}. 

We compare our models within two bias parametrization choices, ``minimal'' and ``maximal'' freedom, where the first assumes tidal and third-order biases to be fixed to coevolution and the latter allows all four bias parameters to vary independently with wide, uninformative priors. We show that when fitting Abacus simulations provided in 8 ($h^{-1}$Gpc)$^3$ cubic boxes with three types of tracers (LRG, $z = 0.8$; ELG, $z = 1.1$; QSO, $z = 1.4$), the cosmological constraints are consistent between models in both minimal and maximal freedom cases. We perform this comparison using both parameter compression (``ShapeFit'') and direct fitting (``Full-Modeling'') methods. Despite finding close agreement between the models, we note a degeneracy between the third-order bias and the ShapeFit parameter, m, in the \pybird\ model that results in a tail in the posterior distribution of $m$ for the fits to the quasar sample. By including a Gaussian prior on the $b_3$ parameter, we can minimize this effect without biasing the cosmological constraints. Our conclusions on model consistency are in agreement with those of previous work \cite{Simon23,Linde24}. In particular, Ref.~\cite{Simon23} performed a comparison between \pybird\ and \texttt{CLASS-PT} using BOSS data and found that the primary effect driving the differences between the two models is the different bias parameterizations, and that controlling these differences results in a significant improvement in model consistency. In this paper we come to a similar conclusion, but given the more substantial theoretical differences between theories considered here, do not have a direct mapping of bias parameters between all models available to us. However, the minimal and maximal freedom choices were sufficient in controlling the effect of having different parameterizations between models. The effect of applying Gaussian vs flat priors on $b_3$ in \pybird\ is not directly explored as the comparison used Guassian priors on higher order biases by default. In Ref.~\cite{Linde24} comparisons between \texttt{CLASS-ONELOOP} and \texttt{CLASS-PT} were only made at the level of power spectrum predictions and not constraints on a data set. However, their comparison of $P_\ell(k)$ predictions show similar levels of agreement as the model predictions shown in Fig.~\ref{fig:mock_data} in this paper.\footnote{To be precise, the two models compared in Ref.~\cite{Linde24} employed the exact same bias expansions and, by setting counterterms and stochastic terms to zero, $P_\ell(k)$ predictions were compared under the exact same parameter values. In our case the model predictions used the same $\Lambda$CDM parameters but all other parameters values were independently chosen by fitting to data. So our theoretical spectra differ slightly more than in their tests, but differences are $\mathcal{O}(1\%)$ for $k\leq 0.25 \ \ihmpc$  in both cases.}

In addition to fitting cubic mock samples, we generate noiseless data using best-fitting predictions from each model. For each data vector generated by one model, we run fits using the other models for a variety of $k_{\rm max}$ values. We show that significant deviations of model predictions only occur when raising the scale cuts to $k_{\rm max} \sim 0.26 \ihmpc$ or higher. Since the physics at such small scales (high $k$) is already known to be sensitive to two-loop effects that are not included in our models, we conclude that the stronger restriction on $k_{\rm max}$ comes from the limitation of one-loop perturbation theory and not from disagreements of the theories themselves. For the range of scales, i.e. $0.02 < k \lesssim 0.2 \ihmpc$, that will be the fiducial setting for the DESI Y1 analysis, the EFT models all give consistent cosmological constraints. 

While the settings (priors, bias parametrization, etc.) chosen in this paper allow for better comparison between models, they are not necessarily the optimal settings with which one may use the pipelines in future analyses. Our aim in this work is simply to demonstrate that these theories agree when consistent assumptions are made and pave the way towards the study of modeling systematics for Full-Shape analysis with DESI. A similar comparison of these models along with \texttt{EFT-GSM} is also being performed in configuration space\cite{KP5s5-Ramirez}.
We have thus far focused our analysis on $\Lambda$CDM models only with $n_{\rm s}$ fixed. Additional tests have been performed that open up the parameter space beyond $\Lambda$CDM and study/mitigate the parameter projection effects that arise in more exotic models. These will be presented in future publications(\cite{KP5s2-Maus,KP5s3-Noriega,KP5s4-Lai,KP5s5-Ramirez}). 

\section{Data availability}

Chains from the plots in this paper are available on Zenodo as part of DESI’s Data Management Plan (DOI: \href{https://doi.org/10.5281/zenodo.10823205}{https://doi.org/10.5281/zenodo.10823205}). The data used in this analysis will be made public along the Data Release 1 (details in \href{https://data.desi.lbl.gov/doc/releases/}{https://data.desi.lbl.gov/doc/releases/})

\section*{Acknowledgements}
We thank Arnaud de Mattia for implementing our findings into the DESI pipeline and also thank the members of the Galaxy and Quasar Clustering working group within DESI for helpful discussions pertaining to this work. 

MM is supported by the U.S. Department of Energy (DOE). YL is supported by  the Australian government through the Research Training Program scholarship, Australian Research Council’s Laureate Fellowship (project FL180100168) and Discovery Project (project DP20220101395) funding schemes. HN and SR are supported by Investigacion in Ciencia Basica CONAHCYT grant No. A1-S-13051 and PAPIIT IN108321 and IN116024, and Proyecto PIFF. HN is supported by Ciencia de Frontera grant No. 319359 and PAPIIT IG102123. This material is based upon work supported by the U.S. Department of Energy (DOE), Office of Science, Office of High-Energy Physics, under Contract No. DE–AC02–05CH11231, and by the National Energy Research Scientific Computing Center, a DOE Office of Science User Facility under the same contract. Additional support for DESI was provided by the U.S. National Science Foundation (NSF), Division of Astronomical Sciences under Contract No. AST-0950945 to the NSF’s National Optical-Infrared Astronomy Research Laboratory; the Science and Technology Facilities Council of the United Kingdom; the Gordon and Betty Moore Foundation; the Heising-Simons Foundation; the French Alternative Energies and Atomic Energy Commission (CEA); the National Council of Humanities, Science and Technology of Mexico (CONAHCYT); the Ministry of Science and Innovation of Spain (MICINN), and by the DESI Member Institutions: \url{https://www.desi.lbl.gov/collaborating-institutions}. Any opinions, findings, and conclusions or recommendations expressed in this material are those of the author(s) and do not necessarily reflect the views of the U. S. National Science Foundation, the U. S. Department of Energy, or any of the listed funding agencies.

The authors are honored to be permitted to conduct scientific research on Iolkam Du’ag (Kitt Peak), a mountain with particular significance to the Tohono O’odham Nation.

\appendix

\section{Tests on noiseless data for ELG and QSO redshift bins}
\label{app: ELG_QSO_kmax}

\begin{figure}
\captionsetup[subfigure]{labelformat=empty}
\begin{subfigure}{1\textwidth}
\centering
\includegraphics[width=1\textwidth]{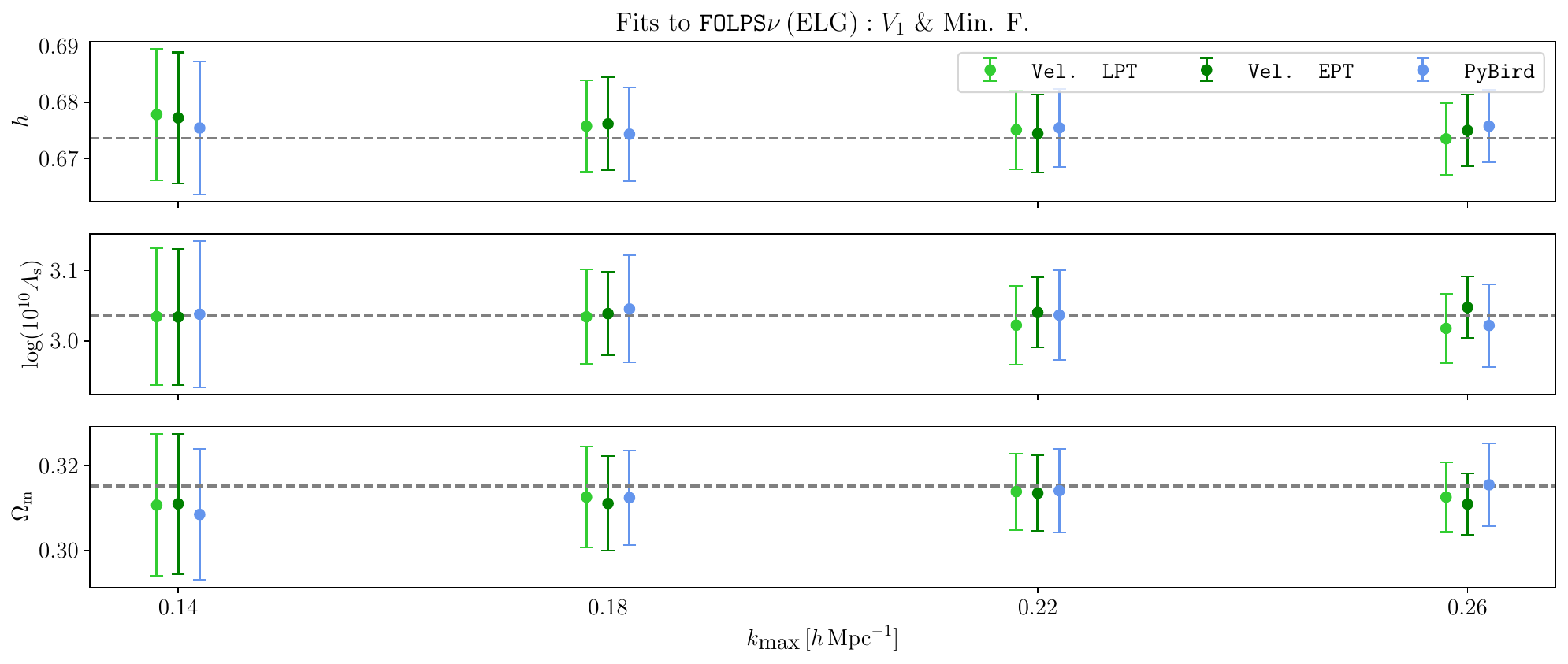}
\end{subfigure}%
\\
\begin{subfigure}{1\textwidth}
\centering
\includegraphics[width=1\textwidth]{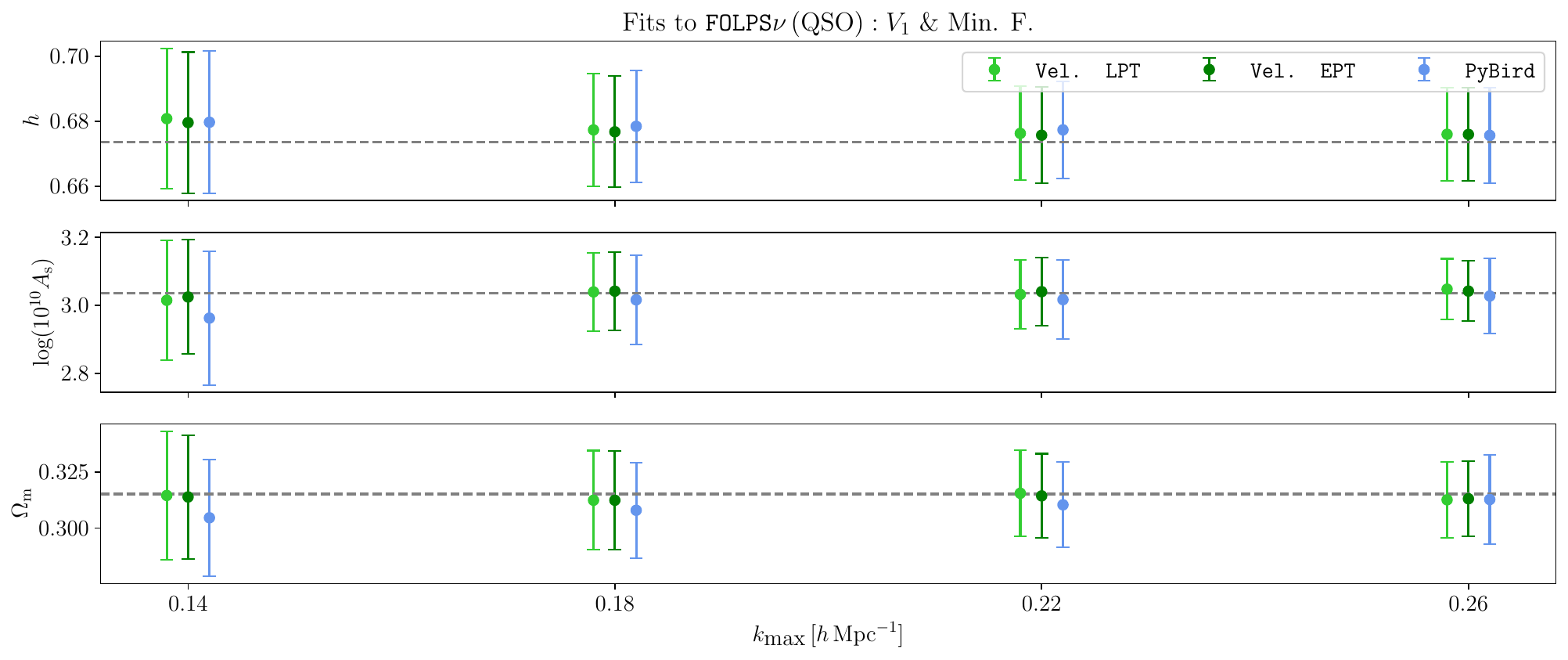}
\end{subfigure}%
\caption{Comparison of 1D constraints of Full-Modeling fits from velocileptors (EPT and LPT) and PyBird to the theoretical model generated by \folps. The theoretical model has $\Lambda$CDM fixed to the true abacus cosmology and nuisance parameters shifted to best-fit the ELG (top) and QSO (bottom) mock data. 
\label{fig: kmax_FOLPS_ELG_QSO}}
\end{figure}

\begin{figure}
\captionsetup[subfigure]{labelformat=empty}
\begin{subfigure}{1\textwidth}
\centering
\includegraphics[width=1\textwidth]{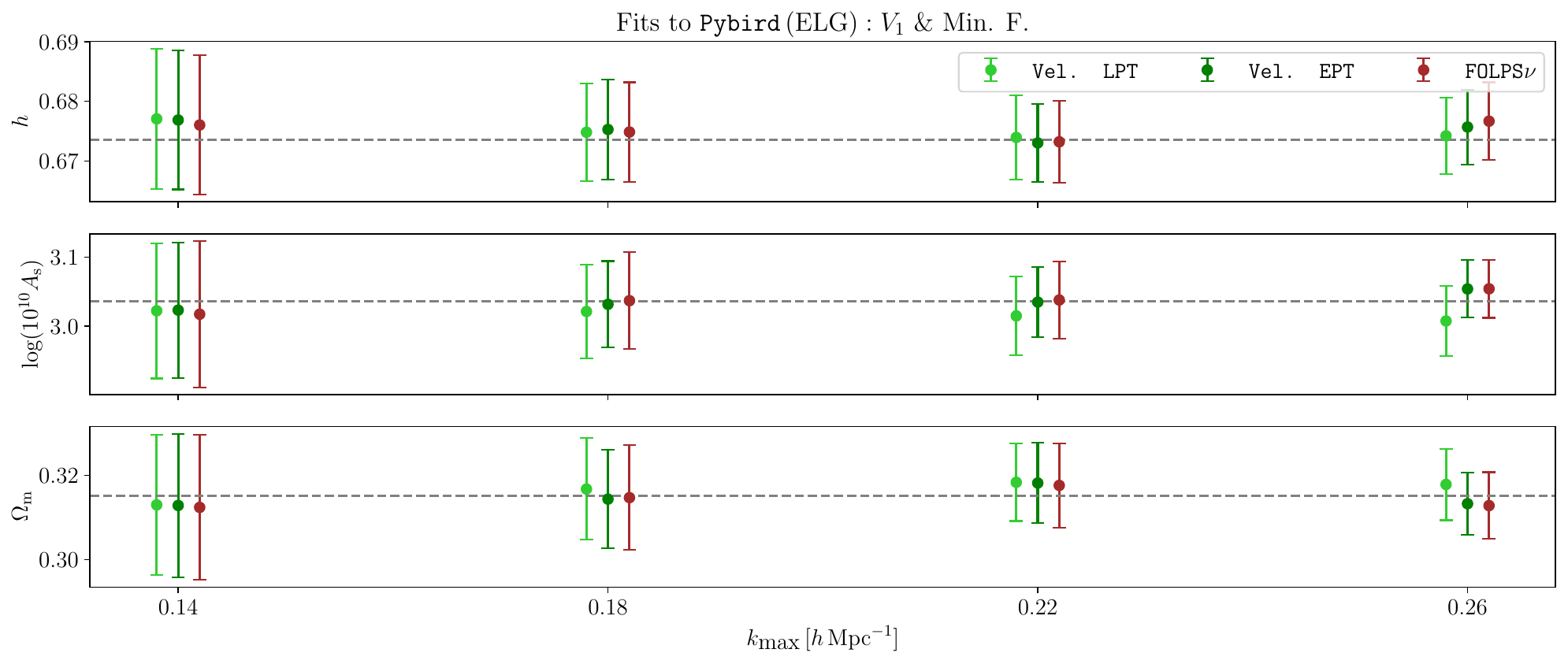}
\end{subfigure}%
\\
\begin{subfigure}{1\textwidth}
\centering
\includegraphics[width=1\textwidth]{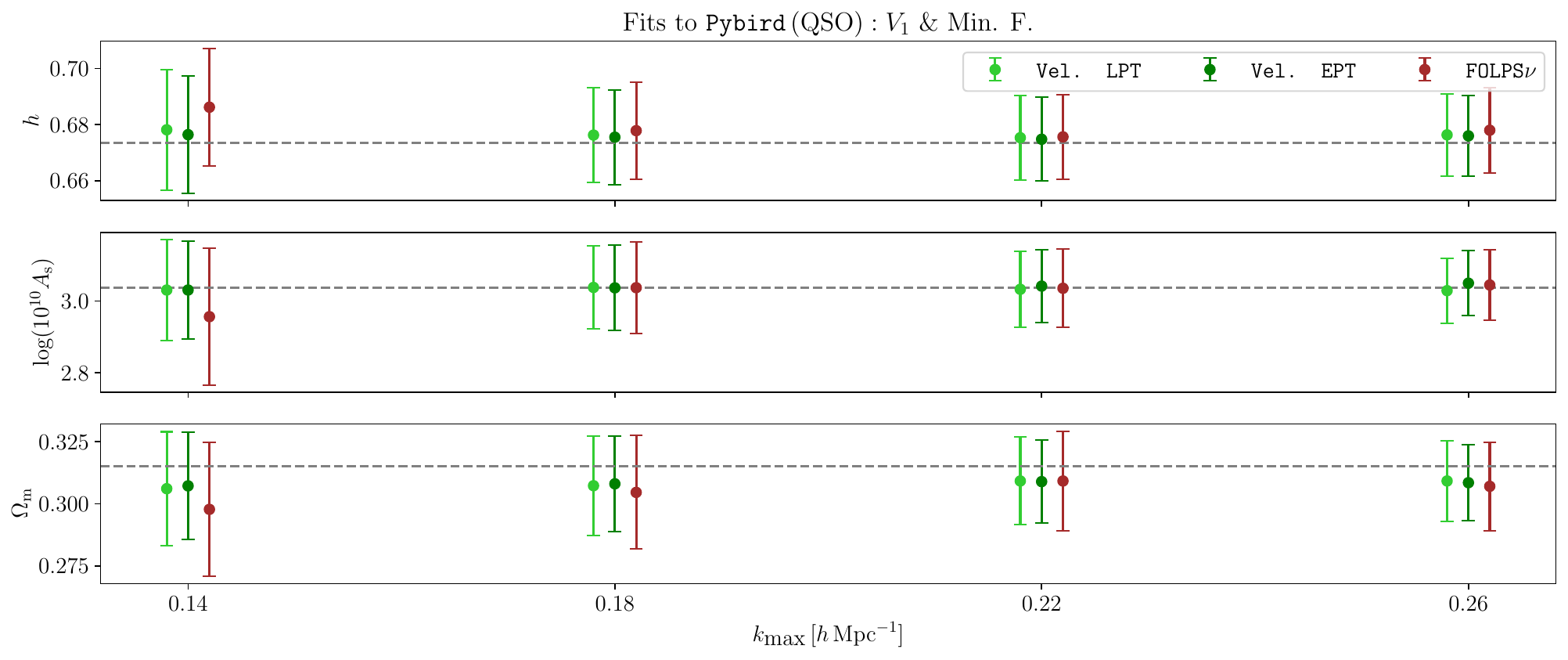}
\end{subfigure}%
\caption{Comparison of 1D constraints of Full-Modeling fits from velocileptors (EPT and LPT) and \folps\, to the theoretical model generated by PyBird. The theoretical model has $\Lambda$CDM fixed to the true abacus cosmology and nuisance parameters shifted to best-fit the ELG (top) and QSO (bottom) mock data. 
\label{fig: kmax_PyBird_ELG_QSO}}
\end{figure}

\begin{figure}
\captionsetup[subfigure]{labelformat=empty}
\begin{subfigure}{1\textwidth}
\centering
\includegraphics[width=1\textwidth]{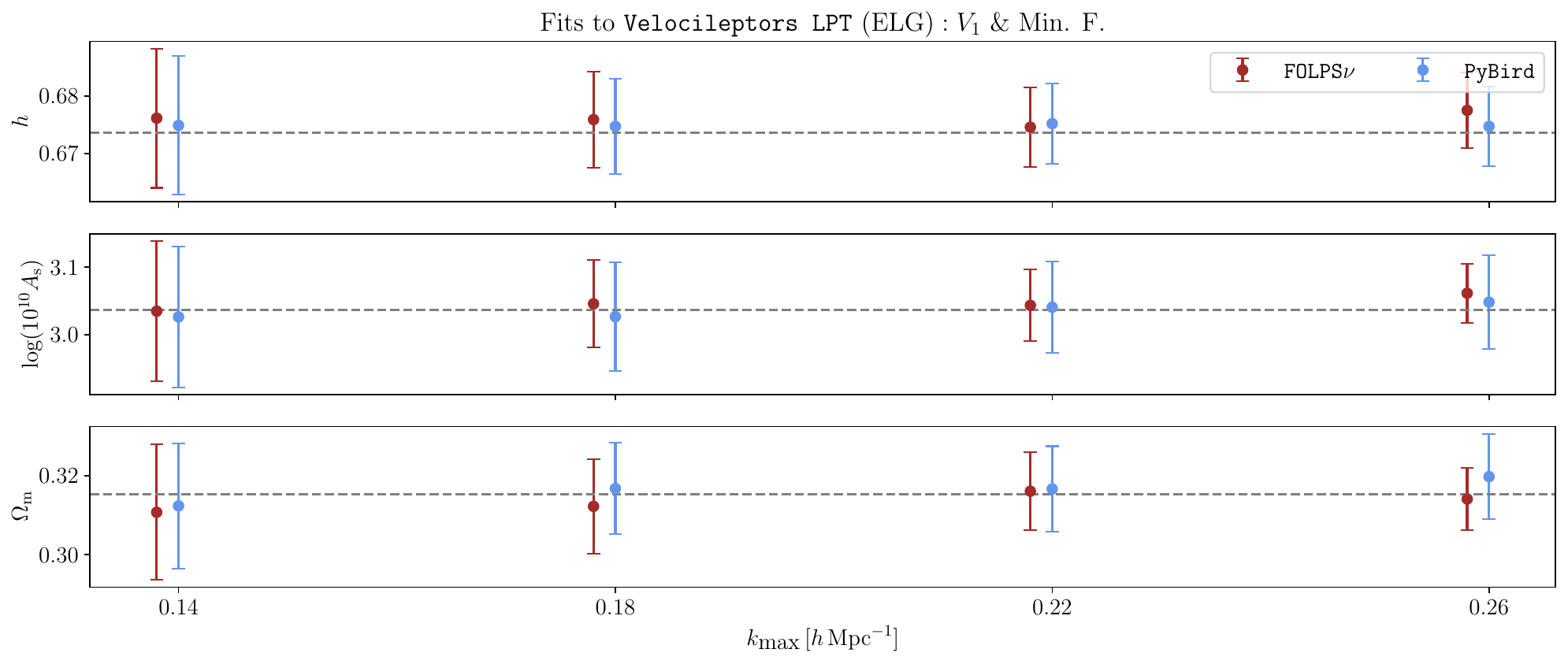}
\end{subfigure}%
\\
\begin{subfigure}{1\textwidth}
\centering
\includegraphics[width=1\textwidth]{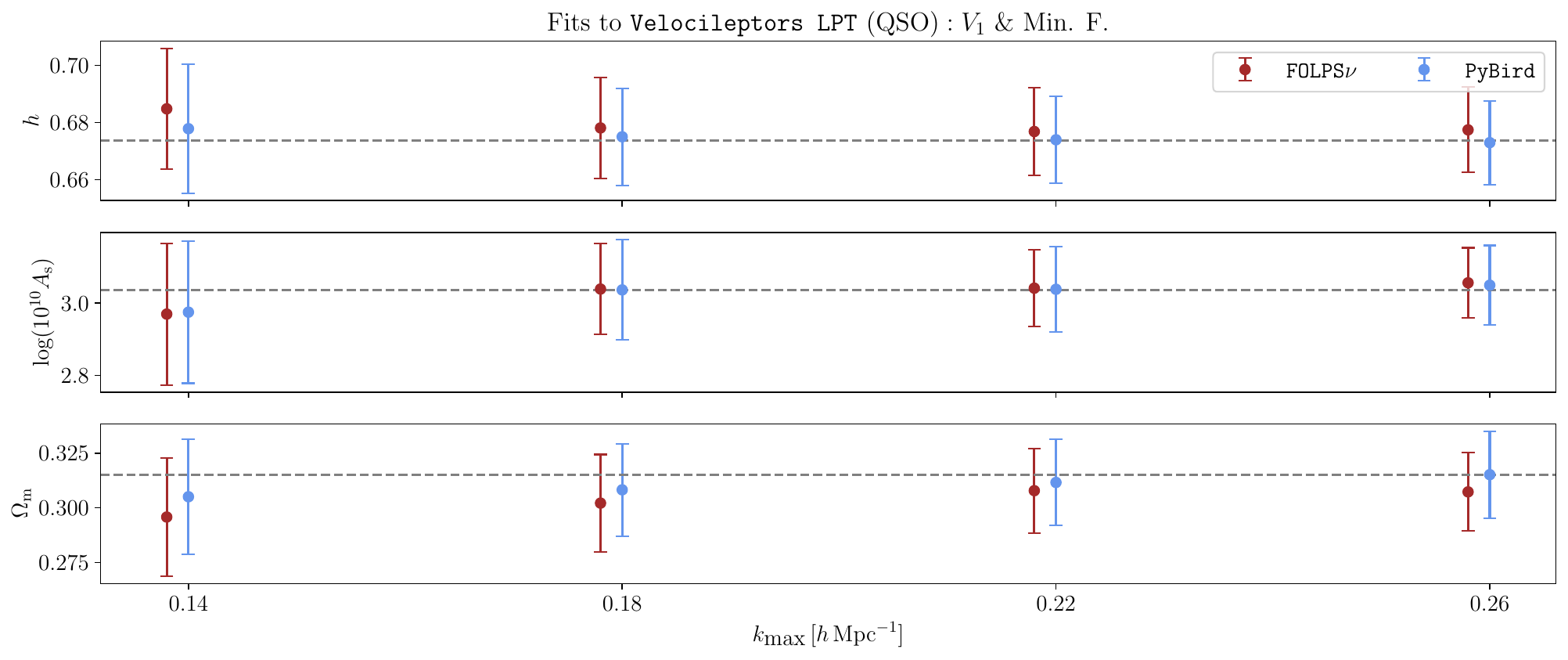}
\end{subfigure}%
\caption{Comparison of 1D constraints of Full-Modeling fits from PyBird and \folps\, to the theoretical model generated by velocileptors LPT. The theoretical model has $\Lambda$CDM fixed to the true abacus cosmology and nuisance parameters shifted to best-fit the ELG (top) and QSO (bottom) mock data. 
\label{fig: kmax_Velo_ELG_QSO}}
\end{figure}

We repeat the above tests with noiseless data mimicking the ELG and QSO mock samples in Figs.~\ref{fig: kmax_FOLPS_ELG_QSO}-\ref{fig: kmax_Velo_ELG_QSO}, this time with fewer $k_{\rm max}$ points just to see if any new behavior shows up at higher redshifts. We observe that all models are more consistent with each other up through $k_{\rm max}=0.26 \ihmpc$ than for the LRG case, and that the maximum $k_{\rm max}$ for which our models agree is not as restricted for ELG and QSO tracers as the LRGs. That being said, even for the LRG tests the models behave consistently for $k_{\rm max}\lesssim 0.24 \ihmpc$, which is a looser bound on $k_{\rm max}$ than what we observe when fitting each model to the actual abacus mock data sets (see individual papers \cite{KP5s2-Maus,KP5s3-Noriega,KP5s4-Lai,KP5s5-Ramirez}). We therefore conclude from this section that the EFT models discussed in this paper are consistent for the range of scales that will be used in the analysis of real DESI data, and that any significant deviations between models only occur at scales where real data is already known to not be well-described by one-loop theories. 

\section{Author Affiliations}
\label{sec:affiliations}

\begin{hangparas}{.5cm}{1}

$^{1}${Department of Physics, University of California, Berkeley, CA 94720, USA}

$^{2}${Lawrence Berkeley National Laboratory, 1 Cyclotron Road, Berkeley, CA 94720, USA}

$^{3}${School of Mathematics and Physics, University of Queensland, 4072, Australia}

$^{4}${Instituto de Ciencias F\'{\i}sicas, Universidad Aut\'onoma de M\'exico, Cuernavaca, Morelos, 62210, (M\'exico)}

$^{5}${Instituto de F\'{\i}sica, Universidad Nacional Aut\'{o}noma de M\'{e}xico,  Cd. de M\'{e}xico  C.P. 04510,  M\'{e}xico}

$^{6}${Instituto Avanzado de Cosmolog\'{\i}a A.~C., San Marcos 11 - Atenas 202. Magdalena Contreras, 10720. Ciudad de M\'{e}xico, M\'{e}xico}

$^{7}${Institute for Advanced Study, 1 Einstein Drive, Princeton, NJ 08540, USA}

$^{8}${Departament de F\'{\i}sica Qu\`{a}ntica i Astrof\'{\i}sica, Universitat de Barcelona, Mart\'{\i} i Franqu\`{e}s 1, E08028 Barcelona, Spain}

$^{9}${Institut d'Estudis Espacials de Catalunya (IEEC), 08034 Barcelona, Spain}

$^{10}${Institut de Ci\`encies del Cosmos (ICCUB), Universitat de Barcelona (UB), c. Mart\'i i Franqu\`es, 1, 08028 Barcelona, Spain.}

$^{11}${Sorbonne Universit\'{e}, CNRS/IN2P3, Laboratoire de Physique Nucl\'{e}aire et de Hautes Energies (LPNHE), FR-75005 Paris, France}

$^{12}${}

$^{13}${Physics Dept., Boston University, 590 Commonwealth Avenue, Boston, MA 02215, USA}

$^{14}${University of Michigan, Ann Arbor, MI 48109, USA}

$^{15}${Institute for Astronomy, University of Edinburgh, Royal Observatory, Blackford Hill, Edinburgh EH9 3HJ, UK}

$^{16}${Department of Physics \& Astronomy, University College London, Gower Street, London, WC1E 6BT, UK}

$^{17}${IRFU, CEA, Universit\'{e} Paris-Saclay, F-91191 Gif-sur-Yvette, France}

$^{18}${Institute for Computational Cosmology, Department of Physics, Durham University, South Road, Durham DH1 3LE, UK}

$^{19}${Department of Physics and Astronomy, The University of Utah, 115 South 1400 East, Salt Lake City, UT 84112, USA}

$^{20}${Korea Astronomy and Space Science Institute, 776, Daedeokdae-ro, Yuseong-gu, Daejeon 34055, Republic of Korea}

$^{21}${University of California, Berkeley, 110 Sproul Hall \#5800 Berkeley, CA 94720, USA}

$^{22}${Institute of Cosmology and Gravitation, University of Portsmouth, Dennis Sciama Building, Portsmouth, PO1 3FX, UK}

$^{23}${Departamento de F\'isica, Universidad de los Andes, Cra. 1 No. 18A-10, Edificio Ip, CP 111711, Bogot\'a, Colombia}

$^{24}${Observatorio Astron\'omico, Universidad de los Andes, Cra. 1 No. 18A-10, Edificio H, CP 111711 Bogot\'a, Colombia}

$^{25}${Institute of Space Sciences, ICE-CSIC, Campus UAB, Carrer de Can Magrans s/n, 08913 Bellaterra, Barcelona, Spain}

$^{26}${Department of Astrophysical Sciences, Princeton University, Princeton NJ 08544, USA}

$^{27}${Center for Cosmology and AstroParticle Physics, The Ohio State University, 191 West Woodruff Avenue, Columbus, OH 43210, USA}

$^{28}${Department of Physics, The Ohio State University, 191 West Woodruff Avenue, Columbus, OH 43210, USA}

$^{29}${The Ohio State University, Columbus, 43210 OH, USA}

$^{30}${Department of Physics, The University of Texas at Dallas, Richardson, TX 75080, USA}

$^{31}${Departament de F\'{i}sica, Serra H\'{u}nter, Universitat Aut\`{o}noma de Barcelona, 08193 Bellaterra (Barcelona), Spain}

$^{32}${Institut de F\'{i}sica d’Altes Energies (IFAE), The Barcelona Institute of Science and Technology, Campus UAB, 08193 Bellaterra Barcelona, Spain}

$^{33}${Instituci\'{o} Catalana de Recerca i Estudis Avan\c{c}ats, Passeig de Llu\'{\i}s Companys, 23, 08010 Barcelona, Spain}

$^{34}${Department of Physics and Astronomy, University of Sussex, Brighton BN1 9QH, U.K}

$^{35}${Departamento de F\'{i}sica, Universidad de Guanajuato - DCI, C.P. 37150, Leon, Guanajuato, M\'{e}xico}

$^{36}${Department of Physics and Astronomy, University of Waterloo, 200 University Ave W, Waterloo, ON N2L 3G1, Canada}

$^{37}${Perimeter Institute for Theoretical Physics, 31 Caroline St. North, Waterloo, ON N2L 2Y5, Canada}

$^{38}${Waterloo Centre for Astrophysics, University of Waterloo, 200 University Ave W, Waterloo, ON N2L 3G1, Canada}

$^{39}${Space Sciences Laboratory, University of California, Berkeley, 7 Gauss Way, Berkeley, CA  94720, USA}

$^{40}${Instituto de Astrof\'{i}sica de Andaluc\'{i}a (CSIC), Glorieta de la Astronom\'{i}a, s/n, E-18008 Granada, Spain}

$^{41}${Department of Physics, Kansas State University, 116 Cardwell Hall, Manhattan, KS 66506, USA}

$^{42}${Ecole Polytechnique F\'{e}d\'{e}rale de Lausanne, CH-1015 Lausanne, Switzerland}

$^{43}${Department of Physics and Astronomy, Sejong University, Seoul, 143-747, Korea}

$^{44}${CIEMAT, Avenida Complutense 40, E-28040 Madrid, Spain}

$^{45}${Department of Physics, University of Michigan, Ann Arbor, MI 48109, USA}

$^{46}${NSF NOIRLab, 950 N. Cherry Ave., Tucson, AZ 85719, USA}

$^{47}${SLAC National Accelerator Laboratory, Menlo Park, CA 94305, USA}

$^{48}${National Astronomical Observatories, Chinese Academy of Sciences, A20 Datun Rd., Chaoyang District, Beijing, 100012, P.R. China}

\end{hangparas}

\bibliography{main}

\providecommand{\href}[2]{#2}\begingroup\raggedright\begin{thebibliography}{10}

\bibitem{SDSS2000}
D.G.~{York}, J.~{Adelman}, J.~{Anderson}, John~E., S.F.~{Anderson}, J.~{Annis},
  N.A.~{Bahcall} et~al., \emph{{The Sloan Digital Sky Survey: Technical
  Summary}}, \href{https://doi.org/10.1086/301513}{\emph{\aj} {\bfseries 120}
  (2000) 1579} [\href{https://arxiv.org/abs/astro-ph/0006396}{{\ttfamily
  astro-ph/0006396}}].

\bibitem{Dawson13}
K.S.~{Dawson}, D.J.~{Schlegel}, C.P.~{Ahn}, S.F.~{Anderson}, {\'E}.~{Aubourg},
  S.~{Bailey} et~al., \emph{{The Baryon Oscillation Spectroscopic Survey of
  SDSS-III}}, \href{https://doi.org/10.1088/0004-6256/145/1/10}{\emph{\aj}
  {\bfseries 145} (2013) 10} [\href{https://arxiv.org/abs/1208.0022}{{\ttfamily
  1208.0022}}].

\bibitem{Reid16}
B.~{Reid}, S.~{Ho}, N.~{Padmanabhan}, W.J.~{Percival}, J.~{Tinker},
  R.~{Tojeiro} et~al., \emph{{SDSS-III Baryon Oscillation Spectroscopic Survey
  Data Release 12: galaxy target selection and large-scale structure
  catalogues}}, \href{https://doi.org/10.1093/mnras/stv2382}{\emph{\mnras}
  {\bfseries 455} (2016) 1553}
  [\href{https://arxiv.org/abs/1509.06529}{{\ttfamily 1509.06529}}].

\bibitem{Alam17}
S.~{Alam}, M.~{Ata}, S.~{Bailey}, F.~{Beutler}, D.~{Bizyaev}, J.A.~{Blazek}
  et~al., \emph{{The clustering of galaxies in the completed SDSS-III Baryon
  Oscillation Spectroscopic Survey: cosmological analysis of the DR12 galaxy
  sample}}, \href{https://doi.org/10.1093/mnras/stx721}{\emph{\mnras}
  {\bfseries 470} (2017) 2617}
  [\href{https://arxiv.org/abs/1607.03155}{{\ttfamily 1607.03155}}].

\bibitem{SDSSIII2015}
S.~{Alam}, F.D.~{Albareti}, C.~{Allende Prieto}, F.~{Anders}, S.F.~{Anderson},
  T.~{Anderton} et~al., \emph{{The Eleventh and Twelfth Data Releases of the
  Sloan Digital Sky Survey: Final Data from SDSS-III}},
  \href{https://doi.org/10.1088/0067-0049/219/1/12}{\emph{\apjs} {\bfseries
  219} (2015) 12} [\href{https://arxiv.org/abs/1501.00963}{{\ttfamily
  1501.00963}}].

\bibitem{eBOSS:2020yzd}
{\scshape eBOSS} collaboration, \emph{{Completed SDSS-IV extended Baryon
  Oscillation Spectroscopic Survey: Cosmological implications from two decades
  of spectroscopic surveys at the Apache Point Observatory}},
  \href{https://doi.org/10.1103/PhysRevD.103.083533}{\emph{Phys. Rev. D}
  {\bfseries 103} (2021) 083533}
  [\href{https://arxiv.org/abs/2007.08991}{{\ttfamily 2007.08991}}].

\bibitem{Bautista18}
J.E.~{Bautista}, M.~{Vargas-Maga{\~n}a}, K.S.~{Dawson}, W.J.~{Percival},
  J.~{Brinkmann}, J.~{Brownstein} et~al., \emph{{The SDSS-IV Extended Baryon
  Oscillation Spectroscopic Survey: Baryon Acoustic Oscillations at Redshift of
  0.72 with the DR14 Luminous Red Galaxy Sample}},
  \href{https://doi.org/10.3847/1538-4357/aacea5}{\emph{\apj} {\bfseries 863}
  (2018) 110} [\href{https://arxiv.org/abs/1712.08064}{{\ttfamily
  1712.08064}}].

\bibitem{DESI}
{DESI Collaboration}, A.~{Aghamousa}, J.~{Aguilar}, S.~{Ahlen}, S.~{Alam},
  L.E.~{Allen} et~al., \emph{{The DESI Experiment Part I: Science,Targeting,
  and Survey Design}}, {\emph{ArXiv e-prints} (2016) }
  [\href{https://arxiv.org/abs/1611.00036}{{\ttfamily 1611.00036}}].

\bibitem{DESI2016_2}
{DESI Collaboration}, A.~{Aghamousa}, J.~{Aguilar}, S.~{Ahlen}, S.~{Alam},
  L.E.~{Allen} et~al., \emph{{The DESI Experiment Part II: Instrument Design}},
  \href{https://doi.org/10.48550/arXiv.1611.00037}{\emph{arXiv e-prints} (2016)
  arXiv:1611.00037} [\href{https://arxiv.org/abs/1611.00037}{{\ttfamily
  1611.00037}}].

\bibitem{DESI2022}
{DESI Collaboration}, B.~{Abareshi}, J.~{Aguilar}, S.~{Ahlen}, S.~{Alam},
  D.M.~{Alexander} et~al., \emph{{Overview of the Instrumentation for the Dark
  Energy Spectroscopic Instrument}},
  \href{https://doi.org/10.3847/1538-3881/ac882b}{\emph{\aj} {\bfseries 164}
  (2022) 207} [\href{https://arxiv.org/abs/2205.10939}{{\ttfamily
  2205.10939}}].

\bibitem{DESI2023}
{DESI Collaboration}, A.G.~{Adame}, J.~{Aguilar}, S.~{Ahlen}, S.~{Alam},
  G.~{Aldering} et~al., \emph{{Validation of the Scientific Program for the
  Dark Energy Spectroscopic Instrument}},
  \href{https://doi.org/10.48550/arXiv.2306.06307}{\emph{arXiv e-prints} (2023)
  arXiv:2306.06307} [\href{https://arxiv.org/abs/2306.06307}{{\ttfamily
  2306.06307}}].

\bibitem{DESI2023_2}
{DESI Collaboration}, A.G.~{Adame}, J.~{Aguilar}, S.~{Ahlen}, S.~{Alam},
  G.~{Aldering} et~al., \emph{{The Early Data Release of the Dark Energy
  Spectroscopic Instrument}},
  \href{https://doi.org/10.48550/arXiv.2306.06308}{\emph{arXiv e-prints} (2023)
  arXiv:2306.06308} [\href{https://arxiv.org/abs/2306.06308}{{\ttfamily
  2306.06308}}].

\bibitem{Euclid}
R.~{Laureijs}, J.~{Amiaux}, S.~{Arduini}, J..~{Augu{\`e}res}, J.~{Brinchmann},
  R.~{Cole} et~al., \emph{{Euclid Definition Study Report}}, {\emph{ArXiv
  e-prints} (2011) } [\href{https://arxiv.org/abs/1110.3193}{{\ttfamily
  1110.3193}}].

\bibitem{Schlegel22}
D.J.~{Schlegel}, S.~{Ferraro}, G.~{Aldering}, C.~{Baltay}, S.~{BenZvi},
  R.~{Besuner} et~al., \emph{{A Spectroscopic Road Map for Cosmic Frontier:
  DESI, DESI-II, Stage-5}}, {\emph{arXiv e-prints} (2022) arXiv:2209.03585}
  [\href{https://arxiv.org/abs/2209.03585}{{\ttfamily 2209.03585}}].

\bibitem{Bernardeau02}
F.~{Bernardeau}, S.~{Colombi}, E.~{Gazta{\~n}aga} and R.~{Scoccimarro},
  \emph{{Large-scale structure of the Universe and cosmological perturbation
  theory}},
  \href{https://doi.org/10.1016/S0370-1573(02)00135-7}{\emph{\physrep}
  {\bfseries 367} (2002) 1}
  [\href{https://arxiv.org/abs/astro-ph/0112551}{{\ttfamily
  astro-ph/0112551}}].

\bibitem{Ivanov_EFT22}
M.M.~{Ivanov}, \emph{{Effective Field Theory for Large Scale Structure}},
  \href{https://doi.org/10.48550/arXiv.2212.08488}{\emph{arXiv e-prints} (2022)
  arXiv:2212.08488} [\href{https://arxiv.org/abs/2212.08488}{{\ttfamily
  2212.08488}}].

\bibitem{Vlah15}
Z.~{Vlah}, M.~{White} and A.~{Aviles}, \emph{{A Lagrangian effective field
  theory}}, \href{https://doi.org/10.1088/1475-7516/2015/09/014}{\emph{\jcap}
  {\bfseries 9} (2015) 014} [\href{https://arxiv.org/abs/1506.05264}{{\ttfamily
  1506.05264}}].

\bibitem{ChenVlahWhite20b}
S.-F.~{Chen}, Z.~{Vlah} and M.~{White}, \emph{{Modeling features in the
  redshift-space halo power spectrum with perturbation theory}},
  \href{https://doi.org/10.1088/1475-7516/2020/11/035}{\emph{\jcap} {\bfseries
  2020} (2020) 035} [\href{https://arxiv.org/abs/2007.00704}{{\ttfamily
  2007.00704}}].

\bibitem{Bharadwaj96}
S.~{Bharadwaj}, \emph{{The Evolution of Correlation Functions in the Zeldovich
  Approximation and Its Implications for the Validity of Perturbation Theory}},
  \href{https://doi.org/10.1086/178036}{\emph{\apj} {\bfseries 472} (1996) 1}
  [\href{https://arxiv.org/abs/arXiv:astro-ph/9606121}{{\ttfamily
  arXiv:astro-ph/9606121}}].

\bibitem{Meiksin99}
A.~{Meiksin}, M.~{White} and J.A.~{Peacock}, \emph{{Baryonic signatures in
  large-scale structure}},
  \href{https://doi.org/10.1046/j.1365-8711.1999.02369.x}{\emph{\mnras}
  {\bfseries 304} (1999) 851}
  [\href{https://arxiv.org/abs/astro-ph/9812214}{{\ttfamily
  astro-ph/9812214}}].

\bibitem{ESW07}
D.J.~{Eisenstein}, H.-J.~{Seo} and M.~{White}, \emph{{On the Robustness of the
  Acoustic Scale in the Low-Redshift Clustering of Matter}},
  \href{https://doi.org/10.1086/518755}{\emph{\apj} {\bfseries 664} (2007) 660}
  [\href{https://arxiv.org/abs/astro-ph/0604361}{{\ttfamily
  astro-ph/0604361}}].

\bibitem{Smith08}
R.E.~{Smith}, R.~{Scoccimarro} and R.K.~{Sheth}, \emph{{Motion of the acoustic
  peak in the correlation function}},
  \href{https://doi.org/10.1103/PhysRevD.77.043525}{\emph{\prd} {\bfseries 77}
  (2008) 043525} [\href{https://arxiv.org/abs/astro-ph/0703620}{{\ttfamily
  astro-ph/0703620}}].

\bibitem{Mat08a}
T.~{Matsubara}, \emph{{Resumming cosmological perturbations via the Lagrangian
  picture: One-loop results in real space and in redshift space}},
  \href{https://doi.org/10.1103/PhysRevD.77.063530}{\emph{\prd} {\bfseries 77}
  (2008) 063530} [\href{https://arxiv.org/abs/0711.2521}{{\ttfamily
  0711.2521}}].

\bibitem{Kaiser87}
N.~{Kaiser}, \emph{{Clustering in real space and in redshift space}},
  \href{https://doi.org/10.1093/mnras/227.1.1}{\emph{\mnras} {\bfseries 227}
  (1987) 1}.

\bibitem{Hamilton98}
A.J.S.~{Hamilton}, \emph{{Linear Redshift Distortions: a Review}},  in
  \emph{The Evolving Universe}, D.~{Hamilton}, ed., vol.~231 of
  \emph{Astrophysics and Space Science Library}, p.~185, 1998,
  \href{https://doi.org/10.1007/978-94-011-4960-0_17}{DOI}
  [\href{https://arxiv.org/abs/astro-ph/9708102}{{\ttfamily
  astro-ph/9708102}}].

\bibitem{Kaiser84}
N.~{Kaiser}, \emph{{On the spatial correlations of Abell clusters}},
  \href{https://doi.org/10.1086/184341}{\emph{\apjl} {\bfseries 284} (1984)
  L9}.

\bibitem{Desjacques16}
V.~{Desjacques}, D.~{Jeong} and F.~{Schmidt}, \emph{{Large-Scale Galaxy Bias}},
  {\emph{ArXiv e-prints} (2016) }
  [\href{https://arxiv.org/abs/1611.09787}{{\ttfamily 1611.09787}}].

\bibitem{DESI2023b.KP1.EDR}
{DESI Collaboration}, A.G.~{Adame}, J.~{Aguilar}, S.~{Ahlen}, S.~{Alam},
  G.~{Aldering} et~al., \emph{{The Early Data Release of the Dark Energy
  Spectroscopic Instrument}},
  \href{https://doi.org/10.48550/arXiv.2306.06308}{\emph{arXiv e-prints} (2023)
  arXiv:2306.06308} [\href{https://arxiv.org/abs/2306.06308}{{\ttfamily
  2306.06308}}].

\bibitem{DESI2024.I.DR1}
{DESI Collaboration}, \emph{{DESI 2024 I: Data Release 1 of the Dark Energy
  Spectroscopic Instrument}}, {\emph{in preparation} (2025) }.

\bibitem{DESI2024.II.KP3}
{DESI Collaboration}, \emph{{DESI 2024 II: Sample definitions, characteristics
  and two-point clustering statistics}}, {\emph{in preparation} (2024) }.

\bibitem{DESI2024.III.KP4}
{DESI Collaboration}, A.G.~Adame, J.~Aguilar, S.~Ahlen, S.~Alam, D.M.~Alexander
  et~al., \emph{{DESI 2024 III: Baryon Acoustic Oscillations from Galaxies and
  Quasars}}, \href{https://doi.org/10.48550/arXiv.2404.03000}{\emph{arXiv
  e-prints} (2024) arXiv:2404.03000}
  [\href{https://arxiv.org/abs/2404.03000}{{\ttfamily 2404.03000}}].

\bibitem{DESI2024.V.KP5}
{DESI Collaboration}, \emph{{DESI 2024 V: Analysis of the full shape of
  two-point clustering statistics from galaxies and quasars}}, {\emph{in
  preparation} (2024) }.

\bibitem{DESI2024.IV.KP6}
{DESI Collaboration}, A.G.~Adame, J.~Aguilar, S.~Ahlen, S.~Alam, D.M.~Alexander
  et~al., \emph{{DESI 2024 IV: Baryon Acoustic Oscillations from the Lyman
  Alpha Forest}}, \href{https://doi.org/10.48550/arXiv.2404.03001}{\emph{arXiv
  e-prints} (2024) arXiv:2404.03001}
  [\href{https://arxiv.org/abs/2404.03001}{{\ttfamily 2404.03001}}].

\bibitem{DESI2024.VI.KP7A}
{DESI Collaboration}, A.G.~Adame, J.~Aguilar, S.~Ahlen, S.~Alam, D.M.~Alexander
  et~al., \emph{{DESI 2024 VI: Cosmological Constraints from the Measurements
  of Baryon Acoustic Oscillations}},
  \href{https://doi.org/10.48550/arXiv.2404.03002}{\emph{arXiv e-prints} (2024)
  arXiv:2404.03002} [\href{https://arxiv.org/abs/2404.03002}{{\ttfamily
  2404.03002}}].

\bibitem{DESI2024.VII.KP7B}
{DESI Collaboration}, \emph{{DESI 2024 VII: Cosmological constraints from
  full-shape analyses of the two-point clustering statistics measurements}},
  {\emph{in preparation} (2024) }.

\bibitem{DESI2024.VIII.KP7C}
{DESI Collaboration}, \emph{{DESI 2024 VIII: Constraints on Primordial
  Non-Gaussianities}}, {\emph{in preparation} (2024) }.

\bibitem{Simon23}
T.~{Simon}, P.~{Zhang}, V.~{Poulin} and T.L.~{Smith}, \emph{{Consistency of
  effective field theory analyses of the BOSS power spectrum}},
  \href{https://doi.org/10.1103/PhysRevD.107.123530}{\emph{\prd} {\bfseries
  107} (2023) 123530} [\href{https://arxiv.org/abs/2208.05929}{{\ttfamily
  2208.05929}}].

\bibitem{Linde24}
D.~{Linde}, A.~{Moradinezhad Dizgah}, C.~{Radermacher}, S.~{Casas} and
  J.~{Lesgourgues}, \emph{{CLASS-OneLoop: Accurate and Unbiased Inference from
  Spectroscopic Galaxy Surveys}}, {\emph{arXiv e-prints} (2024)
  arXiv:2402.09778} [\href{https://arxiv.org/abs/2402.09778}{{\ttfamily
  2402.09778}}].

\bibitem{KP5s5-Ramirez}
S.~Ramirez-Solano, M.~Icaza-Lizaola, H.E.~Noriega, M.~Vargas-Magaña,
  S.~Fromenteau, A.~Aviles et~al., \emph{Full modeling and parameter
  compression methods in configuration space for desi 2024 and beyond},
  \href{https://doi.org/10.48550/arXiv.2404.07268}{\emph{arXiv e-prints} (2024)
  arXiv:2404.07268} [\href{https://arxiv.org/abs/2404.07268}{{\ttfamily
  2404.07268}}].

\bibitem{Maksimova21}
N.A.~Maksimova, L.H.~Garrison, D.J.~Eisenstein, B.~Hadzhiyska, S.~Bose and
  T.P.~Satterthwaite, \emph{{AbacusSummit: a massive set of high-accuracy,
  high-resolution N-body simulations}},
  \href{https://doi.org/10.1093/mnras/stab2484}{\emph{Monthly Notices of the
  Royal Astronomical Society} {\bfseries 508} (2021) 4017}
  [\href{https://arxiv.org/abs/https://academic.oup.com/mnras/article-pdf/508/3/4017/40811763/stab2484.pdf}{{\ttfamily
  https://academic.oup.com/mnras/article-pdf/508/3/4017/40811763/stab2484.pdf}}].

\bibitem{Garrison21}
L.H.~Garrison, D.J.~Eisenstein, D.~Ferrer, N.A.~Maksimova and P.A.~Pinto,
  \emph{{The abacus cosmological N-body code}},
  \href{https://doi.org/10.1093/mnras/stab2482}{\emph{Monthly Notices of the
  Royal Astronomical Society} {\bfseries 508} (2021) 575}
  [\href{https://arxiv.org/abs/https://academic.oup.com/mnras/article-pdf/508/1/575/40458823/stab2482.pdf}{{\ttfamily
  https://academic.oup.com/mnras/article-pdf/508/1/575/40458823/stab2482.pdf}}].

\bibitem{ELGdesi}
A.~{Rocher}, V.~{Ruhlmann-Kleider}, E.~{Burtin}, S.~{Yuan}, A.~{de Mattia},
  A.J.~{Ross} et~al., \emph{{The DESI One-Percent survey: exploring the Halo
  Occupation Distribution of Emission Line Galaxies with ABACUSSUMMIT
  simulations}},
  \href{https://doi.org/10.1088/1475-7516/2023/10/016}{\emph{\jcap} {\bfseries
  2023} (2023) 016} [\href{https://arxiv.org/abs/2306.06319}{{\ttfamily
  2306.06319}}].

\bibitem{LRGQSOdesi}
S.~{Yuan}, H.~{Zhang}, A.J.~{Ross}, J.~{Donald-McCann}, B.~{Hadzhiyska},
  R.H.~{Wechsler} et~al., \emph{{The DESI One-Percent Survey: Exploring the
  Halo Occupation Distribution of Luminous Red Galaxies and Quasi-Stellar
  Objects with AbacusSummit}},
  \href{https://doi.org/10.48550/arXiv.2306.06314}{\emph{arXiv e-prints} (2023)
  arXiv:2306.06314} [\href{https://arxiv.org/abs/2306.06314}{{\ttfamily
  2306.06314}}].

\bibitem{KP5s2-Maus}
M.~Maus, S.~Chen, M.~White, J.~Aguilar, S.~Ahlen, A.~Aviles et~al., \emph{{An
  analysis of parameter compression and full-modeling techniques with
  Velocileptors for DESI 2024 and beyond}},
  \href{https://doi.org/10.48550/arXiv.2404.07312}{\emph{arXiv e-prints} (2024)
  arXiv:2404.07312} [\href{https://arxiv.org/abs/2404.07312}{{\ttfamily
  2404.07312}}].

\bibitem{KP5s3-Noriega}
H.E.~Noriega, A.~Aviles, H.~Gil-Marín, S.~Ramirez-Solano, S.~Fromenteau,
  M.~Vargas-Magaña et~al., \emph{Comparing compressed and full-modeling
  analyses with folps: Implications for desi 2024 and beyond},
  \href{https://doi.org/10.48550/arXiv.2404.07269}{\emph{arXiv e-prints} (2024)
  arXiv:2404.07269} [\href{https://arxiv.org/abs/2404.07269}{{\ttfamily
  2404.07269}}].

\bibitem{KP5s4-Lai}
Y.~Lai, C.~Howlett, M.~Maus, H.~Gil-Marín, H.E.~Noriega, S.~Ramírez-Solano
  et~al., \emph{{A comparison between Shapefit compression and Full-Modelling
  method with PyBird for DESI 2024 and beyond}},
  \href{https://doi.org/10.48550/arXiv.2404.07283}{\emph{arXiv e-prints} (2024)
  arXiv:2404.07283} [\href{https://arxiv.org/abs/2404.07283}{{\ttfamily
  2404.07283}}].

\bibitem{Chuang2015}
C.-H.~{Chuang}, F.-S.~{Kitaura}, F.~{Prada}, C.~{Zhao} and G.~{Yepes},
  \emph{{EZmocks: extending the Zel'dovich approximation to generate mock
  galaxy catalogues with accurate clustering statistics}},
  \href{https://doi.org/10.1093/mnras/stu2301}{\emph{\mnras} {\bfseries 446}
  (2015) 2621}.

\bibitem{Wadekar2020}
D.~{Wadekar} and R.~{Scoccimarro}, \emph{{Galaxy power spectrum multipoles
  covariance in perturbation theory}},
  \href{https://doi.org/10.1103/PhysRevD.102.123517}{\emph{\prd} {\bfseries
  102} (2020) 123517} [\href{https://arxiv.org/abs/1910.02914}{{\ttfamily
  1910.02914}}].

\bibitem{Angulo22}
R.E.~{Angulo} and O.~{Hahn}, \emph{{Large-scale dark matter simulations}},
  \href{https://doi.org/10.1007/s41115-021-00013-z}{\emph{Living Reviews in
  Computational Astrophysics} {\bfseries 8} (2022) 1}
  [\href{https://arxiv.org/abs/2112.05165}{{\ttfamily 2112.05165}}].

\bibitem{Grove22}
C.~{Grove}, C.-H.~{Chuang}, N.C.~{Devi}, L.~{Garrison}, B.~{L'Huillier},
  Y.~{Feng} et~al., \emph{{The DESI N-body simulation project - I. Testing the
  robustness of simulations for the DESI dark time survey}},
  \href{https://doi.org/10.1093/mnras/stac1947}{\emph{\mnras} {\bfseries 515}
  (2022) 1854} [\href{https://arxiv.org/abs/2112.09138}{{\ttfamily
  2112.09138}}].

\bibitem{Dodelson03}
S.~{Dodelson}, \emph{{Modern Cosmology}} (2003).

\bibitem{Donath+:2020}
Y.~Donath and L.~Senatore, \emph{{Biased Tracers in Redshift Space in the
  EFTofLSS with exact time dependence}},
  \href{https://doi.org/10.1088/1475-7516/2020/10/039}{\emph{JCAP} {\bfseries
  10} (2020) 039} [\href{https://arxiv.org/abs/2005.04805}{{\ttfamily
  2005.04805}}].

\bibitem{Fasiello+:2016}
M.~Fasiello and Z.~Vlah, \emph{{Nonlinear fields in generalized cosmologies}},
  \href{https://doi.org/10.1103/PhysRevD.94.063516}{\emph{Phys. Rev. D}
  {\bfseries 94} (2016) 063516}
  [\href{https://arxiv.org/abs/1604.04612}{{\ttfamily 1604.04612}}].

\bibitem{McQWhi16}
M.~{McQuinn} and M.~{White}, \emph{{Cosmological perturbation theory in 1+1
  dimensions}},
  \href{https://doi.org/10.1088/1475-7516/2016/01/043}{\emph{\jcap} {\bfseries
  1} (2016) 043} [\href{https://arxiv.org/abs/1502.07389}{{\ttfamily
  1502.07389}}].

\bibitem{Chen20}
S.-F.~{Chen}, Z.~{Vlah} and M.~{White}, \emph{{Consistent modeling of velocity
  statistics and redshift-space distortions in one-loop perturbation theory}},
  \href{https://doi.org/10.1088/1475-7516/2020/07/062}{\emph{\jcap} {\bfseries
  2020} (2020) 062} [\href{https://arxiv.org/abs/2005.00523}{{\ttfamily
  2005.00523}}].

\bibitem{Chen21}
S.-F.~{Chen}, Z.~{Vlah}, E.~{Castorina} and M.~{White}, \emph{{Redshift-space
  distortions in Lagrangian perturbation theory}},
  \href{https://doi.org/10.1088/1475-7516/2021/03/100}{\emph{\jcap} {\bfseries
  2021} (2021) 100} [\href{https://arxiv.org/abs/2012.04636}{{\ttfamily
  2012.04636}}].

\bibitem{McDonald_2009}
P.~McDonald and A.~Roy, \emph{Clustering of dark matter tracers: generalizing
  bias for the coming era of precision {LSS}},
  \href{https://doi.org/10.1088/1475-7516/2009/08/020}{\emph{Journal of
  Cosmology and Astroparticle Physics} {\bfseries 2009} (2009) 020}.

\bibitem{Hamann2010}
J.~{Hamann}, S.~{Hannestad}, J.~{Lesgourgues}, C.~{Rampf} and Y.Y.Y.~{Wong},
  \emph{{Cosmological parameters from large scale structure - geometric versus
  shape information}},
  \href{https://doi.org/10.1088/1475-7516/2010/07/022}{\emph{\jcap} {\bfseries
  2010} (2010) 022} [\href{https://arxiv.org/abs/1003.3999}{{\ttfamily
  1003.3999}}].

\bibitem{Wallisch2018}
B.~{Wallisch}, \emph{{Cosmological probes of light relics}}, Ph.D. thesis,
  University of Cambridge, UK, Jan., 2018.

\bibitem{ClassPT_2020}
A.~{Chudaykin}, M.M.~{Ivanov}, O.H.E.~{Philcox} and M.~{Simonovi{\'c}},
  \emph{{Nonlinear perturbation theory extension of the Boltzmann code CLASS}},
  \href{https://doi.org/10.1103/PhysRevD.102.063533}{\emph{\prd} {\bfseries
  102} (2020) 063533} [\href{https://arxiv.org/abs/2004.10607}{{\ttfamily
  2004.10607}}].

\bibitem{Chudaykin:2020ghx}
A.~Chudaykin, K.~Dolgikh and M.M.~Ivanov, \emph{{Constraints on the curvature
  of the Universe and dynamical dark energy from the Full-shape and BAO data}},
  \href{https://doi.org/10.1103/PhysRevD.103.023507}{\emph{Phys. Rev. D}
  {\bfseries 103} (2021) 023507}
  [\href{https://arxiv.org/abs/2009.10106}{{\ttfamily 2009.10106}}].

\bibitem{Perko2016}
A.~Perko, L.~Senatore, E.~Jennings and R.H.~Wechsler, \emph{{Biased Tracers in
  Redshift Space in the EFT of Large-Scale Structure}},
  \href{https://arxiv.org/abs/1610.09321}{{\ttfamily 1610.09321}}.

\bibitem{d_Amico_2020}
G.~d{\textquotesingle}Amico, J.~Gleyzes, N.~Kokron, K.~Markovic, L.~Senatore,
  P.~Zhang et~al., \emph{The cosmological analysis of the {SDSS}/{BOSS} data
  from the effective field theory of large-scale structure},
  \href{https://doi.org/10.1088/1475-7516/2020/05/005}{\emph{Journal of
  Cosmology and Astroparticle Physics} {\bfseries 2020} (2020) 005}.

\bibitem{Scoccimarro99}
R.~{Scoccimarro}, H.M.P.~{Couchman} and J.A.~{Frieman}, \emph{{The Bispectrum
  as a Signature of Gravitational Instability in Redshift Space}},
  \href{https://doi.org/10.1086/307220}{\emph{\apj} {\bfseries 517} (1999) 531}
  [\href{https://arxiv.org/abs/astro-ph/9808305}{{\ttfamily
  astro-ph/9808305}}].

\bibitem{Nishimichi20}
T.~{Nishimichi}, G.~{D'Amico}, M.M.~{Ivanov}, L.~{Senatore},
  M.~{Simonovi{\'c}}, M.~{Takada} et~al., \emph{{Blinded challenge for
  precision cosmology with large-scale structure: Results from effective field
  theory for the redshift-space galaxy power spectrum}},
  \href{https://doi.org/10.1103/PhysRevD.102.123541}{\emph{\prd} {\bfseries
  102} (2020) 123541} [\href{https://arxiv.org/abs/2003.08277}{{\ttfamily
  2003.08277}}].

\bibitem{Holm23}
E.B.~{Holm}, L.~{Herold}, T.~{Simon}, E.G.M.~{Ferreira}, S.~{Hannestad},
  V.~{Poulin} et~al., \emph{{Bayesian and frequentist investigation of prior
  effects in EFT of LSS analyses of full-shape BOSS and eBOSS data}},
  \href{https://doi.org/10.1103/PhysRevD.108.123514}{\emph{\prd} {\bfseries
  108} (2023) 123514} [\href{https://arxiv.org/abs/2309.04468}{{\ttfamily
  2309.04468}}].

\bibitem{Noriega:2022nhf}
H.E.~Noriega, A.~Aviles, S.~Fromenteau and M.~Vargas-Maga\~na, \emph{{Fast
  computation of non-linear power spectrum in cosmologies with massive
  neutrinos}}, \href{https://doi.org/10.1088/1475-7516/2022/11/038}{\emph{JCAP}
  {\bfseries 11} (2022) 038}
  [\href{https://arxiv.org/abs/2208.02791}{{\ttfamily 2208.02791}}].

\bibitem{Aviles:2020cax}
A.~Aviles and A.~Banerjee, \emph{{A Lagrangian Perturbation Theory in the
  presence of massive neutrinos}},
  \href{https://doi.org/10.1088/1475-7516/2020/10/034}{\emph{JCAP} {\bfseries
  10} (2020) 034} [\href{https://arxiv.org/abs/2007.06508}{{\ttfamily
  2007.06508}}].

\bibitem{Aviles:2021que}
A.~Aviles, A.~Banerjee, G.~Niz and Z.~Slepian, \emph{{Clustering in massive
  neutrino cosmologies via Eulerian Perturbation Theory}},
  \href{https://doi.org/10.1088/1475-7516/2021/11/028}{\emph{JCAP} {\bfseries
  11} (2021) 028} [\href{https://arxiv.org/abs/2106.13771}{{\ttfamily
  2106.13771}}].

\bibitem{Rodriguez-Meza:2023rga}
M.A.~Rodriguez-Meza, A.~Aviles, H.E.~Noriega, C.-Z.~Ruan, B.~Li,
  M.~Vargas-Maga\~na et~al., \emph{{fkPT: constraining scale-dependent modified
  gravity with the full-shape galaxy power spectrum}},
  \href{https://doi.org/10.1088/1475-7516/2024/03/049}{\emph{JCAP} {\bfseries
  03} (2024) 049} [\href{https://arxiv.org/abs/2312.10510}{{\ttfamily
  2312.10510}}].

\bibitem{McDonald:2006mx}
P.~McDonald, \emph{{Clustering of dark matter tracers: Renormalizing the bias
  parameters}}, \href{https://doi.org/10.1103/PhysRevD.74.129901}{\emph{Phys.
  Rev. D} {\bfseries 74} (2006) 103512}
  [\href{https://arxiv.org/abs/astro-ph/0609413}{{\ttfamily
  astro-ph/0609413}}].

\bibitem{Aviles:2020wme}
A.~Aviles, G.~Valogiannis, M.A.~Rodriguez-Meza, J.L.~Cervantes-Cota, B.~Li and
  R.~Bean, \emph{{Redshift space power spectrum beyond Einstein-de Sitter
  kernels}}, \href{https://doi.org/10.1088/1475-7516/2021/04/039}{\emph{JCAP}
  {\bfseries 04} (2021) 039}
  [\href{https://arxiv.org/abs/2012.05077}{{\ttfamily 2012.05077}}].

\bibitem{Senatore:2014via}
L.~Senatore and M.~Zaldarriaga, \emph{{The IR-resummed Effective Field Theory
  of Large Scale Structures}},
  \href{https://doi.org/10.1088/1475-7516/2015/02/013}{\emph{JCAP} {\bfseries
  02} (2015) 013} [\href{https://arxiv.org/abs/1404.5954}{{\ttfamily
  1404.5954}}].

\bibitem{Ivanov:2018gjr}
M.M.~Ivanov and S.~Sibiryakov, \emph{{Infrared Resummation for Biased Tracers
  in Redshift Space}},
  \href{https://doi.org/10.1088/1475-7516/2018/07/053}{\emph{JCAP} {\bfseries
  07} (2018) 053} [\href{https://arxiv.org/abs/1804.05080}{{\ttfamily
  1804.05080}}].

\bibitem{ChenVlahWhite20}
S.-F.~{Chen}, Z.~{Vlah} and M.~{White}, \emph{{Consistent modeling of velocity
  statistics and redshift-space distortions in one-loop perturbation theory}},
  \href{https://doi.org/10.1088/1475-7516/2020/07/062}{\emph{\jcap} {\bfseries
  2020} (2020) 062} [\href{https://arxiv.org/abs/2005.00523}{{\ttfamily
  2005.00523}}].

\bibitem{Chan:2012jj}
K.C.~Chan, R.~Scoccimarro and R.K.~Sheth, \emph{{Gravity and Large-Scale
  Non-local Bias}},
  \href{https://doi.org/10.1103/PhysRevD.85.083509}{\emph{Phys. Rev. D}
  {\bfseries 85} (2012) 083509}
  [\href{https://arxiv.org/abs/1201.3614}{{\ttfamily 1201.3614}}].

\bibitem{Baldauf:2012hs}
T.~Baldauf, U.~Seljak, V.~Desjacques and P.~McDonald, \emph{{Evidence for
  Quadratic Tidal Tensor Bias from the Halo Bispectrum}},
  \href{https://doi.org/10.1103/PhysRevD.86.083540}{\emph{Phys. Rev. D}
  {\bfseries 86} (2012) 083540}
  [\href{https://arxiv.org/abs/1201.4827}{{\ttfamily 1201.4827}}].

\bibitem{Saito:2014qha}
S.~Saito, T.~Baldauf, Z.~Vlah, U.~Seljak, T.~Okumura and P.~McDonald,
  \emph{{Understanding higher-order nonlocal halo bias at large scales by
  combining the power spectrum with the bispectrum}},
  \href{https://doi.org/10.1103/PhysRevD.90.123522}{\emph{Phys. Rev.}
  {\bfseries D90} (2014) 123522}
  [\href{https://arxiv.org/abs/1405.1447}{{\ttfamily 1405.1447}}].

\bibitem{Fujita_2020}
T.~Fujita and Z.~Vlah, \emph{Perturbative description of biased tracers using
  consistency relations of {LSS}},
  \href{https://doi.org/10.1088/1475-7516/2020/10/059}{\emph{Journal of
  Cosmology and Astroparticle Physics} {\bfseries 2020} (2020) 059}.

\bibitem{EH1998}
D.J.~{Eisenstein} and W.~{Hu}, \emph{{Baryonic Features in the Matter Transfer
  Function}}, \href{https://doi.org/10.1086/305424}{\emph{\apj} {\bfseries 496}
  (1998) 605} [\href{https://arxiv.org/abs/astro-ph/9709112}{{\ttfamily
  astro-ph/9709112}}].

\bibitem{Hinton2017}
S.R.~{Hinton}, E.~{Kazin}, T.M.~{Davis}, C.~{Blake}, S.~{Brough}, M.~{Colless}
  et~al., \emph{{Measuring the 2D baryon acoustic oscillation signal of
  galaxies in WiggleZ: cosmological constraints}},
  \href{https://doi.org/10.1093/mnras/stw2725}{\emph{\mnras} {\bfseries 464}
  (2017) 4807} [\href{https://arxiv.org/abs/1611.08040}{{\ttfamily
  1611.08040}}].

\bibitem{Briedan21}
S.~{Brieden}, H.~{Gil-Mar{\'\i}n} and L.~{Verde}, \emph{{ShapeFit: extracting
  the power spectrum shape information in galaxy surveys beyond BAO and RSD}},
  \href{https://doi.org/10.1088/1475-7516/2021/12/054}{\emph{\jcap} {\bfseries
  2021} (2021) 054} [\href{https://arxiv.org/abs/2106.07641}{{\ttfamily
  2106.07641}}].

\bibitem{Alcock79}
C.~{Alcock} and B.~{Paczynski}, \emph{{An evolution free test for non-zero
  cosmological constant}}, \href{https://doi.org/10.1038/281358a0}{\emph{\nat}
  {\bfseries 281} (1979) 358}.

\bibitem{Diego_Blas_2011}
D.~Blas, J.~Lesgourgues and T.~Tram, \emph{The cosmic linear anisotropy solving
  system ({CLASS}). part {II}: Approximation schemes},
  \href{https://doi.org/10.1088/1475-7516/2011/07/034}{\emph{Journal of
  Cosmology and Astroparticle Physics} {\bfseries 2011} (2011) 034}.

\bibitem{Lewis_2011}
A.~{Lewis} and A.~{Challinor}, ``{CAMB: Code for Anisotropies in the Microwave
  Background}.'' Astrophysics Source Code Library, record ascl:1102.026, Feb.,
  2011.

\bibitem{Carrilho23}
P.~{Carrilho}, C.~{Moretti} and A.~{Pourtsidou}, \emph{{Cosmology with the
  EFTofLSS and BOSS: dark energy constraints and a note on priors}},
  \href{https://doi.org/10.1088/1475-7516/2023/01/028}{\emph{\jcap} {\bfseries
  2023} (2023) 028} [\href{https://arxiv.org/abs/2207.14784}{{\ttfamily
  2207.14784}}].

\bibitem{DESI2023a.KP1.SV}
{DESI Collaboration}, A.G.~{Adame}, J.~{Aguilar}, S.~{Ahlen}, S.~{Alam},
  G.~{Aldering} et~al., \emph{{Validation of the Scientific Program for the
  Dark Energy Spectroscopic Instrument}},
  \href{https://doi.org/10.3847/1538-3881/ad0b08}{\emph{\aj} {\bfseries 167}
  (2024) 62} [\href{https://arxiv.org/abs/2306.06307}{{\ttfamily 2306.06307}}].

\end{thebibliography}\endgroup
\bibliographystyle{jhep}

\end{document}